\documentclass{aastex631}
\usepackage{rotating}
\usepackage{booktabs}

\begin{document}

\title{Starspot distribution  and flare events in two young  low-mass stars using TESS data}

\author[0000-0001-7277-2577]{Rajib Kumbhakar}
\affiliation{S.N. Bose National Centre for Basic Sciences, Kolkata 700106}

\author[0000-0003-1457-0541]{Soumen Mondal}
\affiliation{S.N. Bose National Centre for Basic Sciences, Kolkata 700106}


\author[0000-0003-3354-850X]{Samrat Ghosh}
\affiliation{Aryabhatta Research Institute of Observational Sciences (ARIES), 
Nainital 263001, India}

\author[0009-0008-7884-3741]{Diya Ram}
\affiliation{S.N. Bose National Centre for Basic Sciences, Kolkata 700106}

\begin{abstract}
Wide-field high-precision photometric observations such as \textit{Transiting Exoplanet Survey Satellite (TESS)} allowed the investigation of the stellar magnetic activity of cool stars. M-dwarf's starspots and stellar flares are the main indicators of magnetic activity. The present study focuses on modeling light curves (LCs) to analyze the distribution and characteristics of starspots e.g., location, temperature, and spot size. The \textit{TESS} light curves of two selected young M-dwarfs i.e. GJ~182 and 2MASS~J05160212+2214528 were reconstructed using the \textsc{BASSMAN} software, obtaining a  three-spot model for GJ~182 and two-spot model for  2MASS~J05160212+2214528, describing their light curves. For GJ~182, the mean spot temperature was estimated to be  approximately 3279~K, covering  5-8.5\% of the stellar surface while for 2MASS~J05160212+2214528 the average spot temperature was approximately  2631~K, with a mean spottedness of about  5.4\%. Using the 2-min cadence LC data, we identified and analyzed  48 flare events from GJ~182, while no flares were detected in 2MASS~J05160212+2214528. The estimated bolometric flare energy ranged from $10^{32} - 10^{35}$ erg, and 10$^{31}$ - 10$^{33}$ erg in the TESS bandpass. We derived the power-law index of  -1.53 $\pm$ 0.12 and -1.86 $\pm$ 0.22 for flare frequency distribution in sectors 5 and 32  respectively in the flare energy 10$^{33}$ to 10$^{35}$ erg, consistent with previous studies for M-dwarfs. A positive linear correlation between flare energy and duration was found with a slope of $0.67 \pm 0.02$, suggesting a similar mechanism followed by stellar superflares and solar flares. By assuming the similarities with solar flares, we also estimated the lower limit of the magnetic field strength around  12 - 232~G to produce such superflare events.
\end{abstract}

\keywords{}

\section{Introduction} \label{sec:intro}
Starspots are local concentrations of magnetic fields on the stellar surface just like Sunspots. Starspots are cool and dark regions and are generated due to the stellar dynamo \citep{Davenport2015PhDT.......177D}. The starspots' formation is due to the local suppression of convective motion by the magnetic flux tubes that block or redirect the energy flow to the surface and as a result, the region appears cool and darker than the bright photosphere (see reviews, \citealt{Strassmeier2009A&ARv..17..251S}). In this area magnetic flux tubes aligned almost vertically \citep{Davenport2015PhDT.......177D}.  Generally, M-dwarfs are magnetically active objects, exhibiting a complex and multi-scale nature of magnetic fields. This complexity arises from their fully convective nature in the case of later-type M-dwarfs (spectral type later than M4.0) or the presence of convective layers in the upper parts of their interiors for early-type M-dwarfs \citep{Kochukhov2021A&ARv..29....1K}.  Being magnetically active plays a vital role in generating the stellar spots in young M-dwarfs and triggers magnetic reconnection, which serves as a primary mechanism of the energy release during stellar flares \citep{Lin2019ApJ...873...97L, Bicz2022ApJ...935..102B}. Therefore, starspots analysis can provide some insights into the internal dynamo activity as well as the magnetic field structure of the stars \citep{Strassmeier2009A&ARv..17..251S}.
We can also measure precisely the stellar rotation period by photometric analysis of the light curve \citep{Strassmeier2009A&ARv..17..251S, Davenport2015PhDT.......177D}. The larger and darker stellar spots generate light curves with larger amplitude and vice-versa. \citet{Strassmeier2009A&ARv..17..251S} pointed out that so far, more than 200 billion stars have been spotted stars in our galaxy and most of them are not detected using current techniques.

 Stellar flares are intense, fast-occurring phenomena triggered by magnetic reconnection in stellar coronae and dramatic releases of magnetic energy in the form of electromagnetic radiation across all wavelength \citep{Ilin2022MNRAS.513.4579I, Pietras2022ApJ...935..143P}. The flares in low-mass M-dwarfs are analogous to solar flares and originate due to analogous magnetic reconnection events \citep{Pettersen_1989}. But flares in M-dwarfs are more energetic and frequent than the Sun-like stars because the Sun-like stars spin down quickly whereas M-dwarfs are rapid rotators for longer time \citep{Irwin2011ApJ...727...56I, Newton2016ApJ...821...93N}.  Moreover, due to the low surface temperature of these M-type stars weak flares are easily detectable while detecting weak flares in Sun-like stars is very challenging due to their higher surface temperature and the parameters of weak flares on M-dwarfs are very much similar to the solar flares \citep{Pietras2023ApJ...954...19P}. This rapid rotation can also enhance the magnetic activity and trigger more powerful and frequent flare events in M-dwarfs \citep{Ilin2021JOSS....6.2845I}. M-dwarfs often emit stronger superflares than typical solar flares with total energy $\geq$ 10$^{34}$ erg \citep{Tu2020ApJ...890...46T}. Previously, \citet{Ilin2021JOSS....6.2845I} found that flares happened in the relatively higher latitude compared to the Sun for fully convective stars which indicates the magnetic field concentrates to the stellar rotational poles. 

We can not spatially resolve the starspots for the distant object, so analyzing the variation of light curves can give hints about the presence of starspots on the stellar surface. However, only examining the light curve is not enough to extract information of the location of the starspots. Therefore, it is necessary to model the light curve to get the distribution of the starspots, which are the main source of the periodic modulation of the light curve. In this work, we focus on modeling the starspots in young M-dwarfs to investigate their spatial distribution and its implications for stellar activity. The starspots' size, temperature, numbers and location could be the main indicator to understand how the superflares are triggered \citep{Namekata2019ApJ...871..187N, Namekata2022NatAs...6..241N}. The temporal evolution of spots can guide the temporal change of magnetic field structure on the stellar surface \citep{Namekata2019ApJ...871..187N}. In the earlier studies, several researchers investigated the starspots activity and lifetime of starspots of active young stars, cool stars and RS CVn-type stars through ground-based and space-based observations \citep{Henry1995ApJS...97..513H, Messina2002A&A...393..225M, Maehara2021PASJ...73...44M, Namekata2019ApJ...871..187N}. They found that the lifetimes of starspots are proportional to the starspots area for small spots domain and for the domain in larger spots lifetime decreases as spot area increases, because of differential rotation \citep{Henry1995ApJS...97..513H}. Moreover, the \textsc{Kepler} light curve of GJ~1243 was modeled by two starspot models where one spot is located at a higher latitude and another at the stellar equator \citep{Davenport2015ApJ...806..212D}. From the \textsc{TESS} light curve of GJ~1243, YZ CMi and V374 Peg, \citet{Bicz2022ApJ...935..102B} also performed starspot modeling using \textsc{BASSMAN} package and compared the results with previous analysis. \citet{Ikuta2023ApJ...948...64I} conducted starspot modeling using the \textsc{TESS} light curves of three M-dwarfs i.e., AU Mic, YZ CMi and EV Lac using adaptive parallel tempering algorithm.  There are various methods to analyze the starspots distribution. Using direct imaging observation from the Hubble Space Telescope (HST) of the Betelgeuse ($\alpha$ Orionis) \citet{Gilliland1996ApJ...463L..29G} found a bright spot in the Ultraviolet wavelengths. Apart from this, microlensing observations \citep{Heyrovsky2000ApJ...529...69H, Hendry2002MNRAS.335..539H}, Doppler Imaging \citep{Strassmeier2009A&ARv..17..251S}, Polarimetry method \citep{Valenti1995ApJ...439..939V, Johns-Krull1996ApJ...459L..95J} are also used to probe the starspots on the surfaces of stars. 

The objects in this study, GJ~182, and  2MASS~J05160212+221452 (hereafter, 2M0516+2214), are both young M-dwarf stars. GJ~182 is a M0.5  dwarf star \citep{Torres2006A&A...460..695T} located at a distance of 26.7 $\pm$1.7 pc \citep{Donati2008MNRAS.390..545D}. It has a mass of 0.60M$_{\odot}$, a radius of 0.87R$_{\odot}$ and an effective temperature of 3866 $\pm$ 143 K \citep{Stassun2019AJ....158..138S}. Moreover, it has a large lithium abundance, suggesting it is a very young object of age around 20 Myrs, and an active star with a high surface magnetic field \citep{Favata1998A&A...335..218F}. 2M0516+2214is also a young Taurus member with spectral type of M4.5, initially identified  by \citet{Slesnick2006AJ....132.2665S}.  More details are mentioned in the Table~\ref{tab:obj_det}. The selection of these objects is based on the smooth variation in the TESS light curve. GJ~182 has shown a strong starspot signature as well as a flare. Unfortunately, 2M0516+2214 has no flare events but displayed a smooth and variable phase light curve, which strongly indicates the presence of starspot signatures.

In this paper, we presented the spatial distribution of starspots of two young M-dwarfs, GJ~182 and 2MASS~J05160212+2214528 (object details in Table~\ref{tab:obj_det}) for the first time and identified flare events of GJ~182 using TESS photometry. This paper is organized as follows, in Section~\ref{sec:TODA}, we describe the TESS observation and measuring the rotation period, flare detection method, calculation of flare energies, and starspot modeling to the light curves. The obtained results of starspot modeling are described in Section~\ref{sec:result} and we also presented the analysis of flare. Finally, we discussed this paper in Section~\ref{sec:diss} and summarised in Section~\ref{sec:summ}.

\begin{deluxetable*}{cccc}
\tablecaption{Properties of the studied objects}
\tablehead{
\colhead{Parameters} & \colhead{Object~1}  & \colhead{Object~2} & \colhead{References} \\
}
\startdata
TIC ID & 452763353 & 5800708  & - \\ 
Other Name & GJ~182 & 2MASS~J05160212+2214528 & -\\
RA (hh:mm:sec) & $04:59:34.8$ & $05:16:02.1$ & \citep{Cutri2003yCat.2246....0C} \\ 
Dec (deg:mm:sec) & $+01:47:00.7$ & $+22:14:52.8$ & \citep{Cutri2003yCat.2246....0C}\\ 
TESS Sector & 5,32 & 43,44,45 & - \\
TESS Cadence & 120s & 120s & -  \\
SpT & M0.5 & M4.5 & \citep{Donati2008MNRAS.390..545D,Slesnick2006AJ....132.2665S}\\
Distance (pc) & 26.7 $\pm$ 1.7 & 181.68 & \citep{Donati2008MNRAS.390..545D, Bailer-Jones2018AJ....156...58B}\\
Temperature (K) & 3866 $\pm$ 143 & 3025 $\pm$ 122 & \citep{Stassun2019AJ....158..138S}\\
Radius (R$_{\odot}$) & 0.87 & 0.82 & \citep{Stassun2019AJ....158..138S, Herczeg2014ApJ...786...97H}\\
Mass~(M$_{\odot}$) & 0.60 & 0.09 &  \citep{Stassun2019AJ....158..138S}\\  
 Age (Myr) &  25 &---&  \citep{Donati2008MNRAS.390..545D}\\
$ vsini (km/s)$ &  10 &  15.7 &  \citep{Donati2008MNRAS.390..545D,Kraus2017ApJ...838..150K} \\
inclination angle (deg) & 60 & 24.48$^{*}$  & \citep{Donati2008MNRAS.390..545D}\\
\enddata
\tablecomments{* computed using rotation period, radius and rotational velocity ($vsini$)}
\label{tab:obj_det}
\end{deluxetable*}

\section{TESS Observations and Data Analysis}\label{sec:TODA}
In this paper, we studied the stellar activities of the selected objects (Table \ref{tab:obj_det}) using the time-series photometry data from TESS due to its high precision, long duration, and continuity. TESS is a space-based NASA Explorer program telescope, launched in April 2018 and it is an all-sky transit survey equipped with four 10.5 cm telescopes with a combined field of view 24 $\times$ 96 degrees, also known as a sector (see \citealt{Ricker2015JATIS...1a4003R} for details). 
A given sector is observed approximately in 27 days with roughly a day gap. The primary mission of TESS, completed in July 2020, covered 26 sectors in both hemispheres that cover about 85\% of the sky. The next 29 sectors (from 27 to 55) are observed in the first extended part of the mission. During the primary mission, two cadences were realized: a 2-minute short cadence and a 30-minute long cadence for full-frame images (FFIs). But during the first extended part of the mission, two short-time cadences, 2-min, and 20-s for a few selected sources, while a 10-min cadence is available for FFIs with an angular resolution of 21 arcsecs per pixel. TESS covers the wavelength range from 600-1000 $nm$ centered at 786.5 $nm$ with slightly redder than the Kepler band. We have selected two young M-dwarfs and analyzed the TESS 2-min cadence data. TESS observed 2M0516+2214 in sectors 43, 44, and 45 while 5 and 32 sectors were used to observe the GJ~182. Using the TESS Science Processing Operations Center pipeline (SPOC; \citealp{Jenkins2016SPIE.9913E..3EJ}), light curves are automatically generated for all 2-min cadence TESS targets and made publicly available on Mikulski Archive for Space Telescopes (MAST) \footnote{https://mast.stsci.edu/portal/Mashup/Clients/Mast/Portal.html}. The light curve data products of TESS contain both Simple Aperture Photometry (SAP) and Pre-Search Data Conditioned (PDCSAP) flux data. Here, we used the `lightkurve' \citep{Lightkurve_Collaboration2018ascl.soft12013L} package to download the TESS light curve from MAST. In our analysis, we used PDCSAP data as PDCSAP light curves are free from instrumental effects and long-term trends due to possible systematic effects. Also, the SPOC pipeline corrects for the dilution from other nearby stars in and around the TESS aperture, as denoted by the CROWDSAP value in the TESS header files. The values of CROWDSAP indicate how much flux is due to the object only in the selected aperture. In addition, we chose to use only the data with QUALITY= 0 as the non-zero value of QUALITY denotes the data has been compromised to some degree by instrumental effects. We also filtered the data by removing NaNs. Furthermore, we removed the outliers and normalized each light curve by dividing each flux by the target's mean flux. 

\subsection{Measuring Rotation Periods in TESS }
Stellar rotation is an important physical characteristic for understanding the physical properties of individual stars and their populations. Rotation drives the stellar dynamo which can give rise to stellar activity e.g. starspots, and flares \citep{Choudhuri2017SCPMA..60a9601C}. Such starspots also co-rotate with the stars into and out of view, producing periodicity in the light curve. If the starspots do not evolve in time, then we would get a perfect periodicity over a full rotation of stars. But starspots are not stationary; rather they change shape, appear, and then vanish over time as the star rotates \citep{Namekata2019ApJ...871..187N}. As a result, the light curves exhibit a quasi-periodic variability rather than a fully periodic variation.
This implies that it is difficult to infer rotation periods using only straightforward sinusoidal variability models which are precise up to a limit. In this work, we used two different non-inference-based methods to estimate the possible periodicity in the light curves such as Lomb-Scragle periodograms \citep{Lomb1976Ap&SS..39..447L}, and Gaussian process \citep{Angus2018MNRAS.474.2094A, Scargle1982ApJ...263..835S}.  Here we briefly discussed the two methods below: \\
(i) \textit{Lomb-Scargle (LS) Periodogram}: The Lomb-Scargle (LS) \citep{Lomb1976Ap&SS..39..447L, Scargle1982ApJ...263..835S} is a well-known algorithm in observational astronomy to detect and characterize periodic signals from unevenly sample time-series data. It uses a Fourier-like power spectrum estimator, in which the time series is decomposed into a linear combination of sinusoidal functions, and the data is transformed from the time domain to the frequency domain based on sinusoidal functions. We estimated the rotation periods using the LS method implemented by NASA Exoplanet Archive Periodogram Service\footnote{https://exoplanetarchive.ipac.caltech.edu/cgi-bin/Pgram/nph-pgram}\citep{Akenson2013PASP..125..989A}. Additionally, we also looked for periods using Astropy \citep{Astropy_collaboration2013A&A...558A..33A, astropy2018AJ....156..123A} package. The power spectrums of the selected objects were shown in the middle panel in Fig \ref{fig:lc_ls_phs} and Fig \ref{fig:2214_lc_ls_phs}.

\begin{figure*}
    \centering
    \includegraphics[width=.45\linewidth]{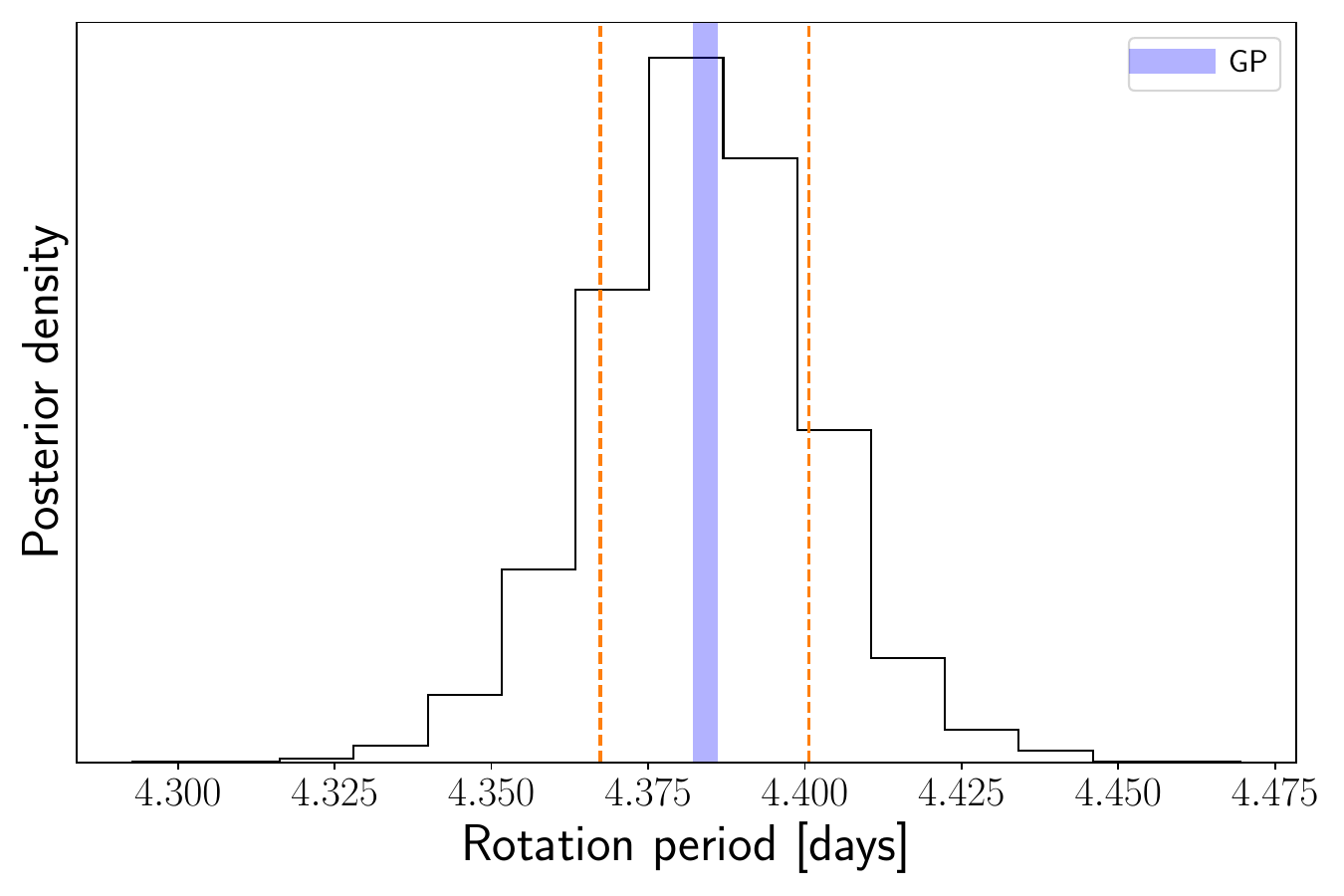}
    \quad
    \includegraphics[width=.45\linewidth]{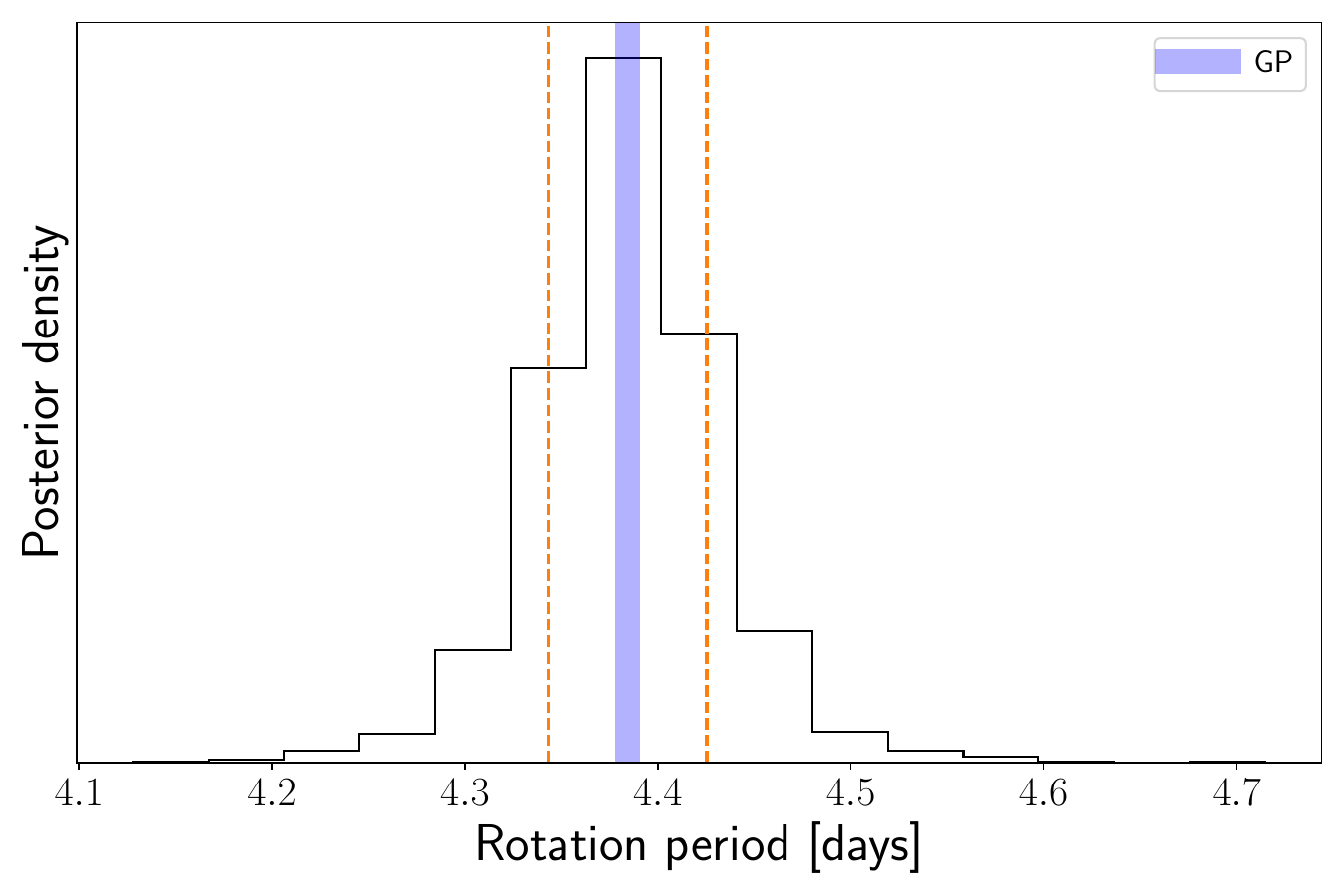}   
    \caption{Posterior models of rotation periods of GJ~182, measured from TESS 2-min cadence data in sector~5 (left) and sector~32 (right) are shown here. The violet-blue line represents the rotation periods and the orange lines are the uncertainties of the periods.}
    \label{fig:gp_182}
\end{figure*}

(ii) \textit{Guassian Process (GP) Regression}: We also estimated the rotation periods using the GP method as described in \citet{Angus2018MNRAS.474.2094A}. This method is slower than other approaches (e.g. LS, PDM, etc.) to infer the rotation period, but the GP method can deal with the non-sinusoidal, unevenly spaced, and more complex variation in the time-series data. It offers a probabilistic framework for rotation period estimation. In addition, it allows us to assess the posterior distribution across time and, as a result, to get useful error estimates. Furthermore, \citet{Angus2018MNRAS.474.2094A} tested GP and found that it provides slightly more accurate rotation periods 
than the periodogram or autocorrelation function methods and can be applied to non-uniformly sampled time-series data. We used \texttt{STARSPOT}\footnote{https://github.com/RuthAngus/starspot}, a \textsc{Python} module to measure the rotation period using Gaussian Process. It used fast and scalable \texttt{exoplanet} \citep{Foreman-Mackey2021exoplanet:joss} and \texttt{celerite} \citep{Foreman-Mackey2017celerite1, Foreman-Mackey2018celerite2} package to fit stellar variability and modeled the rotation periods in each TESS sectors individually. Here, \texttt{PyMC3} supports a variety of general GP models. Then the systematic-corrected light curves using a RotationTerm Gaussian Process kernel.
As an example, we have shown a posterior distribution of the rotation period for GJ 182, measured from TESS time-series data of sectors 5 and 32 in figure \ref{fig:gp_182}. The estimated rotation periods using the GP method for our selected objects are consistent with the rotation periods derived from other methods. For more details on Gaussian processes in the field of astronomy, see \citet{Foreman_guassian_2017AJ....154..220F}. The rotation periods are tabulated in Table \ref{tab:rot_per} and we adopted the rotation periods estimated using the GP method.

\begin{deluxetable*}{cccccccccccccc}
\tablecaption{Rotation periods in days from TESS data}
\tablehead{
\colhead{Object} & \colhead{Sector}  & \multicolumn{2}{c}{Rotation Periods (this work)} &\multicolumn{6}{c}{Rotaion period (previous studies)}\\
\cline{3-4}
\cline{6-14}
\colhead{} & \colhead{} & \colhead{Period1} & \colhead{Period2} & \colhead{} &\colhead{Wright} &\colhead{Messian}  & \colhead{Magaudda} &\colhead{Bustos} & \colhead{Vach} & \colhead{Donati} & \colhead{Vidotto} & \colhead{Kiraga} &\colhead{Byrne}\\
}
\startdata
GJ 182 & 05 & 4.370 & $4.348 \pm 0.016$ & & 1.86 & 4.43 & 1.86 & 4.41 & 4.4 & 4.35 & 4.35 & 4.41 & 4.56 \\
       & 32 & 4.400 & $4.384 \pm 0.042$   \\
\hline
2M0516+2214 & 43 & 1.103 & $1.101 \pm 0.002$ \\
            & 44 & 1.102 & $1.101 \pm 0.001$  \\
            & 45 & 1.102 & $1.101 \pm 0.004$  \\
\enddata
\tablecomments{{Period1}=Periods determined using {\sc NASA Exoplanet Archive Periodogram Service}. ~
{Period2}= Periods computed using the Gaussian Process. Wright=\citet{Wright2011ApJ...743...48W}, Messina=\citet{Messina2017}, Magaudda=\citet{Magaudda}, Bustos=\citet{bustos2023}, Vach=\citet{Vach2024AJ....167..210V}, Donati=\citet{Donati2008MNRAS.390..545D}, Viddotto=\citet{Vidotto2014MNRAS.441.2361V}.
Kiraga=\citet{Kiraga2007AcA....57..149K}, Byrne=\citet{Byrne1984MNRAS.206..907B}.}
\label{tab:rot_per}
\end{deluxetable*}

\subsection{Flare Detection Method}
We used \textsc{ALTAIPONY} \footnote{https://github.com/ekaterinailin/AltaiPony} \citep{Ilin2021JOSS....6.2845I}, an open source \textit{python} based package to detect the flare events from the \textit{TESS} and \textit{Kepler} lightcurves and to estimate the flare parameters of each event. Initially, we applied a detrending approach to remove astrophysical trend-like modulation due to the starspots and instrumental trends from the time series data. To detrend the PDCSAP lightcurves, we used a custom\_detrending method\footnote{https://altaipony.readthedocs.io/en/latest/tutorials/detrend.html}. A third-order spline function is fitted through non-gap portions of the light curve and subtracted, which removes long-term trends and starspot variability. Then, the strong sinusoidal signal is removed iteratively from the light curves.  This iteration process first masks outliers points using the $\sigma$-clipping method. When the light curve showed a single outlier above 3$\sigma$, it treated the data point as a pure outlier. At the same time, there is a number of series of data points present above 3$\sigma$, it masked as a flare candidate. LS periodogram is calculated in each iteration. A cosine function fit is generated using the dominant frequency determined through a least-square procedure. The light curve was then subtracted from this cosine fit, and this procedure continued until the dominant peak's signal-to-noise ratio (SNR) decreased to 1. Finally, any non-sinusoidal variability was eliminated using two third-order Savitzky-Golay filters \citep{Savitzkydoi:10.1021/ac60214a047}  with window size 6 hours and 3 hours respectively.  To detrend the light curve in sector~5 we used cubic spline with spline\_coarseness value of 8 hours. Spline\_coarseness is how densely the spline points are distributed in the time-series data. For sector~32 we apply the cubic spline order with spline coarseness of 6 hours. In both sectors, we used the 3$\sigma$ threshold for outlier rejection as \textit{`max\_sigma'} in the custom detrending method. and such method is also described in \citet{Ilin2022MNRAS.513.4579I}.

\begin{figure*}
    \centering
    \includegraphics[width=.48\linewidth]{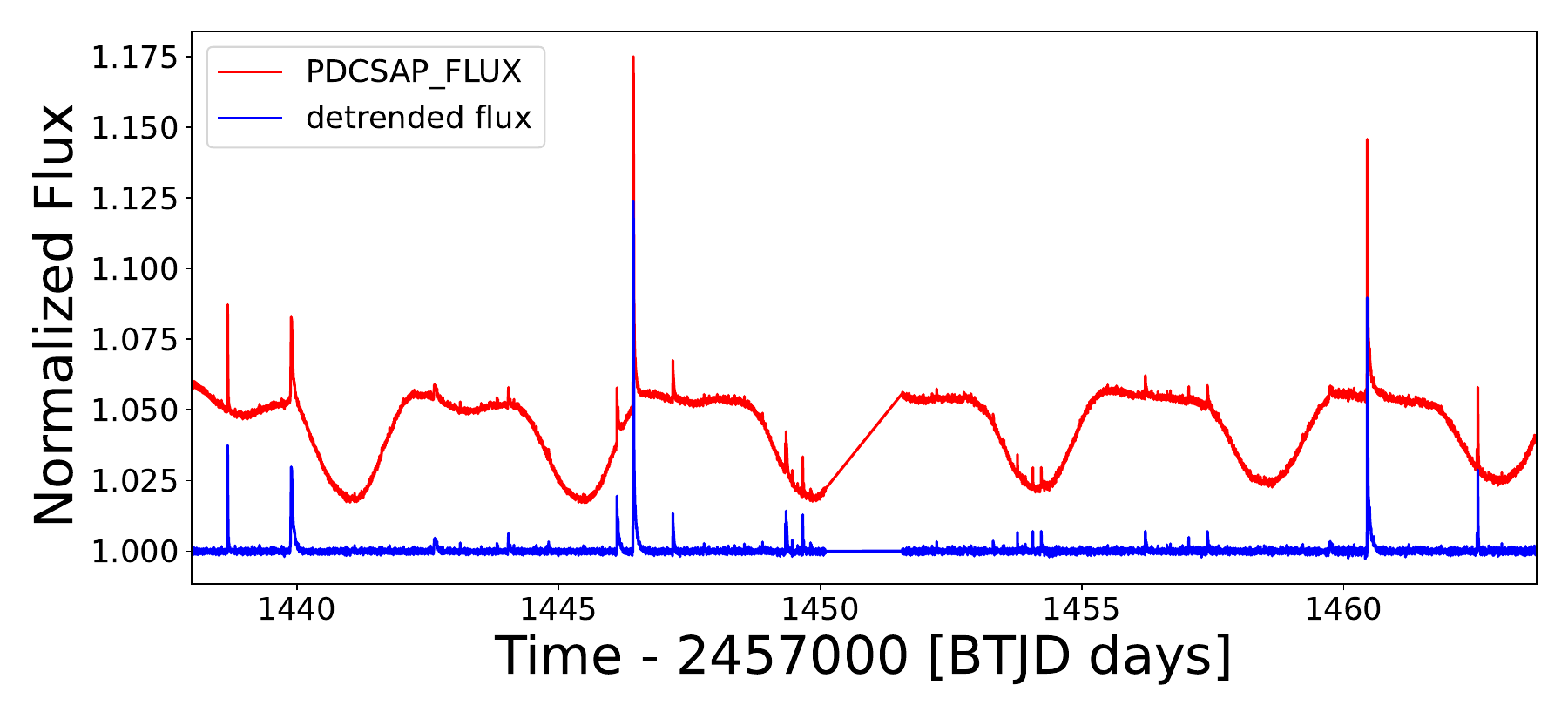}
    \quad
    \includegraphics[width=.48\linewidth]{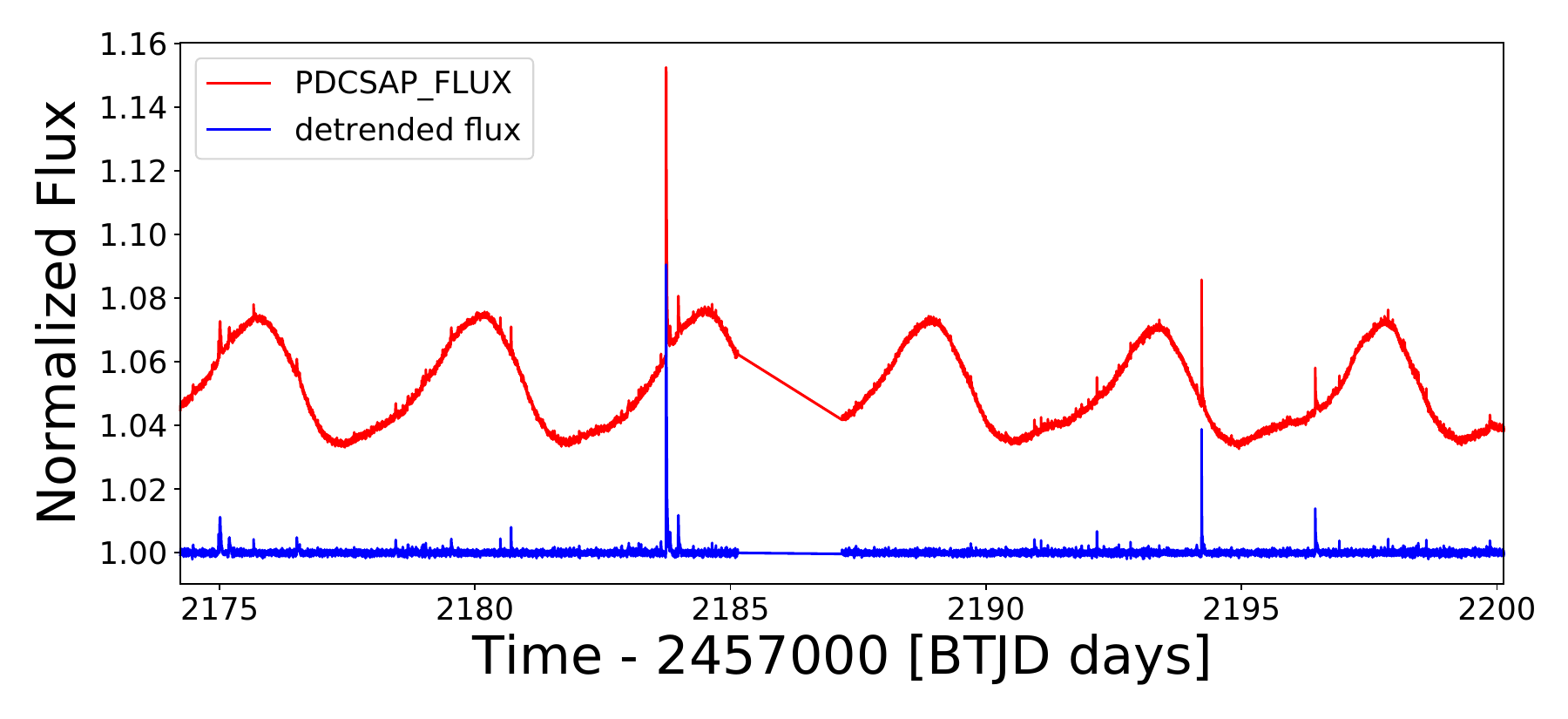}   
    \caption{The PDCSAP and detrended light curves of GJ~182 for sector~5 (left) and sector~32 (right). The x-axis represents time, in Barycentric TESS Julian Days (BTJD), and the y-axis represents the normalized TESS flux. PDCSAP data are shown in red and the detrended flux data is shown in blue.}
    \label{fig:dtrn_flc}
\end{figure*}

To identify the flare events, we followed a similar method as described in \citet{Chang2015ApJ...814...35C}. \textsc{Altaipony} uses \textit{FlareLightCurve.find\_flares()} to identify flare candidate data points. This process involved the use of three parameters, namely N1, N2, and N3, with values set to 3, 2, and 3, respectively, establishing the criteria for flare event detection. Following detrending and flare identification in \textsc{Altaipony}, a visual inspection of all flare lightcurves was conducted to check for any inconsistency. The key characteristics considered were a distinct sharp rise and slow exponential decay. Furthermore, we again zoomed in on the particular flare region and checked visually that three consecutive data points must surpass the $3\sigma$ threshold of the light curve 
and be spared at least 6 minutes above it. We excluded those flare events which do not meet this criteria. The detrended light curves of GJ 182 in both sectors are shown in figure~\ref{fig:dtrn_flc}. In this study, we have detected  48 flare events of GJ 182 in both sectors. Most of the flares have multiple peaks in the decay phase  and complex nature,
further discussed in section \ref{sec:diss_flr_analy}.

\subsection{Flare Energy Calculation}\label{sec:flr_enr}
We have estimated the total energy of the flare by two different methods. In the first method, we determined the total flare energy of each flare using the method described in \citealp{Shibayama2013ApJS..209....5S, Kumbhakar2023ApJ...955...18K}.  The total energy of the flare is calculated using stellar luminosity, amplitude, flare duration, and equivalent duration (ED).   ED is defined as the unit of time, during when a substellar object (in its quiescent state) would have emitted the same amount of energy as the flare emitted. It is measured by the area under the flare light curve \citep{Hawley2014ApJ...797..121H}. The amplitude, flare duration, and ED were obtained directly from the \textsc{altaipony} software. According to previous work \citep{Kumbhakar2023ApJ...955...18K}, the white light flare can be assumed as hot-blackbody radiation with a constant temperature of 10,000~K (denoted as $T_{flare}$) \citep{Kowalski2013ApJS..207...15K, Howard2018ApJ...860L..30H}. Additionally, we used a synthetic photospheric spectrum of GJ 182 to calculate the stellar luminosity. The synthetic photospheric spectrum was generated using the BT-Settl model using the VOSA web service\footnote{http://svo2.cab.inta-csic.es/theory/newov2/}\citep{Allard2012RSPTA.370.2765A}. The bolometric flare luminosity is determined using $T_{flare}$ and the area of the flare ($A_{flare}$) by the following equation:
\begin{equation}
    L_{flare}=\sigma T_{flare}^{4}A_{flare}
    \label{flare_lum}
\end{equation}
where $\sigma$ is the Stefan-Boltzmann constant. The area of the flare is estimated as follows,
\begin{equation}
A_{flare}= (\Delta F/F)\pi R_{star}^{2}\frac{\int R_{\lambda}F(\lambda,T_{eff})~d\lambda}{\int R_{\lambda}B(\lambda,T_{flare})~d\lambda} 
\label{A_flare}
\end{equation}
where $R_{\lambda}$ is the TESS response function, $\lambda$ is the wavelength, $F(\lambda,T_{eff})$ is the flux of the synthetic photospheric spectrum of the flaring object and $B(\lambda,T_{flare})$ is the Planck function at the flare temperature. $\frac{\Delta F}{F}$=$\frac{F_{i}-F_{0}}{F_{0}}$ is referred to as the relative flare amplitude, where $F_{i}$ is the stellar brightness and $F_{0}$ is the  local mean flux in the quiescent state of the object.
Now, the bolometric energy of the flare is obtained by integrating the bolometric flare luminosity ($L_{flare}$) throughout the flare duration as follows,
\begin{equation}
    E_{flare}=\int L_{flare}(t)dt=\sigma T_{flare}^{4} \times\pi R_{star}^{2}\frac{\int R_{\lambda}F(\lambda,T_{eff})~d\lambda}{\int R_{\lambda}B(\lambda,T_{flare})~d\lambda} \times  \int \frac{\Delta F}{F}~dt
    \label{E_flare}
\end{equation}
The last time-integrated factor is generally denoted as the equivalent duration of the flare in which the integration was done of relative flux within the flare duration \citep{Hawley2014ApJ...797..121H, Ikuta2023ApJ...948...64I, Kumbhakar2023ApJ...955...18K}.

In the second method, to estimate the flare energy in the TESS bandpass, we used the modified method proposed by \citet{Kovari2007AN....328..904K} and described in further detail in \citet{Vida2019ApJ...884..160V, Pietras2022ApJ...935..143P}. To determine the energy of the detected flares, we integrated the normalized flare intensity during the flare event,
\begin{equation}
    \epsilon_{f}= \int_{t_1}^{t_2} \left( \frac{I_{0+f} (t)}{I_{0}} -1 \right) dt
\end{equation}
where $t_{1}$ and $t_{2}$ are the begin and end times of the flare event, $I_{0+f}$ and $I_{0}$ are the intensities with and without a flare. This integral gives the relative flare energy or equivalent duration of the flare \citep{Vida2019ApJ...884..160V} which was already calculated using \texttt{ALTAIPONY}. We estimated the quiescent luminosity by using the synthetic photosphere spectrum of the flared objects. The spectrum was generated with the help of the BT-Settl model grid \footnote{http://svo2.cab.inta-csic.es/theory/newov2/}. The parameters $\log(g)$, $T_{eff}$, metalicities of the respective flared objects were taken from \citet{Stassun2019AJ....158..138S} to generate the theoretical spectrum. The quiescent stellar luminosity of the star was determined by multiplying the spectrum of the star $F (\lambda)$ with the TESS bandpass $R_{TESS}(\lambda)$ and the effective area of the stars with Radius R and integration was done in the interval of the wavelengths of the TESS bandpasses ($\lambda_{1}$,~$\lambda_{2}$).
\begin{equation}
    L_{star}= \pi R_{star}^2 \int_{\lambda_1}^{\lambda_2} F(\lambda) R_{TESS}(\lambda) d\lambda
\end{equation}

Now, to calculate the flare energy in the TESS bandpass ($E_{f}$), we had to multiply the relative flare energy or ED by the quiescent stellar luminosity~($L_{star}$) as 
follows
\begin{equation}
    E_{f}=L_{star}\cdot \epsilon_{f}= L_{star}\cdot ED
\end{equation}
The estimated flare energy using this 2nd method was not bolometric like the method proposed by \citet{Shibayama2013ApJS..209....5S}. So, the energy using the 2nd method provides a lower value than the 1st method. The flare parameters and flare energies were estimated using two methods mentioned in  Table \ref{tab:flr_pars_sec5} and \ref{tab:flr_pars_sec32}.

\begin{deluxetable}{cccccccc}
\tablecaption{ The estimated flare parameters of confirmed flare events from the 2-min cadence TESS data of GJ 182 from sector~5 in this work.}
\label{tab:flr_pars_sec5}
\tablehead{
\colhead{$t_s$ [BTJD]}  & \colhead{$t_f$ [BTJD] } &
 \colhead{$a$} & \colhead{$Dur$} & \colhead{$ED$} & \colhead{$E_{bol}^{syn}$} &  \colhead{$E_{tess}$} & \colhead{Magnetic Field} \\
 \colhead{[Time-2457000]} & \colhead{[Time-2457000]} & \colhead{[Rel. Flux]}  & \colhead{[min]} & \colhead{[Sec]} &  \colhead{[erg]} &\colhead{[erg]} &\colhead{[G]}
}
 
\startdata
      1438.6743 &     1438.7104 &  0.038 &      52.0 &   $25.76 \pm 0.18$ &           1.74e+34 &      8.83e+32 &                79.8 \\
      1439.8744 &     1440.0105 &  0.030 &     196.0 &  $120.53 \pm 0.40$ &           8.13e+34 &      4.13e+33 &               172.6 \\
       1442.6147 &     1442.6953 &  0.005 &     116.0 &   $21.36 \pm 0.38$ &           1.44e+34 &      7.32e+32 &                72.7 \\
     1443.1175 &     1443.1217 &  0.003 &       6.0 &    $0.87 \pm 0.09$ &           5.90e+32 &      2.99e+31 &                14.7 \\
      1444.0412 &     1444.0717 &  0.006 &      44.00 &    $8.31 \pm 0.31$ &           5.60e+33 &      2.85e+32 &                45.3 \\
       1444.8078 &     1444.8134 &  0.004 &       8.0 &    $1.20 \pm 0.10$ &           8.06e+32 &      4.10e+31 &                17.2 \\
      1446.1134 &     1446.2065 &  0.020 &     134.0 &   $45.89 \pm 0.34$ &           3.09e+34 &      1.57e+33 &               106.5 \\
      1446.4190 &     1446.5634 &  0.124 &     208.0 &  $218.63 \pm 0.31$ &           1.47e+35 &      7.49e+33 &               232.5 \\
     1447.1801 &     1447.2232 &  0.013 &      62.0 &   $20.09 \pm 0.25$ &           1.35e+34 &      6.88e+32 &                70.5 \\
      1447.7787 &     1447.7829 &  0.003 &       6.0 &    $0.91 \pm 0.09$ &           6.11e+32 &      3.10e+31 &                15.0 \\
      1449.3329 &     1449.3857 &  0.014 &      76.0 &   $28.68 \pm 0.29$ &           1.93e+34 &      9.83e+32 &                84.2 \\
     1449.6621 &     1449.7024 &  0.013 &      58.0 &   $12.17 \pm 0.21$ &           8.20e+33 &      4.17e+32 &                54.9 \\
      1449.8135 &     1449.8176 &  0.003 &       6.0 &    $0.77 \pm 0.09$ &           5.18e+32 &      2.63e+31 &                13.8 \\
     1452.2204 &     1452.2274 &  0.003 &      10.0 &    $1.36 \pm 0.11$ &           9.19e+32 &      4.67e+31 &                18.4 \\
      1453.2996 &     1453.3163 &  0.004 &      24.0 &    $3.68 \pm 0.18$ &           2.48e+33 &      1.26e+32 &                30.2 \\
      1453.5066 &     1453.5135 &  0.002 &      10.0 &    $1.10 \pm 0.11$ &           7.38e+32 &      3.75e+31 &                16.5 \\
     1453.7677 &     1453.7760 &  0.007 &      12.0 &    $2.40 \pm 0.11$ &           1.62e+33 &      8.23e+31 &                24.4 \\
     1454.0607 &     1454.0691 &  0.007 &      12.0 &    $2.62 \pm 0.11$ &           1.77e+33 &      8.99e+31 &                25.5 \\
     1454.2191 &     1454.2357 &  0.007 &      24.0 &    $5.74 \pm 0.16$ &           3.87e+33 &      1.96e+32 &                37.7 \\
      1456.2094 &     1456.2371 &  0.007 &      40.0 &    $8.76 \pm 0.22$ &           5.90e+33 &      3.00e+32 &                46.5 \\
   1457.0371 &     1457.0455 &  0.005 &      12.0 &    $1.95 \pm 0.12$ &           1.32e+33 &      6.69e+31 &                22.0 \\
    1457.3913 &     1457.4260 &  0.007 &      50.0 &    $9.92 \pm 0.23$ &           6.69e+33 &      3.40e+32 &                49.5 \\
     1459.7233 &     1459.7552 &  0.004 &      46.0 &    $6.72 \pm 0.24$ &           4.53e+33 &      2.30e+32 &                40.8 \\
    1460.4482 &     1460.6121 &  0.090 &     236.0 &  $183.44 \pm 0.35$ &           1.24e+35 &      6.29e+33 &               213.0 \\
     1462.5482 &     1462.5857 &  0.029 &      54.0 &   $22.65 \pm 0.19$ &           1.53e+34 &      7.761e+32 &                74.8 \\
\enddata
\tablecomments{$t_{s}$: Start time of flare, $t_{f}$: Stop time of flare, $a$: relative amplitude of flare, $Dur$: duration of flare, $ED$: equivalent duration of flare, $E_{bol}^{syn}$: Bolometric energy of flare by taking the synthetic spectra of the object, $E_{tess}$: Flare energy in TESS bandpass.  }
\end{deluxetable}

\begin{deluxetable}{cccccccc}
\tablecaption{ The estimated flare parameters of confirmed flare events from the 2-min cadence TESS data of GJ 182 from sector~32 in this work.}
\label{tab:flr_pars_sec32}
\tablehead{
\colhead{$t_s$ [BTJD]}  & \colhead{$t_f$ [BTJD] } &
 \colhead{$a$} & \colhead{$Dur$} & \colhead{$ED$} & \colhead{$E_{bol}^{syn}$} &  \colhead{$E_{tess}$} & \colhead{Magnetic Field} \\
 \colhead{[Time-2457000]} & \colhead{[Time-2457000]} & \colhead{[Rel. Flux]}  & \colhead{[min]} & \colhead{[Sec]} &  \colhead{[erg]} &\colhead{[erg]} &\colhead{[G]}
}
 
\startdata
      2174.4825 &     2174.4908 &  0.003 &      12.0 &    $1.41 \pm 0.13$ &           9.51e+32 &      4.83e+31 &                18.7 \\
      2174.9797 &     2174.9936 &  0.006 &      20.0 &    $3.75 \pm 0.15$ &           2.53e+33 &      1.28e+32 &                30.5 \\
     2175.0005 &     2175.0478 &  0.011 &      68.0 &   $25.04 \pm 0.28$ &           1.69e+34 &      8.58e+32 &                78.7 \\
     2175.1811 &     2175.2186 &  0.005 &      54.0 &    $9.89 \pm 0.26$ &           6.67e+33 &      3.39e+32 &                49.4 \\
     2175.6672 &     2175.6742 &  0.004 &      10.0 &    $1.68 \pm 0.11$ &           1.13e+33 &      5.75e+31 &                20.4 \\
     2176.5131 &     2176.5284 &  0.005 &      22.0 &    $3.64 \pm 0.16$ &           2.45e+33 &      1.25e+32 &                30.0 \\
     2178.4423 &     2178.4562 &  0.004 &      20.0 &    $3.00 \pm 0.16$ &           2.02e+33 &      1.03e+32 &                27.2 \\
     2178.6826 &     2178.6951 &  0.003 &      18.0 &    $2.18 \pm 0.16$ &           1.47e+33 &      7.46e+31 &                23.2 \\
      2179.5312 &     2179.5437 &  0.004 &      18.0 &    $3.46 \pm 0.16$ &           2.33e+33 &      1.18e+32 &                29.2 \\
      2180.4965 &     2180.5020 &  0.004 &       8.0 &    $1.33 \pm 0.10$ &           8.94e+32 &      4.55e+31 &                18.1 \\
      2180.7062 &     2180.7132 &  0.008 &      10.0 &    $2.52 \pm 0.11$ &           1.70e+33 &      8.64e+31 &                25.0 \\
      2182.9923 &     2182.9965 &  0.002 &       6.0 &    $0.56 \pm 0.09$ &           3.81e+32 &      1.93e+31 &                11.8 \\
     2183.0048 &     2183.0243 &  0.003 &      28.0 &    $3.13 \pm 0.19$ &           2.11e+33 &      1.07e+32 &                27.8 \\
    2183.7396 &     2183.8340 &  0.091 &     136.0 &  $100.93 \pm 0.26$ &           6.80e+34 &      3.46e+33 &               158.0 \\
   2183.9743 &     2184.0104 &  0.012 &      52.0 &   $14.81 \pm 0.24$ &           9.98e+33 &      5.07e+33 &                60.5 \\
   2189.7007 &     2189.7063 &  0.003 &       8.0 &    $1.04 \pm 0.10$ &           7.01e+32 &      3.56e+31 &                16.0 \\
   2190.3882 &     2190.3924 &  0.002 &       6.0 &    $0.66 \pm 0.09$ &           4.46e+32 &      2.27e+31 &                12.8 \\
  2190.9493 &     2190.9576 &  0.004 &      12.0 &    $1.80 \pm 0.12$ &           1.21e+33 &      6.16e+31 &                21.1 \\
  2192.1715 &     2192.1799 &  0.007 &      12.0 &    $2.05 \pm 0.11$ &           1.38e+33 &      7.01e+31 &                22.5 \\
  2194.1257 &     2194.1326 &  0.002 &       10.0 &    $1.26 \pm 0.12$ &           8.50e+32 &      4.32e+31 &                17.7 \\
  2194.2187 &     2194.2590 &  0.039 &      58.0 &   $22.53 \pm 0.19$ &           1.52e+34 &      7.72e+32 &                74.6 \\
  2196.4437 &     2196.4812 &  0.014 &      54.0 &   $14.49 \pm 0.22$ &           9.77e+33 &      4.96e+32 &                59.8 \\
  2199.8631 &     2199.8728 &  0.004 &      14.0 &    $1.76 \pm 0.13$ &           1.18e+33 &      6.02e+31 &                20.8 \\
\enddata
\tablecomments{$t_{s}$: Start time of flare, $t_{f}$: Stop time of flare, $a$: relative amplitude of flare, $Dur$: duration of flare, $ED$: equivalent duration of flare, $E_{bol}^{syn}$: Bolometric energy of flare by taking the synthetic spectra of the object, $E_{tess}$: Flare energy in TESS bandpass.  }
\end{deluxetable}

\subsection{Estimation of Magnetic Field Strength from Flare Energy}
 GJ~182 is a young M-type object exhibiting significant magnetic activity. Previously, \citet{Lang2012MNRAS.424.1077L} reported a large-scale average magnetic field for this object around 172 G measured from spectropolarimetric measurements, and from unpolarized spectroscopy, they provided an average magnetic field (small+large-scale field) of 2.5~kG \citep{Reiners2009A&A...496..787R}. Interestingly, we also identified most of the flare events (35 out of 48 ) were in the superflare energy range, ranging from 10$^{32}$  to 10$^{34}$ erg which indicated it's magnetic activity nature. 
    Therefore, by assuming the similarities between the physical process in solar flares and the flares in those VLMs and BDs and found a rough estimation of the lower limit of maximum magnetic field strength ($B_{m}$) associated with those flares by using a scaling relation that has been reported by several authors i.e. \citet{Aulanier2013A&A...549A..66A, Notsu2013ApJ...771..127N, Paudel2018ApJ...861...76P}. According to \citet{Aulanier2013A&A...549A..66A}, stellar flares occur when a particular fraction of magnetic energy from the star is released through an active region. The active region is assumed to be associated with positive and negative polarity (bipolar spot) with a linear separation $L_{bi}$ and we considered the corresponding maximum magnetic field strength was $B_{m}$ in the active regions on the stellar surface. In a simplified way, let's assume a portion (f) of $B_{m}$ throughout the volume $V_{f}$ in the atmosphere is reconnected in order to power the flare. Again, we considered the volume $V_{f}$ to be equal to some fraction $f^{\prime}$ of the volume of a cube with sides of length  $L_{bi}$. According to these assumptions, the magnetic energy released in the flare as a whole can be written as $E_{f} = 0.04 \times f^{2} f^{\prime} B^2_{m} L^3_{bi}$ (see \citealt{Aulanier2013A&A...549A..66A, Paudel2020MNRAS.494.5751P} for more details). Furthermore, based on the model calculation performed by \citet{Aulanier2013A&A...549A..66A} to calculate the flare energy for the Sun, they found that f and $f^{\prime}$ to be equal to 0.2. This value indicates that only about 20\%  of the maximum field strength was required to power the flare and only about 20\%  of the available volume needed to participate in the flare events. To set the lower limit of $B_{m}$, we considered the upper limit of $L_{bi}$ and the upper limit of $L_{bi}$ would be set if the two poles (bipolar spots) in the active regions are well separated on the stellar surface. In the limiting case, $L_{bi}$= $\pi \cdot R_{star}$ and the above equation  is modified as follows, $E_{flare} \approx 4 \times 10^{-4} B_{m}^2 (\pi R_{star})^{3}$. Therefore, to generate the bolometric flare energies in the range $ 10^{32-35}$ erg (mentioned in Table \ref{tab:flr_pars_sec5} and \ref{tab:flr_pars_sec32}), we estimated the lower limits on $B_{m}$ of our objects from  15~G to 232~G. We assumed that two bi-poles are placed at both poles or 180 degrees apart on the stellar surface, the field can be considered as the global field of the objects. The lower limit of $B_{m}$  derived in this work is reported in  the last column of Table \ref{tab:flr_pars_sec5} and \ref{tab:flr_pars_sec32}.   Our estimated result is almost similar to those of \citet{Namekata2017ApJ...851...91N, Pietras2022ApJ...935..143P}. For instance, \citet{Namekata2017ApJ...851...91N} also found roughly the same magnetic field strength of about 30~G to 300~G for the solar-type stars using Kepler 30-min cadence data. Similarly,\citet{Pietras2022ApJ...935..143P} reported average magnetic field strength values between 10~G to 200~G for 25,000 stars using TESS 2-minute cadence data.

\subsection{Starspot Modeling}\label{sec:SM}
Starspots, prominent features on the surfaces of young M dwarfs, are recognized as significant contributors to periodic or quasi-periodic variations observed in the light curves of these stellar objects. Here, two selected young M dwarfs exhibited periodic variation in the phase folded light curve which might be the cause of the rotational modulation of starspots co-rotating with the objects, periodically coming into and out of view. However, the significance of these periodic variations fails to provide comprehensive insights into the properties of starspots. To qualitatively trace the sizes and position of starspots we conducted an in-depth analysis of \textsc{TESS} light curves utilizing the starspot modeling software, \textsc{BASSMAN} \footnote{https://github.com/KBicz/BASSMAN} (Best rAndom StarSpots Model calculAtioN) \citep{Bicz2022ApJ...935..102B}. In \textsc{BASSMAN}, the surface map of stars is considered a vector of spherical harmonic coefficients. It can be also expressed as a linear combination of spherical harmonics with an index increasing by degree $l$ and order $m$. During the modeling of the light curve, the assumption of spots is spherical. The only parameter i.e. longitude of the spot varies in time due to the rotation of the stars. Here, we have recorded the longitudes of spots at phase=0. The code employed the Markov Chain Monte Carlo (MCMC) method to recreate the light curve of the spotted stars. As a result, the output provides amplitudes, sizes, latitudes, and longitudes of starspots. Here, the software approximates the ``starspots" as the whole active regions that consist of several individual spots. They are not the same as the sunspot structure.  To better understand the spot evolution it is necessary to divide the light curves into time segments so that each segment contains the full rotation of the object and model each modulation separately. This method is helpful to reveal changes in spot configuration from modulation to modulation. For the goodness of fit of individual models, we check the value of log probability. The higher value of log-probability indicated the better fit of the light curves and finally compared the model estimated parameters with the analytical solution. Actually, BASSMAN provides the values of mean spot temperature and percentage spottedness of the star and also calculates these parameters from the analytic solution, taken from \citet{Notsu2019ApJ...876...58N}. The analytic relations of mean spot temperature, 
\begin{equation}
    T_{spot}=0.751T_{star} - 3.58 \times 10^{-5} T_{star}^{2}+808
\end{equation}
and the spot area can be estimated as follows,
\begin{equation}
    A_{spot}=\frac{\Delta F}{F}A_{star}\Biggl[1-\left(\frac{T_{spot}}{T_{star}}\right)^{4}\Biggr]^{-1}
\end{equation}
 
where $T_{star}$=effective temperature of the object, $T_{spot}$=the mean temperature of starspot, $\frac{A_{spot}}{A_{star}}$ is the percentage of spottedness of the object, $\frac{\Delta F}{F}$ is the normalized amplitude of light curve variations.

\section{Results}\label{sec:result}
\textsc{TESS} observed GJ~182 in sectors 5 and 32 while 2M0512+2214 was observed in sectors 43, 44, and 45. We estimated the rotation period using the LS periodogram and Gaussian process. Both methods almost calculated the same rotational periods within the error bar (Table \ref{tab:rot_per}). When we phase-folded the light curve at the most significant peak, they showed significant variability with sinusoidal or semi-sinusoidal structures. Rotational modulation of cold spots with the object is mainly responsible for producing such periodic or quasi-periodic behavior in the light curve. However, by just inspecting the light curve morphology, we cannot find the location of the starspots. To map the distribution of starspots on the star's surface, a more sophisticated approach is necessary. Starspot modeling or light-curve inversion provides a deeper understanding of the distribution of starspots on the surface of these stars (\citealp{Ikuta2020ApJ...902...73I, Bicz2022ApJ...935..102B, Ikuta2023ApJ...948...64I}).

\begin{figure*}
    \centering
    \includegraphics[width=.31\linewidth]{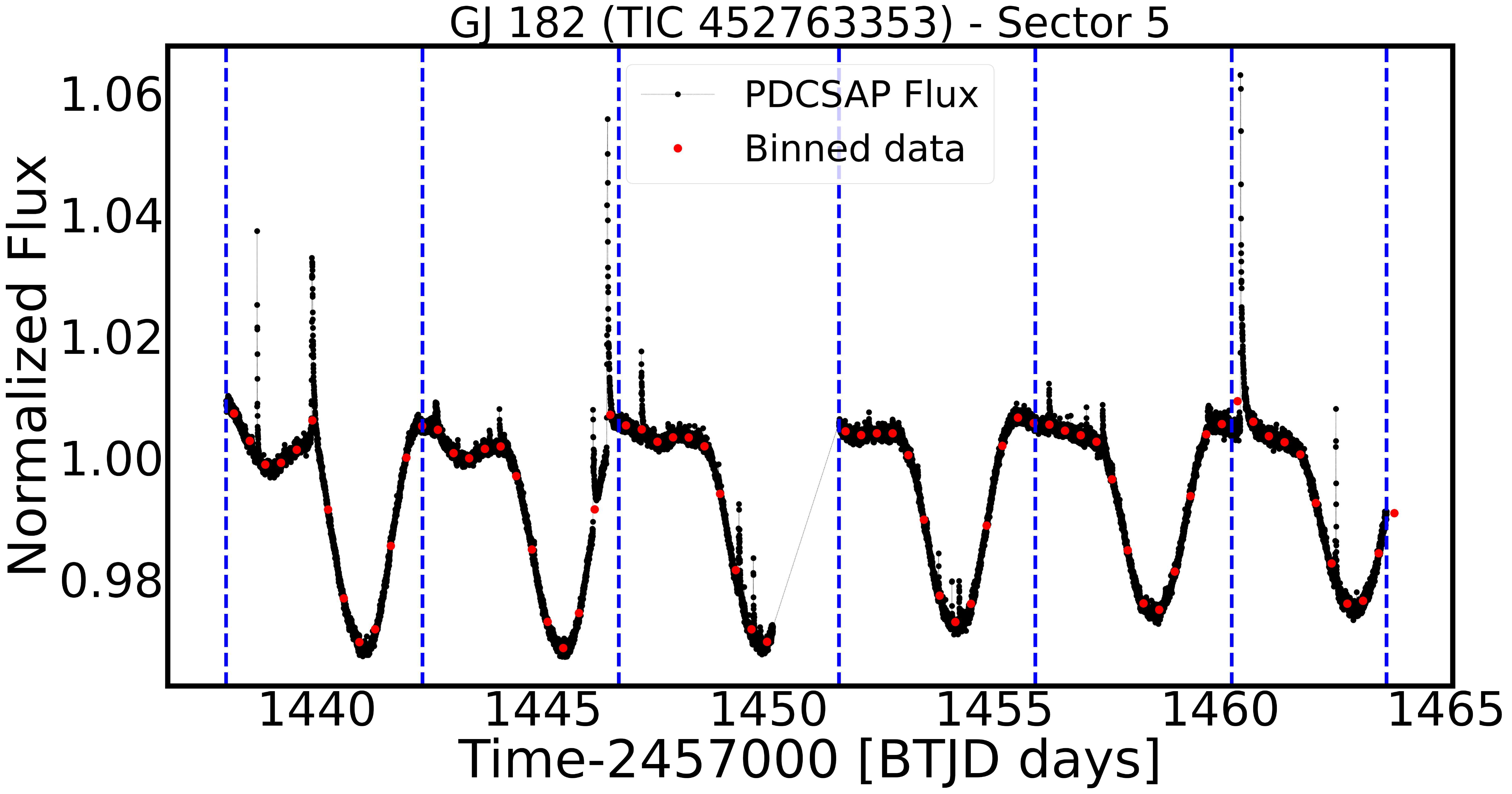}
    \quad
    \includegraphics[width=.31\linewidth]{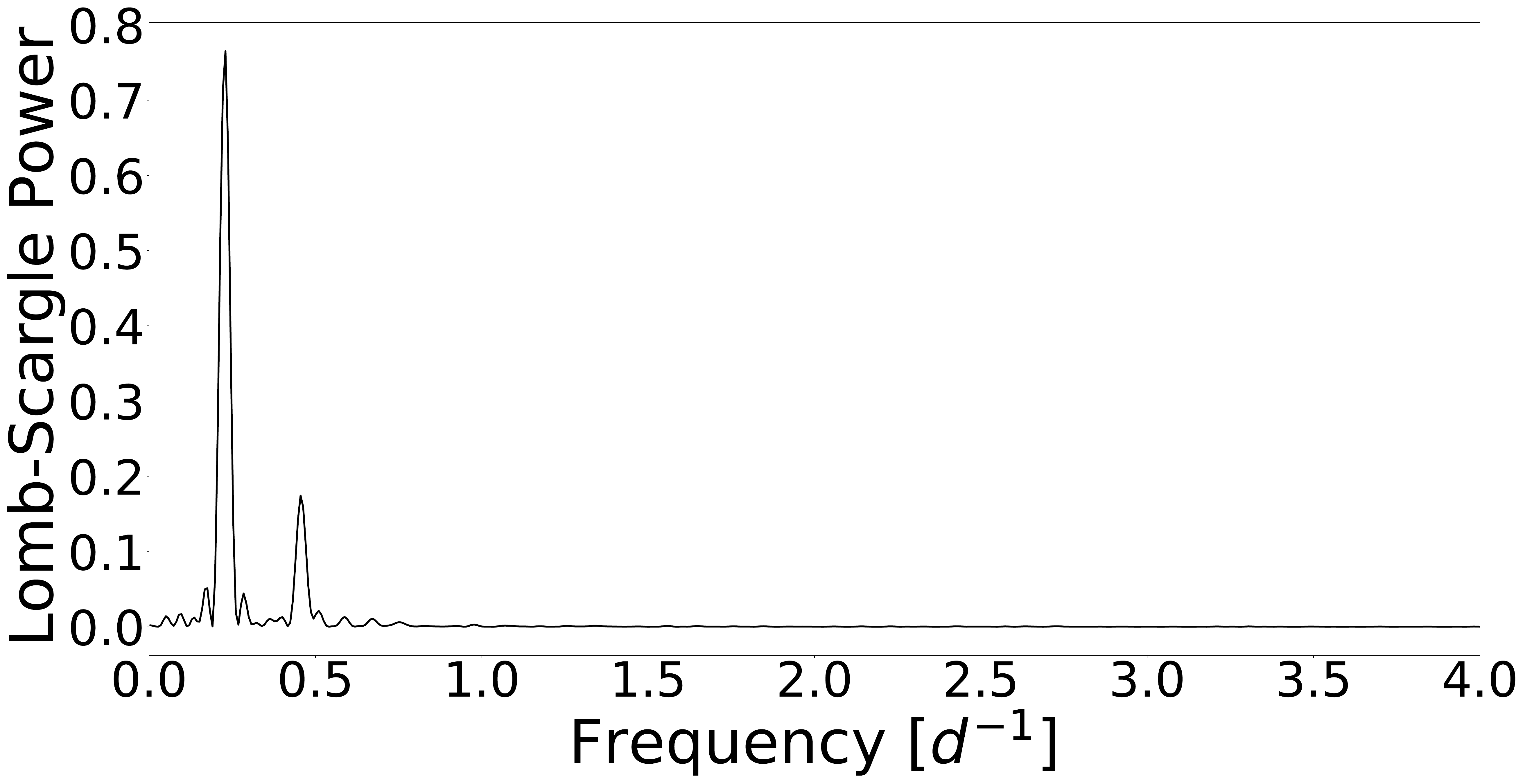}  
    \quad\includegraphics[width=.31\linewidth]{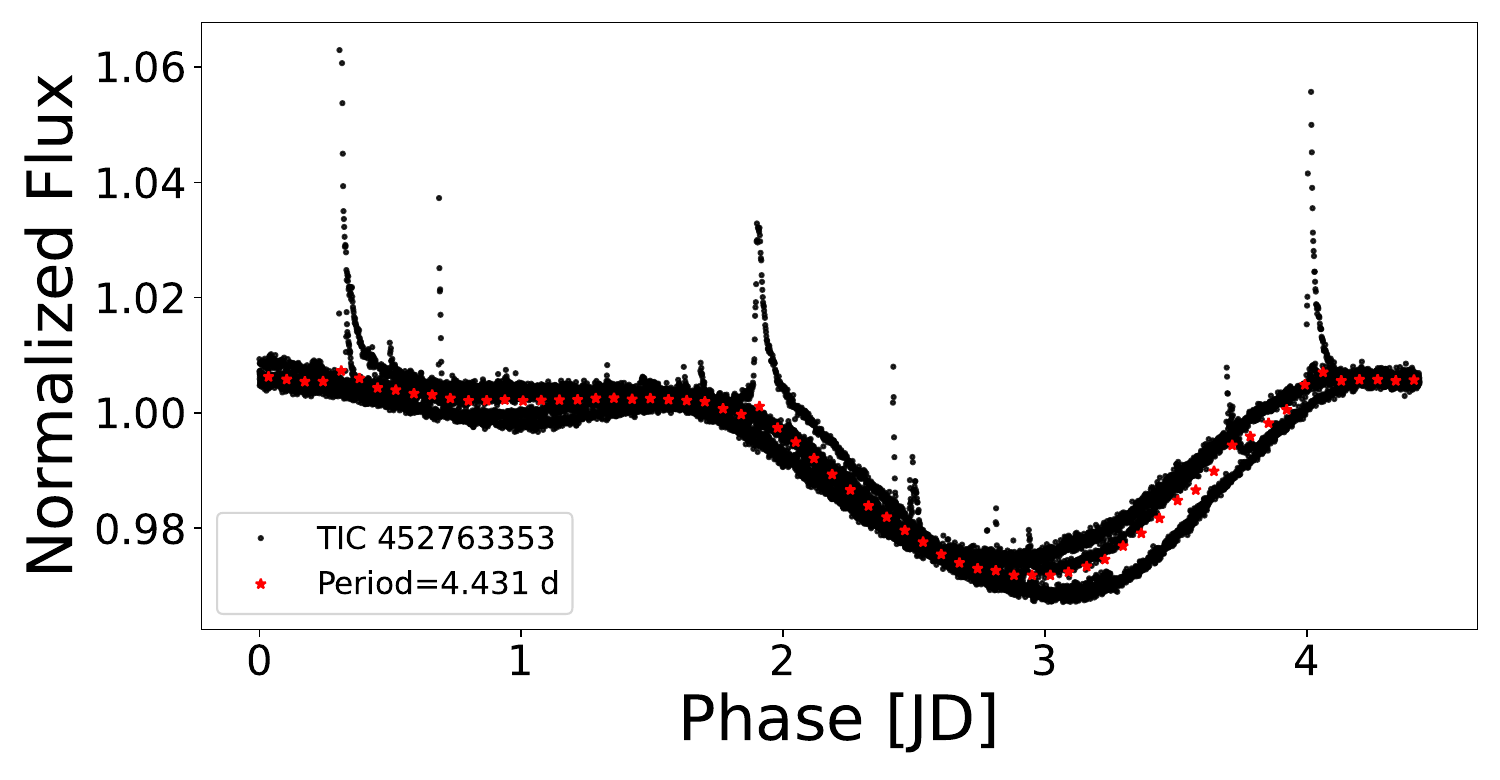}
    \includegraphics[width=.31\linewidth]{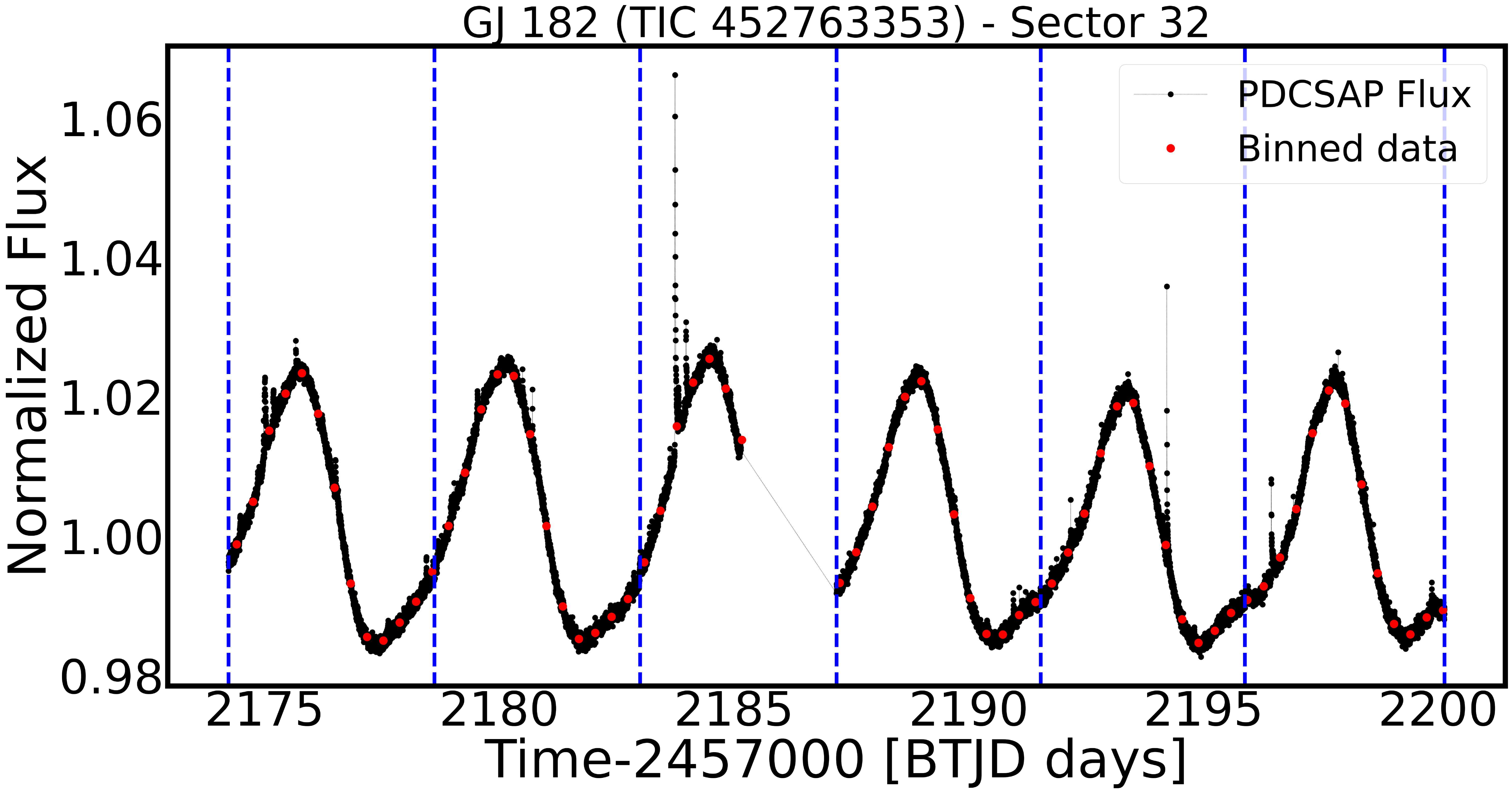}
    \quad
    \includegraphics[width=.31\linewidth]{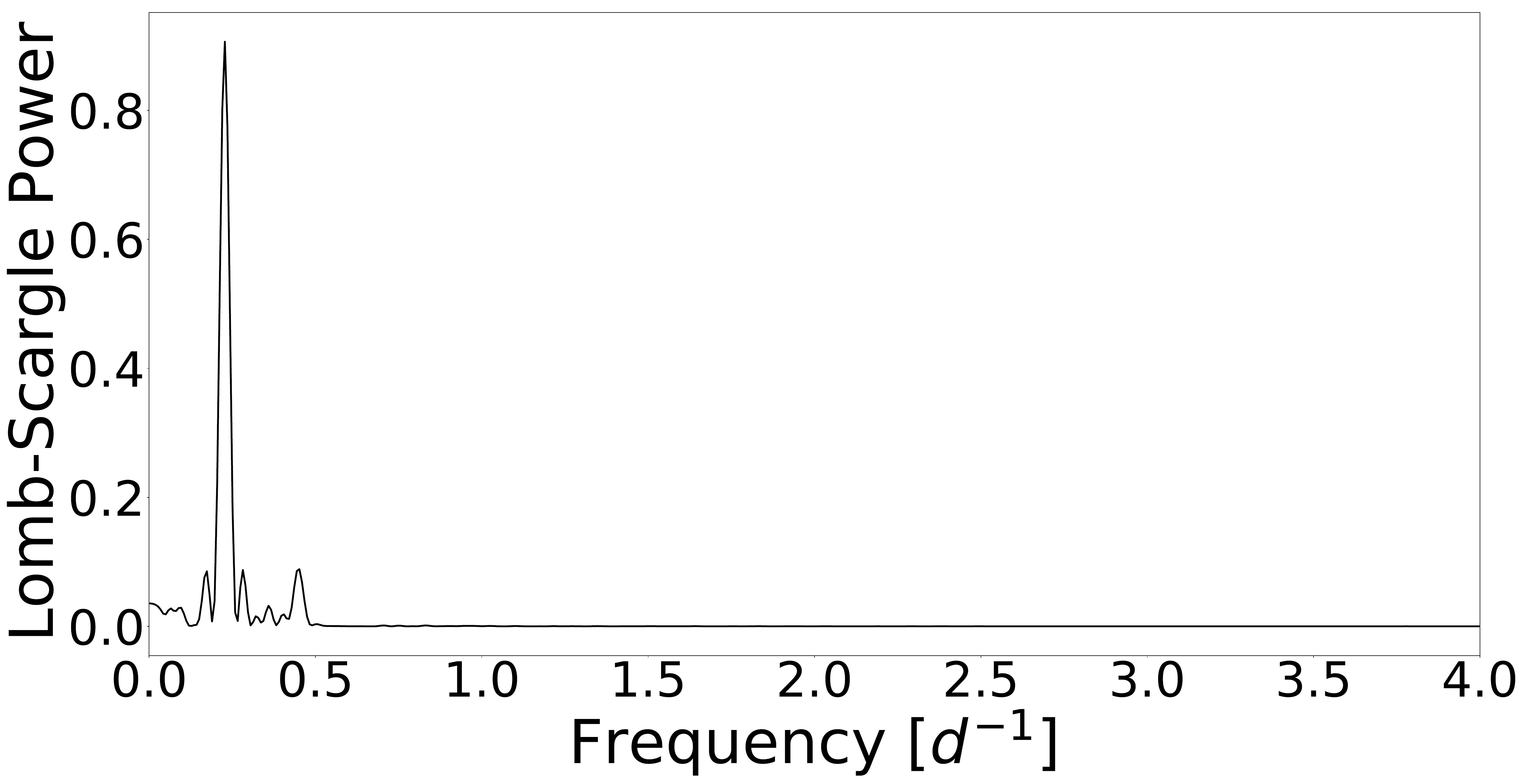}  
    \quad\includegraphics[width=.31\linewidth]{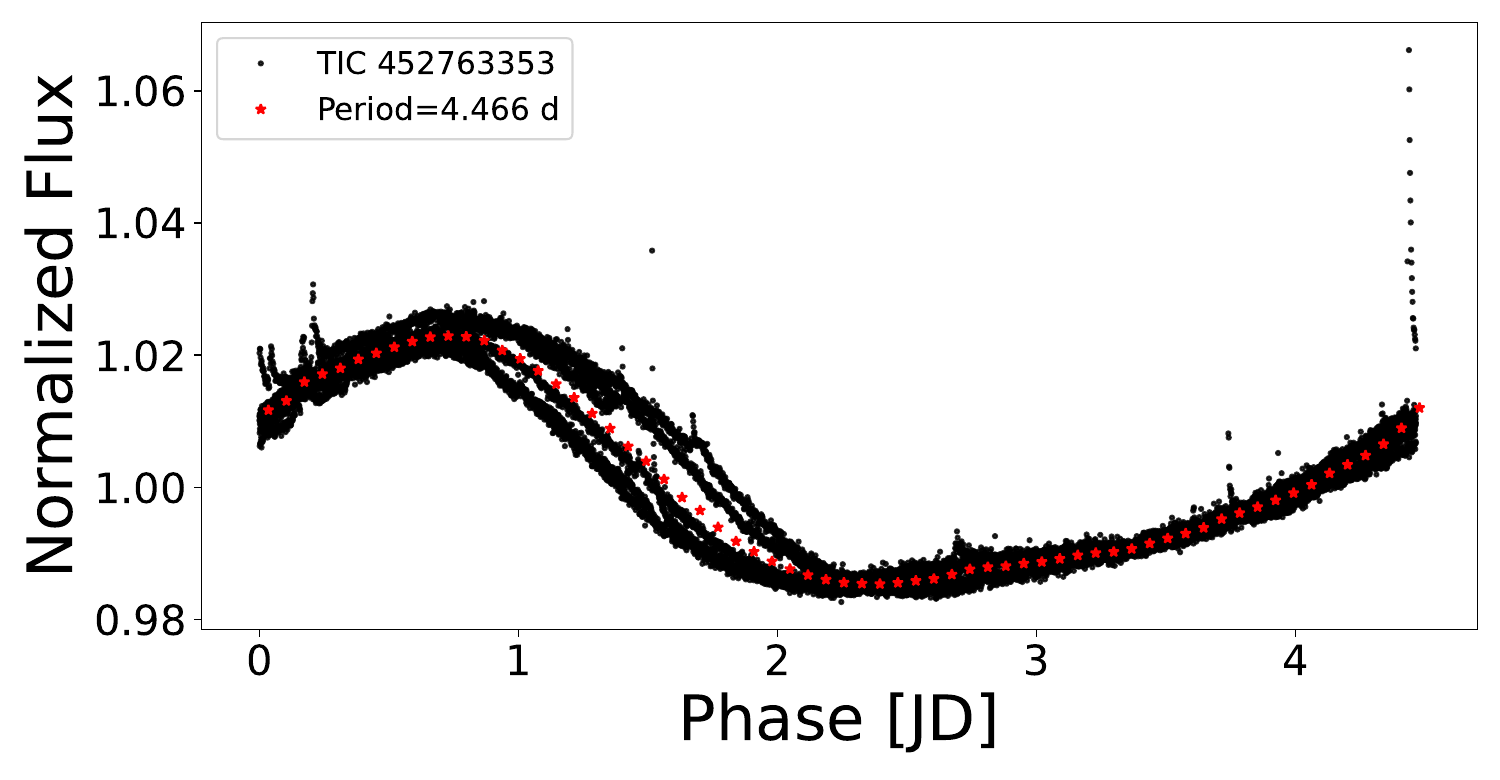}
    \caption{The light curves (left), Lomb-Scargle periodograms (middle) and phase-folded light curves (right)of GJ~182 are shown here for sector 5 (top row) and sector 32 (bottom row). The light curves were binned with intervals of 500 minutes and are shown in red dots  and blue dotted lines represent the segments used for starspot modeling in BASSMAN. The right column displays the phase-folded light curves, where the data are folded with the most significant peak obtained from the LS periodogram. The black stars  in the phase light curve represent the  100 min binned data. The figure titles in the left column include the object names and sectors. Additionally, the rotation periods of the objects are mentioned within the phase light curves.}
    \label{fig:lc_ls_phs}
\end{figure*}

\begin{figure*}
    \centering
    \includegraphics[width=.31\linewidth]{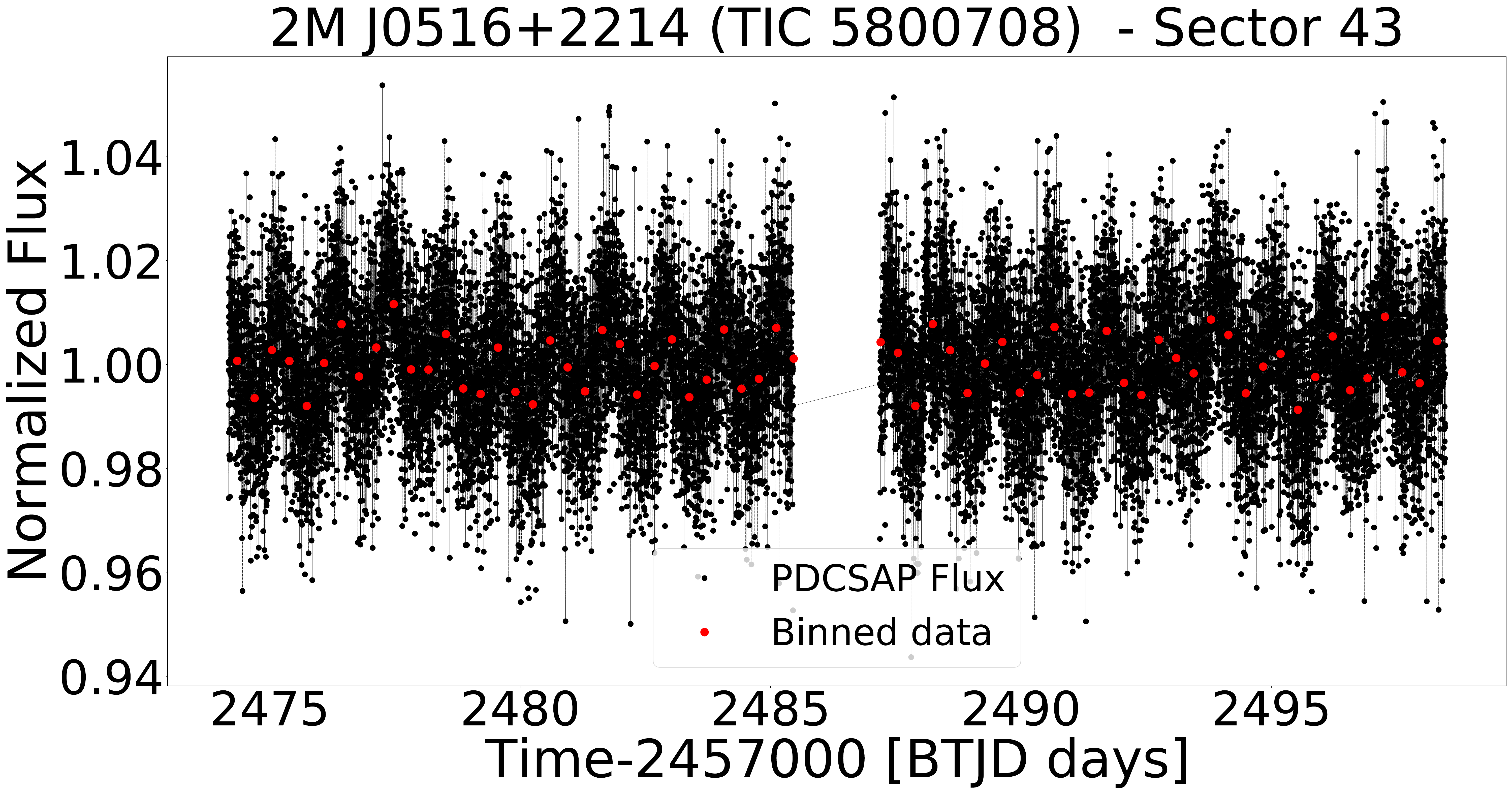}
    \quad
    \includegraphics[width=.31\linewidth]{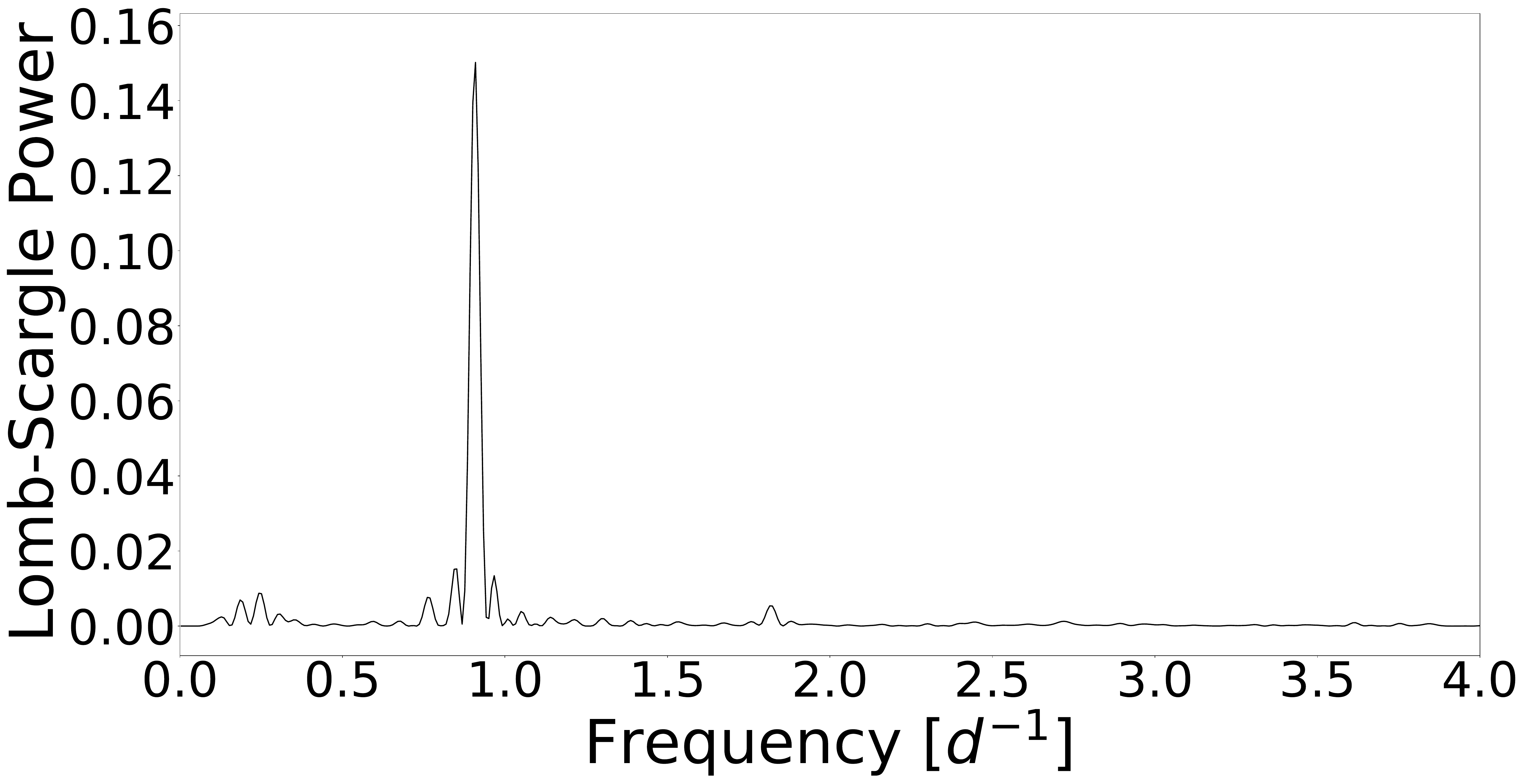}  
    \quad\includegraphics[width=.31\linewidth]{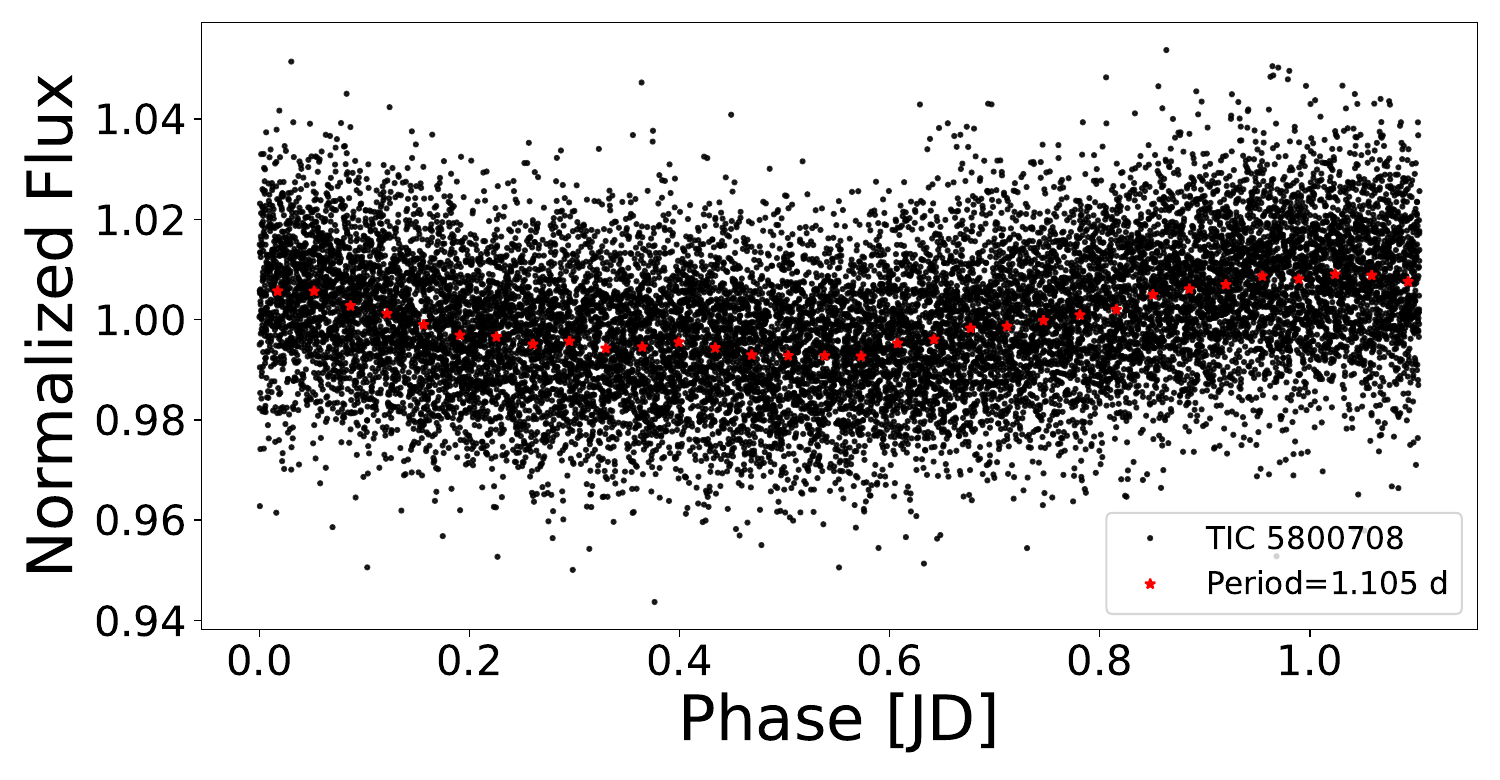}

    \includegraphics[width=.31\linewidth]{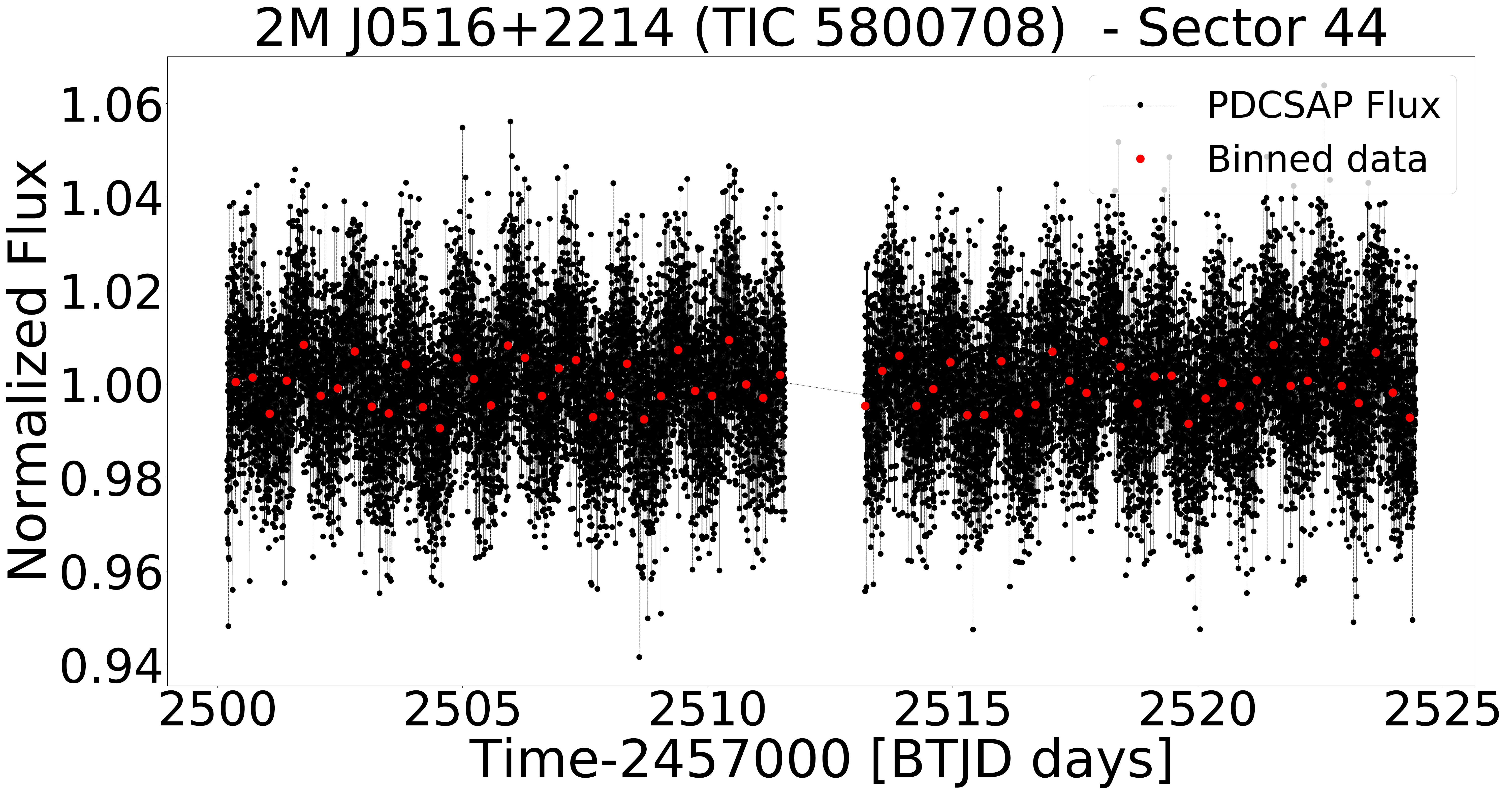}
    \quad
    \includegraphics[width=.31\linewidth]{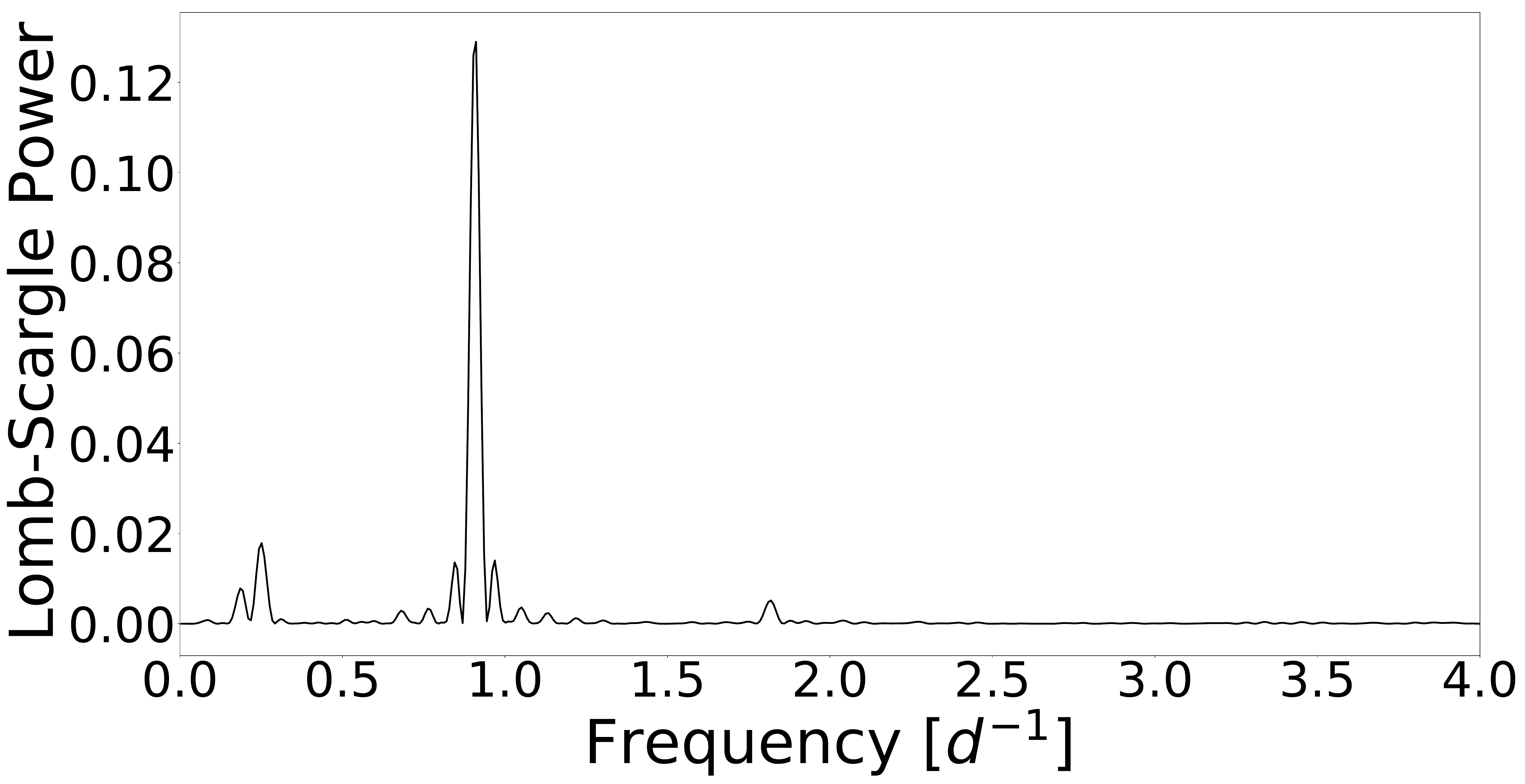}  
    \quad\includegraphics[width=.31\linewidth]{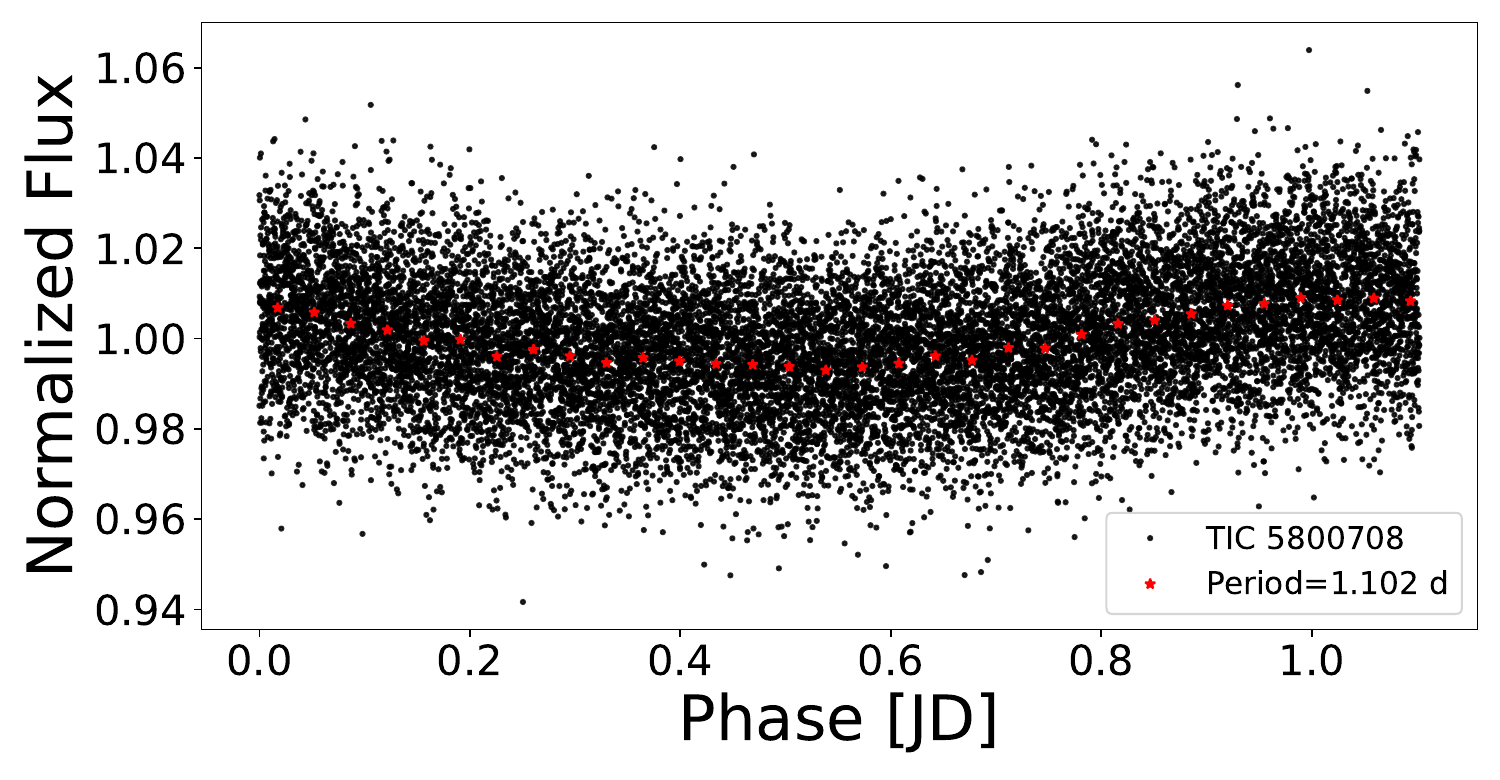}

    \includegraphics[width=.31\linewidth]{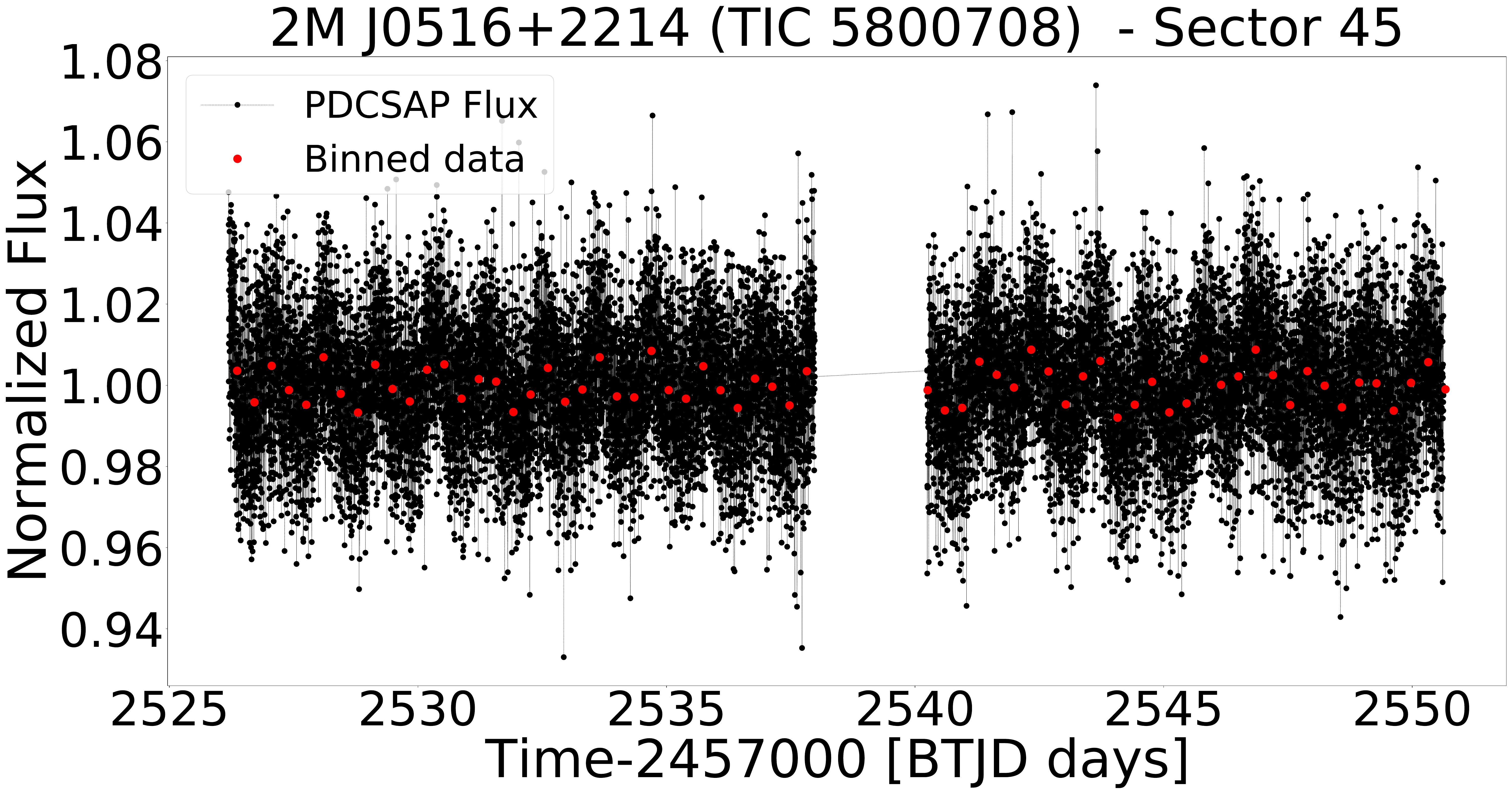}
    \quad
    \includegraphics[width=.31\linewidth]{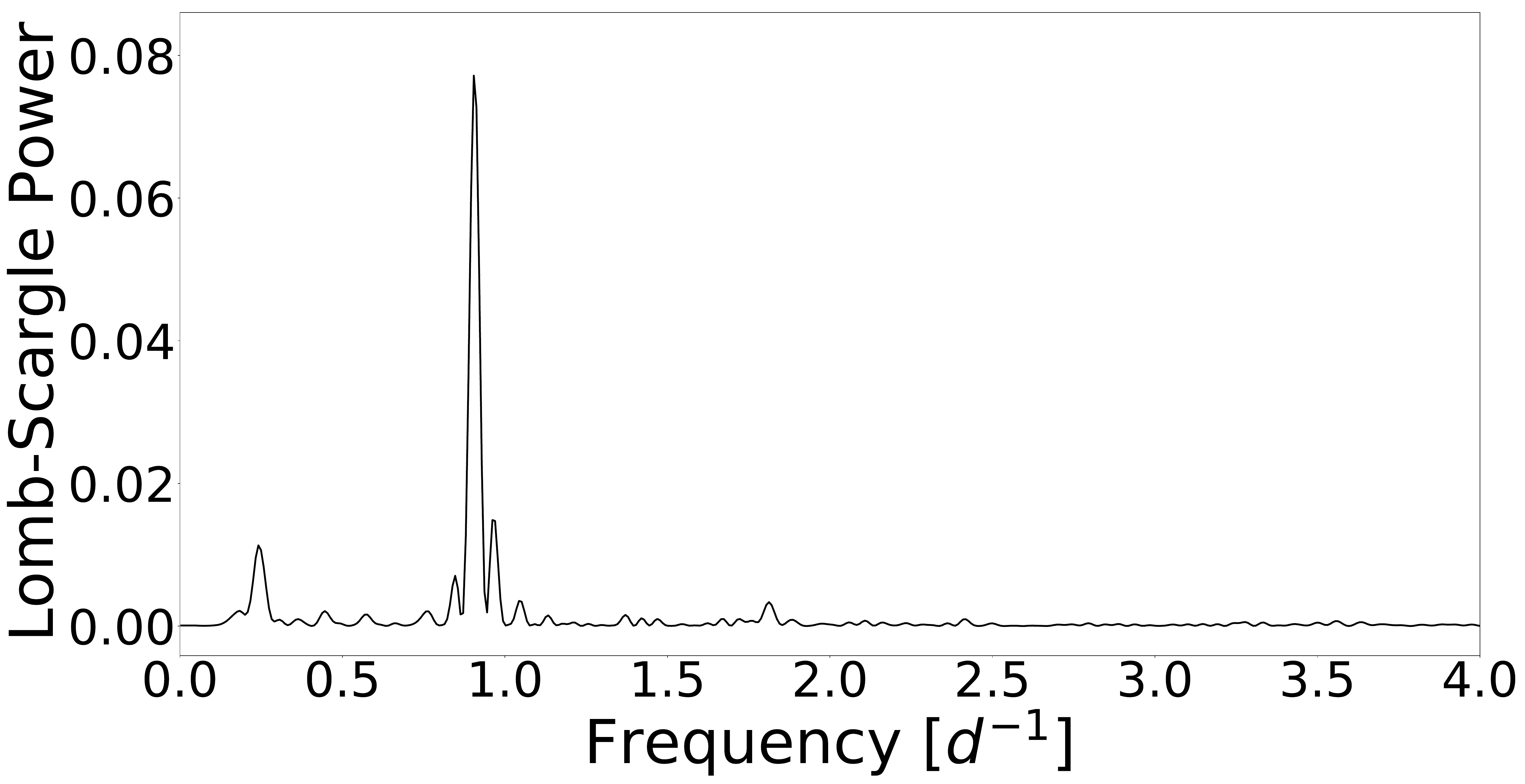}  
    \quad\includegraphics[width=.31\linewidth]{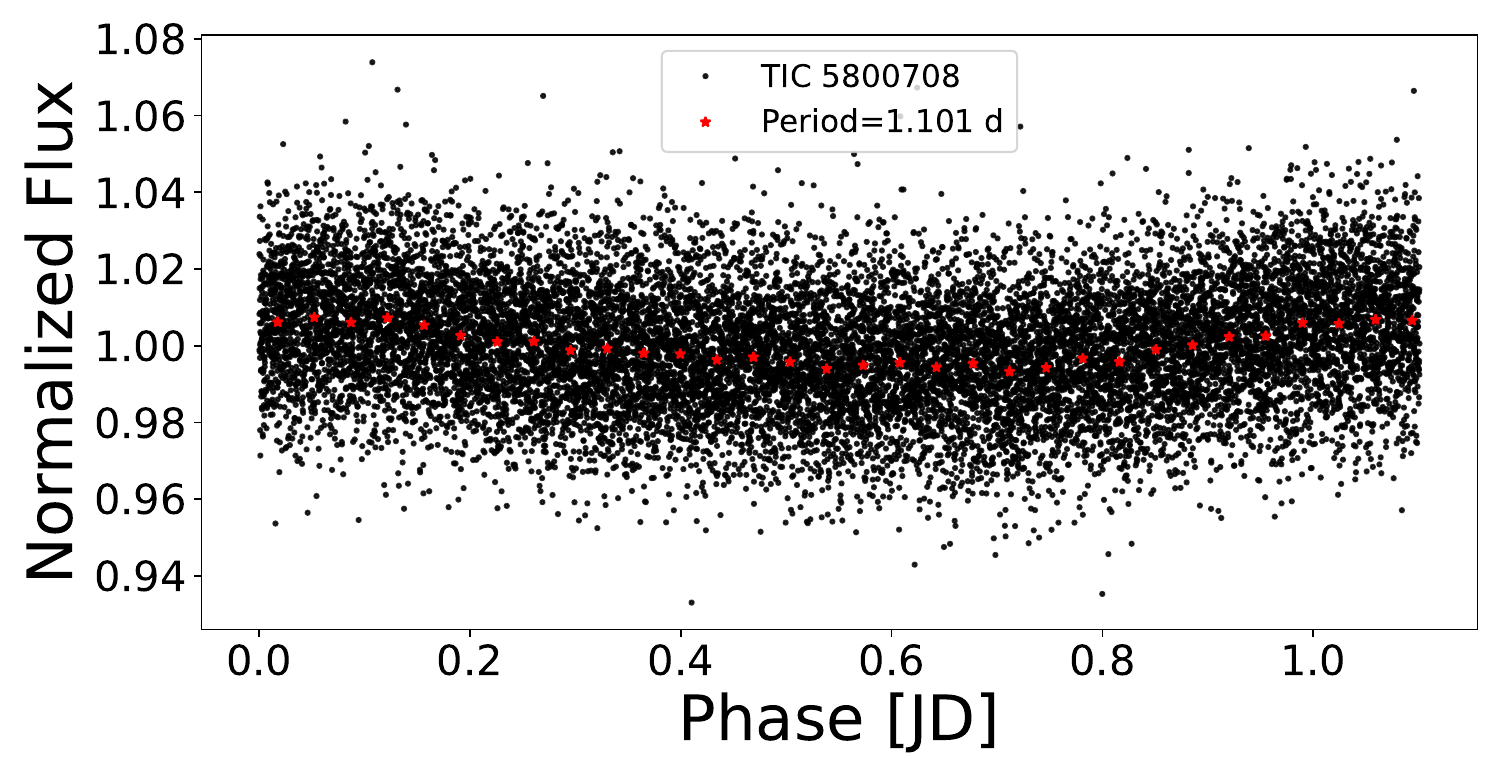}
    
    \caption{The light curves (left), Lomb-Scargle periodograms (middle) and phase folded light curves (right)of 2MASS~J05160212+221452 are shown here for sectors 43 (first row), 44 (second row) and 45 (third row). The light curves were binned with intervals of 500 minutes and are shown in red dots. The right column displays the phase-folded light curves, where the data are folded with the most significant peak obtained from the LS periodogram. The black stars  in phase light curve represent the  50 min binned data. The figure titles in the left column include the object names and sectors. Additionally, the rotation periods of the objects are mentioned within the phase light curves.}
    \label{fig:2214_lc_ls_phs}
\end{figure*}

\begin{figure*}
    \includegraphics[width=0.45\linewidth]{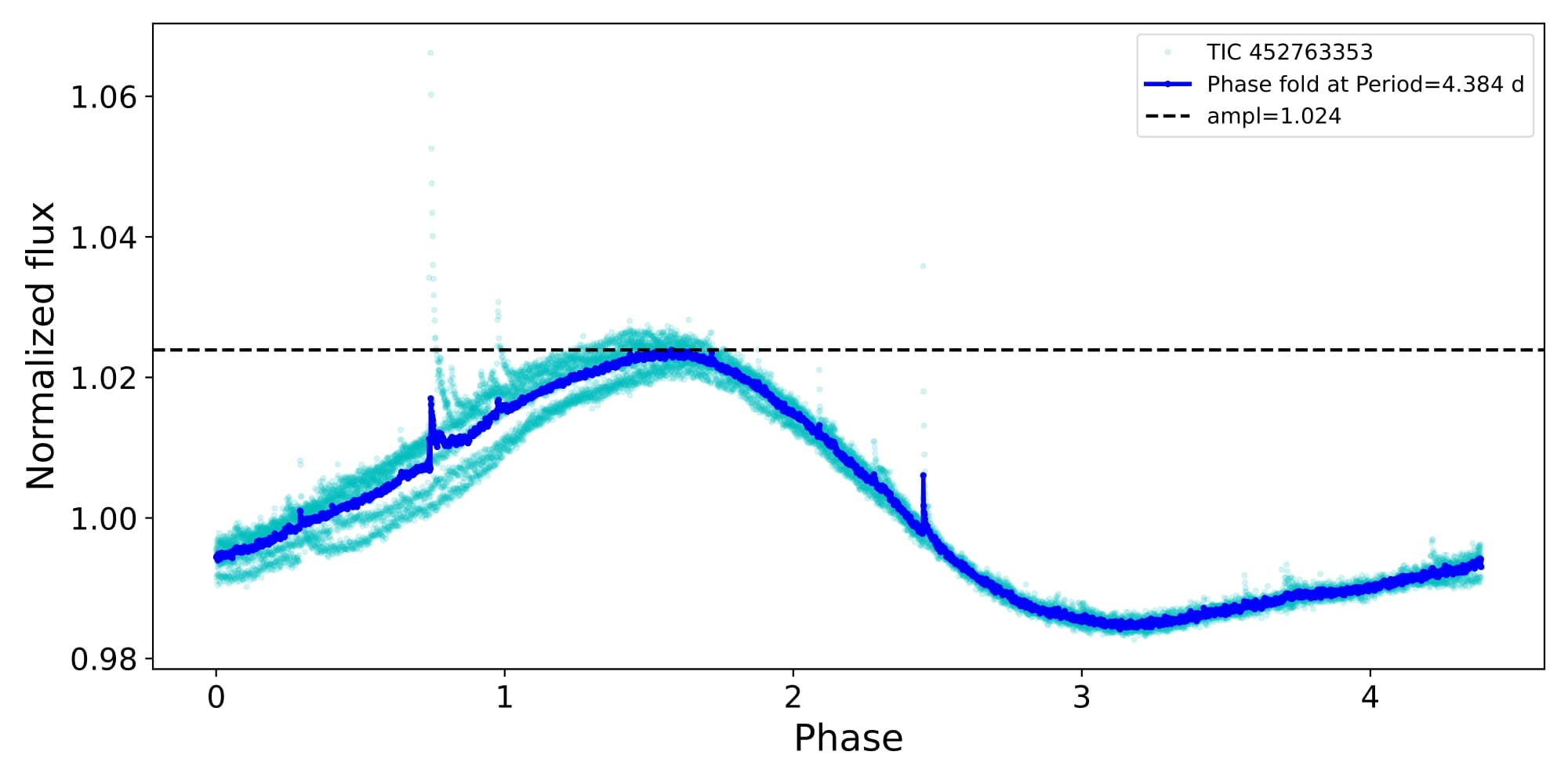}
    \quad\includegraphics[width=0.45\linewidth]{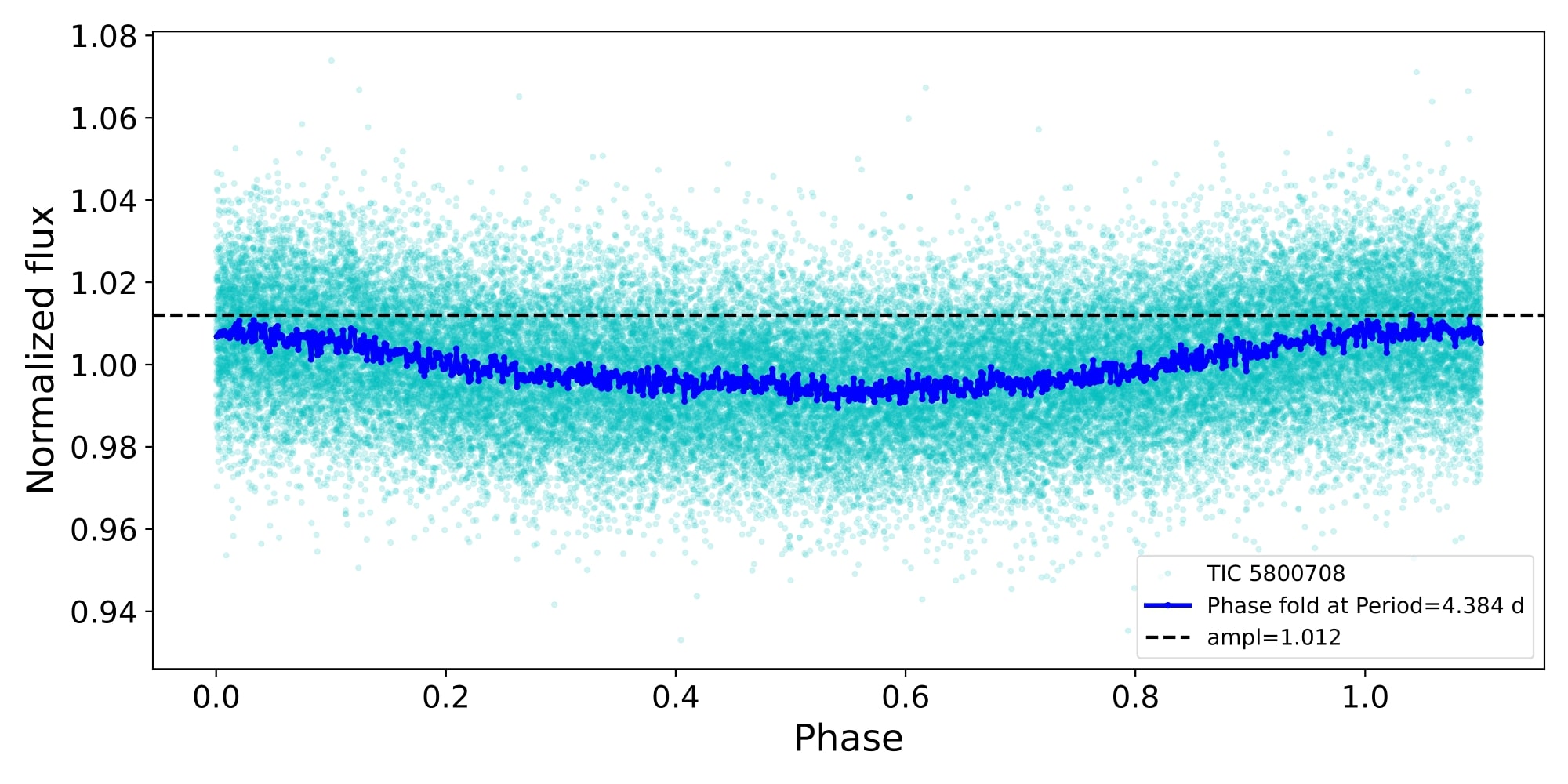}
    \caption{The cyan curve represent the phase light curve of GJ~182 (left) in sector~32 and  2MASS~J056+2214 (right) from all sectors. The blue line indicates the binning light curve and the black dashed line shows the estimated maximal amplitude, used for spot modeling.}
    \label{fig:ampl_fig}
\end{figure*}

\subsection{GJ~182} GJ~182 has been observed in TESS sector 5 (in November 2018; Camera 1 and CCD 2) and sector 32 (in December 2020; Camera 1 and CCD 2) with roughly two years time gap. The observations were conducted in a 2-min cadence mode and from the light curve we determined the rotation period of 4.35 $\pm$ 0.2 and 4.38 $\pm$ 0.4 days for this object in Sector 5 and Sector 32 respectively. Previously, \citet{Kiraga2007AcA....57..149K} reported a rotation period of 4.41 d using ASAS photometric data. A comparison of rotation periods with previous studies was mentioned in Table~\ref{tab:rot_per}.  \citet{Byrne1984MNRAS.206..907B} suggested that the optical variability of GJ~182 is normally due to the presence of large dark spots on the surface. From the TESS observation, we observed the shape of the light curve between sector 5 and sector 32 changes significantly, which reflects the evolutionary changes in the surface phenomena. To find out the distribution of the starspots in both sectors, the light curve was reconstructed by a two-spot model in both sectors with the inclination angle of 60$^{\circ}$ \citep{Donati2008MNRAS.390..545D}. From the light curve, we rejected outlier points exceeding 0.05$\sigma$ above the mean and manually removed the parts of flare events from the light curve. 

\begin{figure*}
    \includegraphics[width=0.45\linewidth]{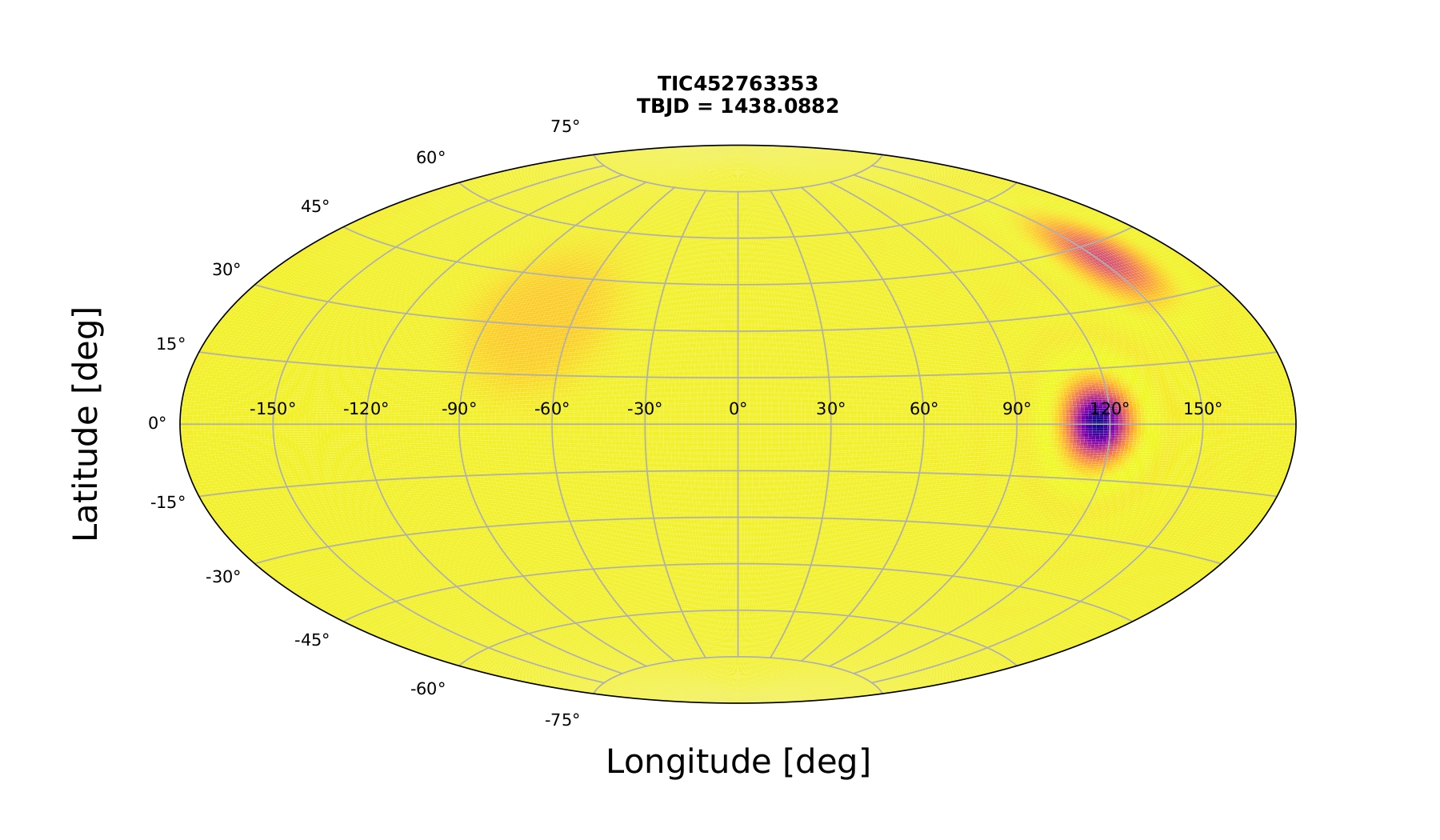}
    \quad\includegraphics[width=0.45\linewidth]{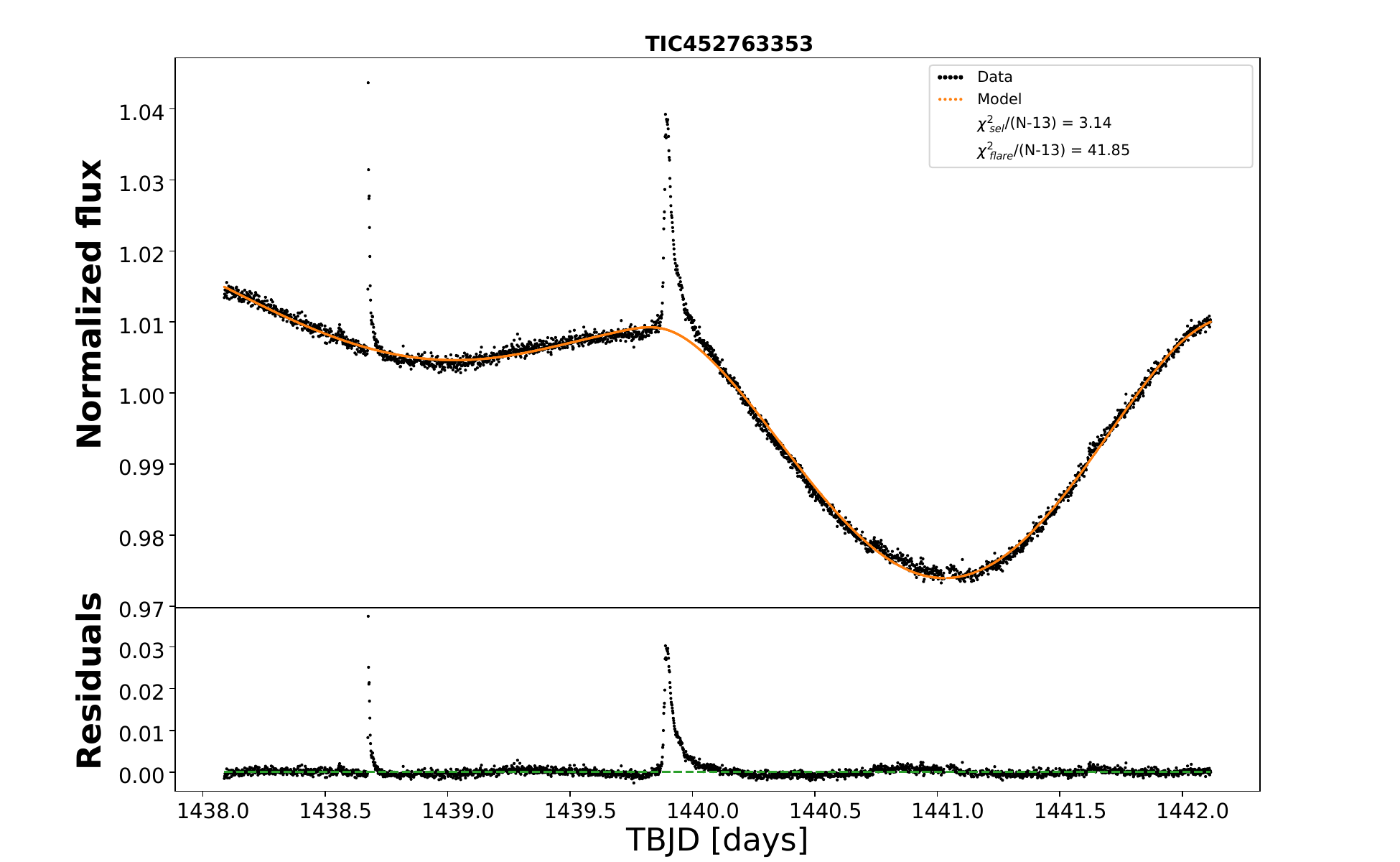}
    
    \quad\includegraphics[width=0.45\linewidth]{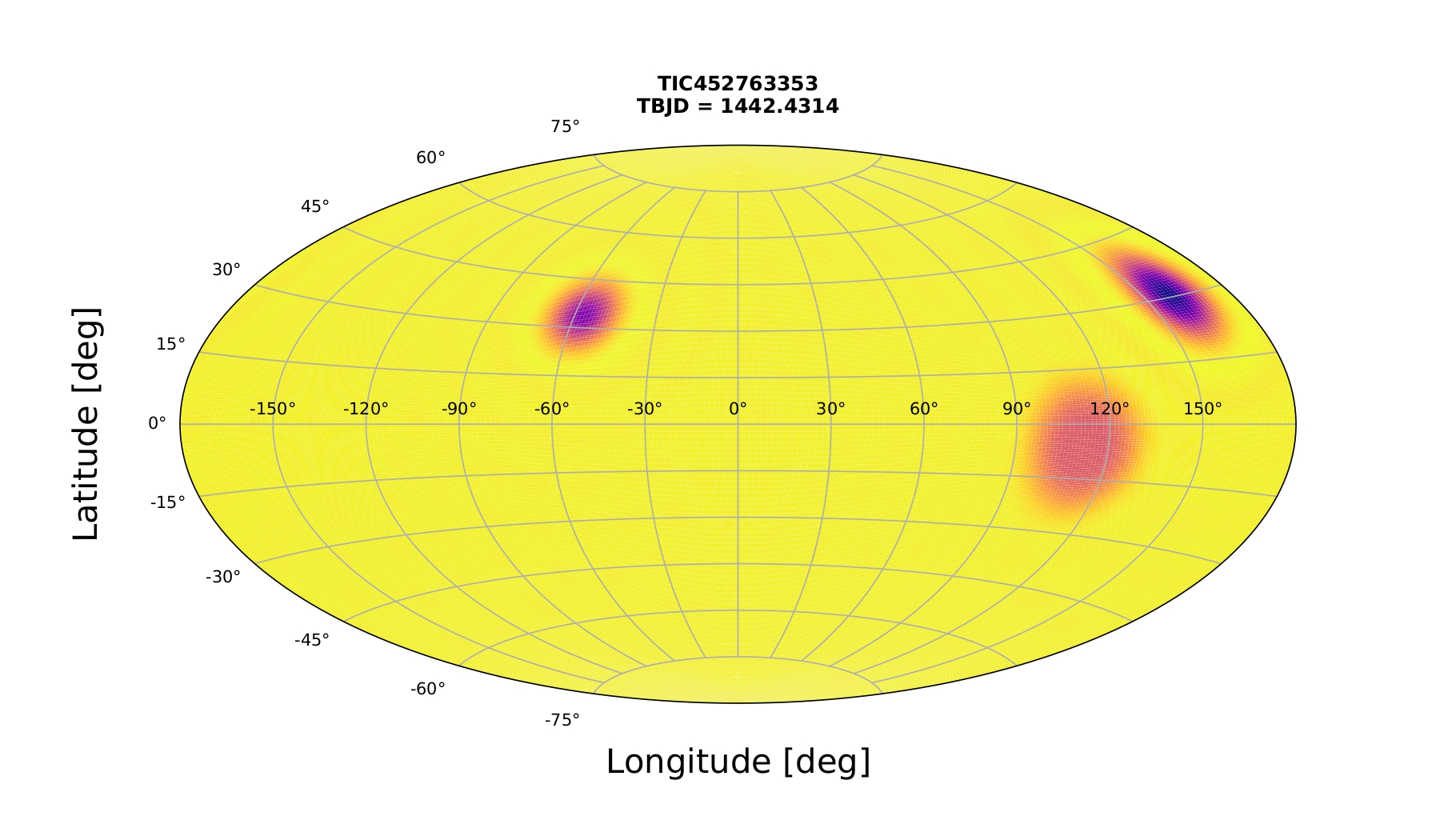}
    \quad\includegraphics[width=0.45\linewidth]{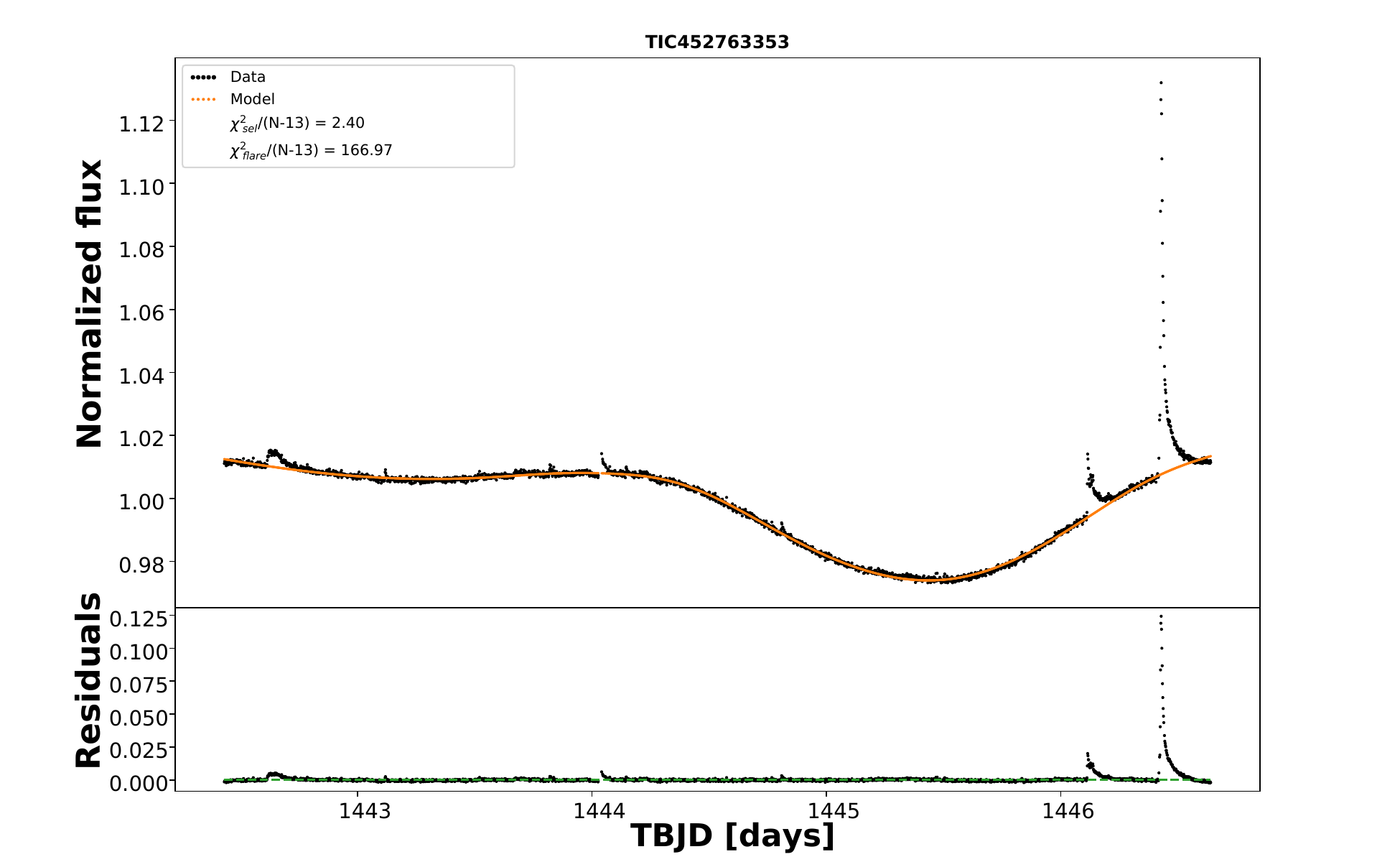}

    \quad\includegraphics[width=0.45\linewidth]{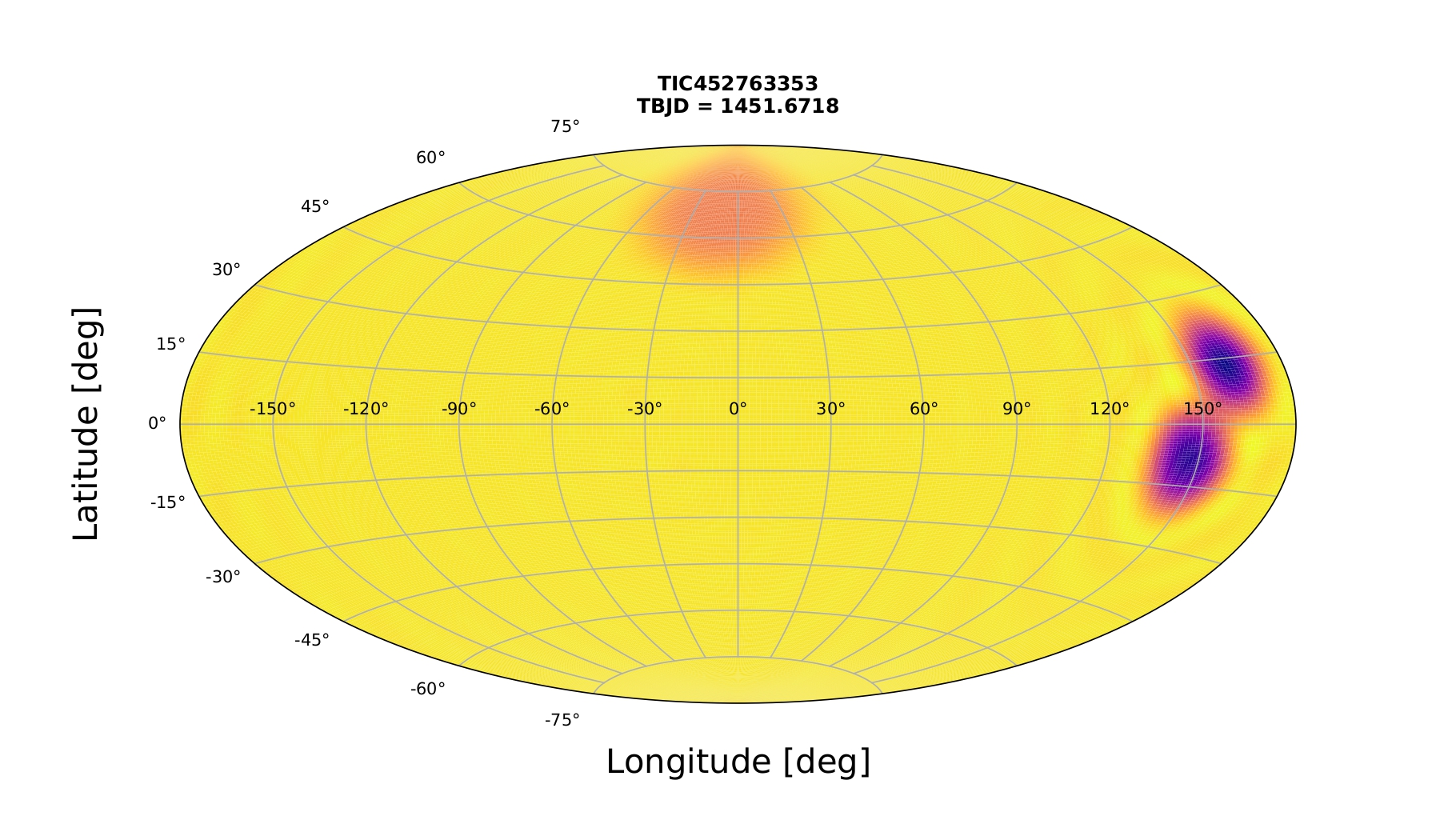}
    \quad\includegraphics[width=0.45\linewidth]{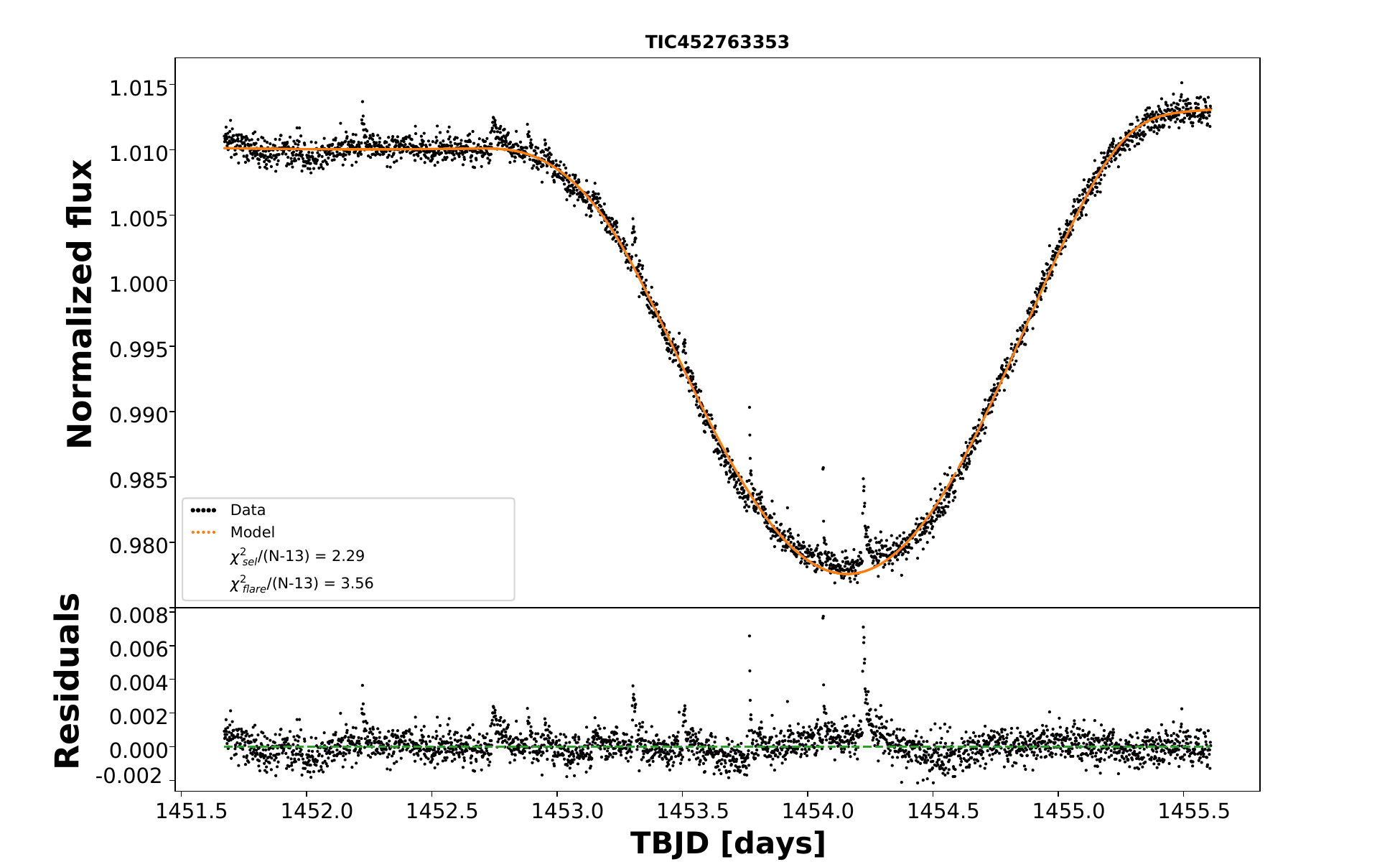}

    \quad\includegraphics[width=0.45\linewidth]{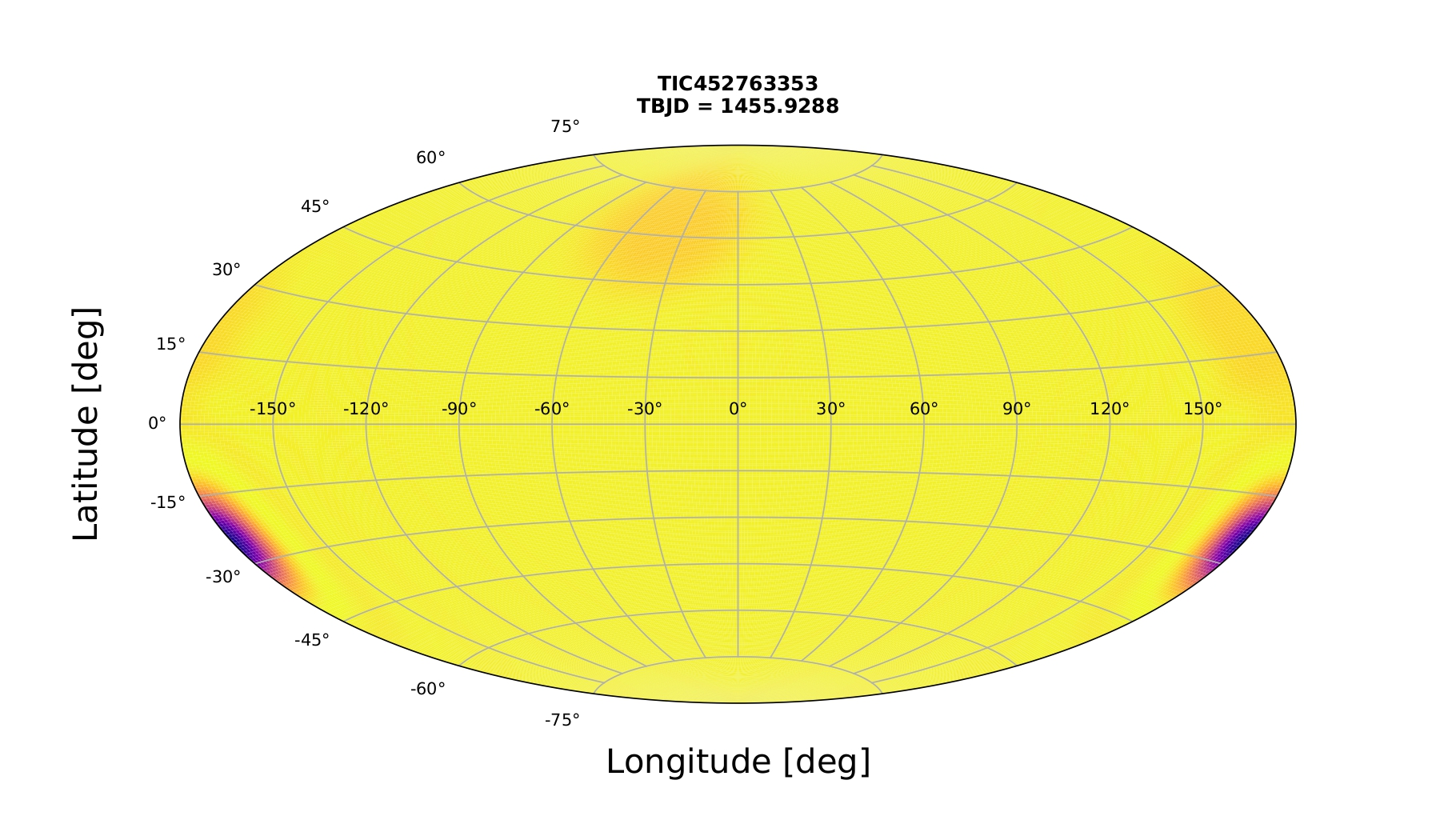}
    \quad\includegraphics[width=0.45\linewidth]{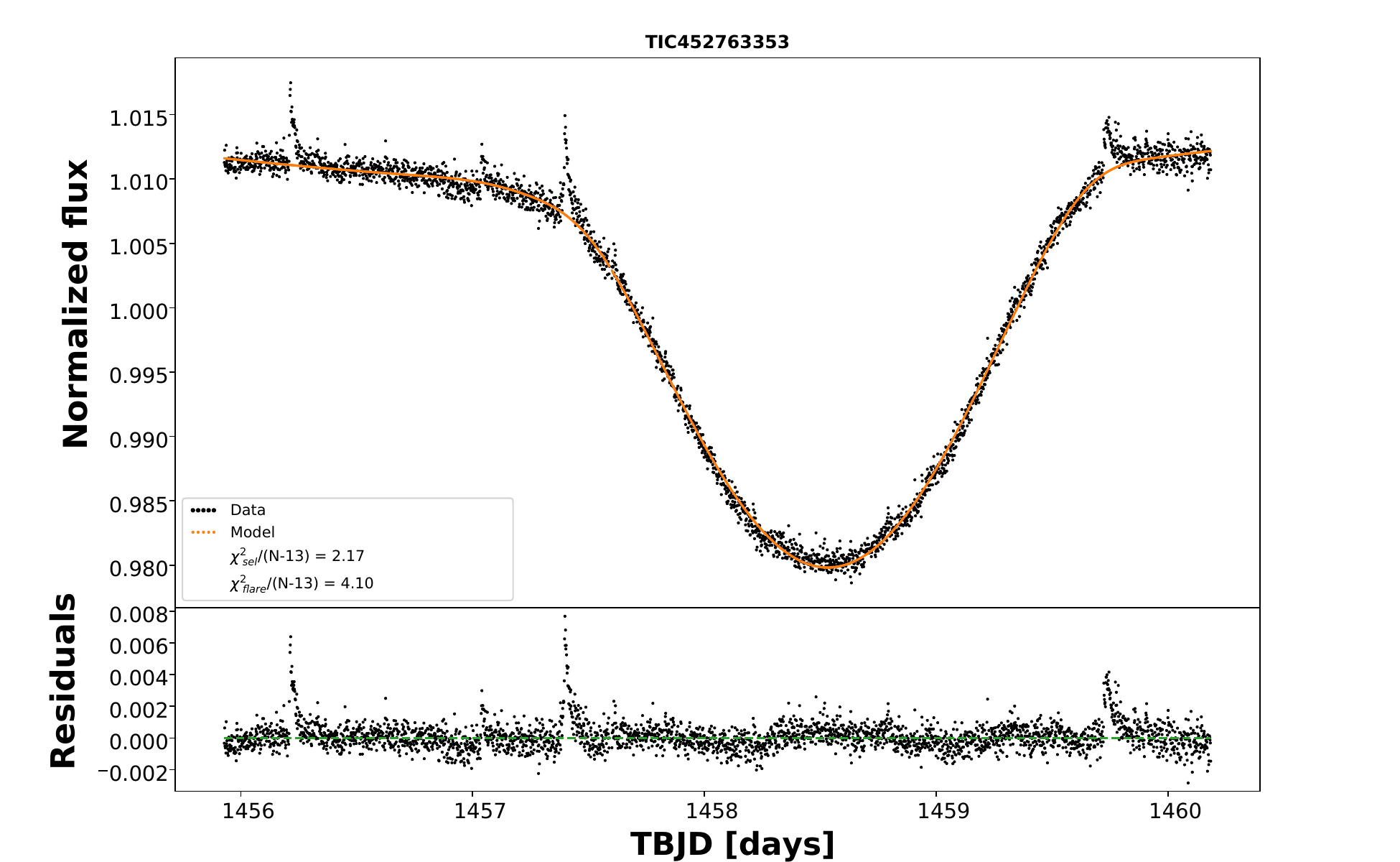}

    \caption{ The left column represents the positions, sizes, and contrast of spots in the Aitoff projection (at phase =0) of GJ 182 in sector 5. The right column shows the observed light curve from TESS (black dots) and the model-fitted light curve (orange curve) along with their residuals and the reduced chi-square written in the corner of the plot. The upper row corresponds to the first modulation of the light curve, while the next bottom row represents the second modulation of the light curve for GJ 182. Additionally, the fourth, and fifth rotations are represented in the third, and fourth rows, respectively.}
    \label{fig:1st_2nd_gjspot_s5}
\end{figure*}

In the case of sector~5, the shape and amplitude of the light curve change significantly in each rotation (Fig \ref{fig:lc_ls_phs}). In the first half of the LC, there is a dip in each rotation and it diminishes in the second half of the LC. As the variability is not constant over time, we modeled the light curve by dividing it into  few individual segments, and each segment contains the full rotation of the object. However, in Sector 5, the 3rd and 6th segments, and in Sector 32, the 3rd
segment, do not contain a full rotation period (see left in Fig \ref{fig:lc_ls_phs}). Consequently, we excluded these incomplete segments from our analysis. It can be allowed to check any spot evolution in the whole LC.   We removed the flare events manually using \textsc{`--rmparts'} command in \textsc{bassman} code from the LC. To estimate the amplitude we used the LC of sector~32 as it has maximum brightness compared to the LC in sector~5. We used the same method as described in \citet{Bicz2022ApJ...935..102B} by phasing the LC and taking the maximum normalized flux value without taking into flare events. In this method, we obtained a maximal amplitude value of 1.024 (see Fig \ref{fig:ampl_fig} (left)) and used this value for our analysis. Furthermore, to better visualize spot evolution during each rotation, we used the model derived from one rotation as a starting point for the next. Additionally, we applied the value of the differential coefficient of 0.06 $\pm$ 0.03 rad/day for our analysis, adopted from \citet{Donati2008MNRAS.390..545D}. In sector~5, each segment of the LC  was reconstructed by a three-spot model with a spot relatively at a higher latitude (ranging  30$^{\circ}$ - 70$^{\circ}$) and others near the equator. We also tested a two-spot model but it had a lower log-probability value compared to the three-spot model. So, we chose the three-spot model to fit the LCs better. Similarly, in sector~32  we modeled each segment of the LC using both two-spot and three-spot models. Again, the three-spot model was well-fitted to the LCs with higher log-probability values compared to the two-spot model. So, the three-spot model is more reliable for capturing the data from the light curve, with one spot located relatively at a lower-latitude region (ranging from -47$^{\circ}$ to -55$^{\circ}$) whereas the other two were situated at the higher mid-latitude region. Thus, the three-spot model is a more reliable choice in both sectors.
The estimated parameters of starspots and comparison with analytic solution are shown in Table \ref{tab:spots_pars_sec5}, Table \ref{tab:gj_spot_comp_s5}  for sector~5 and in Table \ref{tab:spots_pars_sec32}, Table \ref{tab:gj_spot_comp_s32} for sector~32. The figures were decorated in Figure \ref{fig:1st_2nd_gjspot_s5}  (sector~5) and Figure \ref{fig:gjspot_s32} (sector~32). Note that the reduced chi-square values greater than 1.0 in Table \ref{tab:gj_spot_comp_s5} do not indicate underfit but rather indicate the short-term light fluctuation of the TESS light curves.

In addition, we have also detected  48 flare events within the energy range from $10^{32}~to~10^{35}$ erg and the highest number of flares has energy around $10^{33}$ erg. Among them, a few of the flares were lasting more than 2.5 hours. The estimated parameters in these flare events are listed in  Table \ref{tab:flr_pars_sec5} and \ref{tab:flr_pars_sec32}. Previously, \citet{Byrne1984MNRAS.206..907B} recorded four flare events within the energy range from $10^{32}$ to $10^{34}$ erg from the photometric data using the 0.75-m telescope of the South African Astronomical Observatory at Sutherland. We have also discussed the nature of flare events in Section~\ref{sec:diss}.

\begin{figure*}
    
    \includegraphics[width=0.45\linewidth]{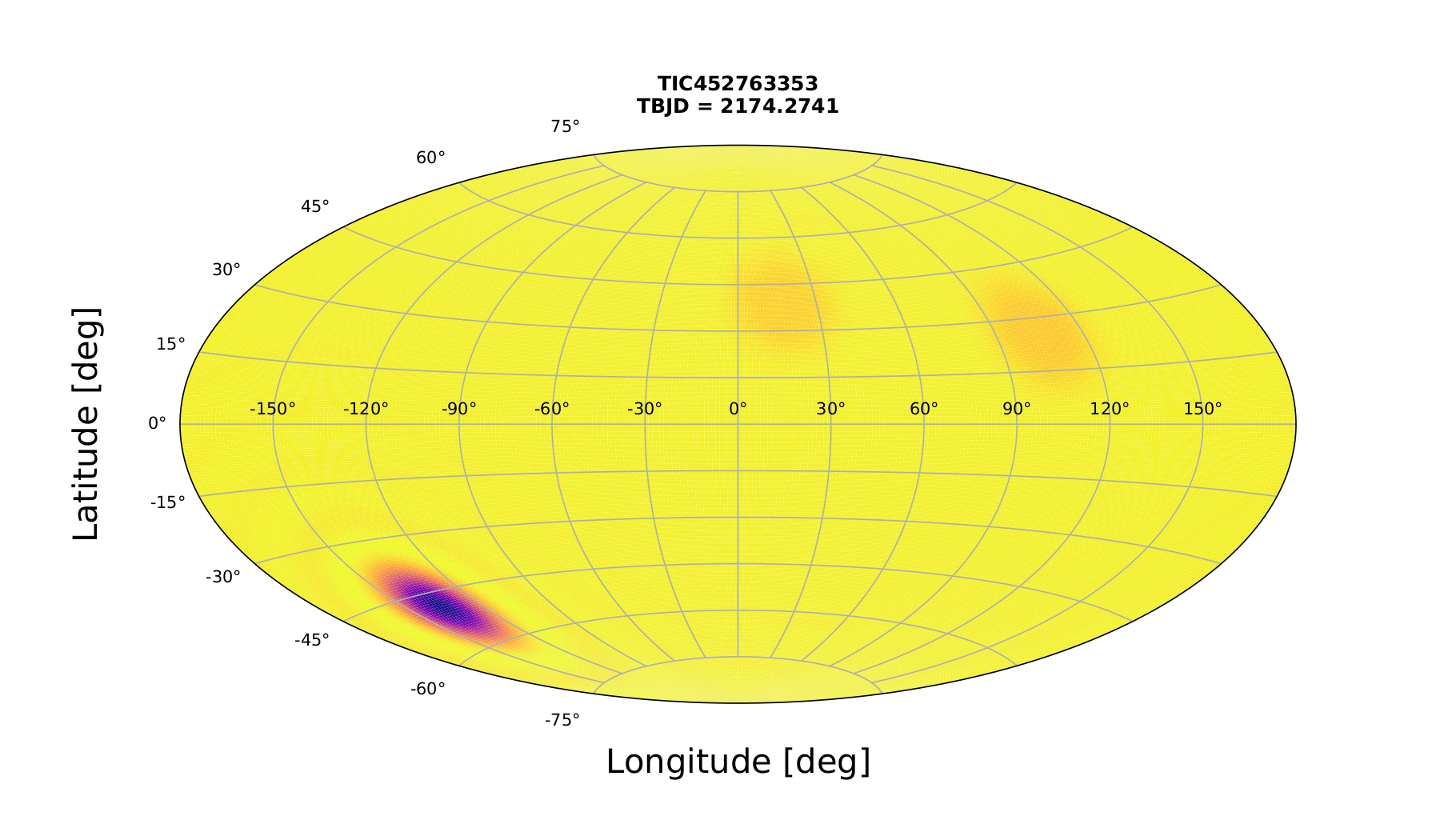}
    \quad\includegraphics[width=0.45\linewidth]{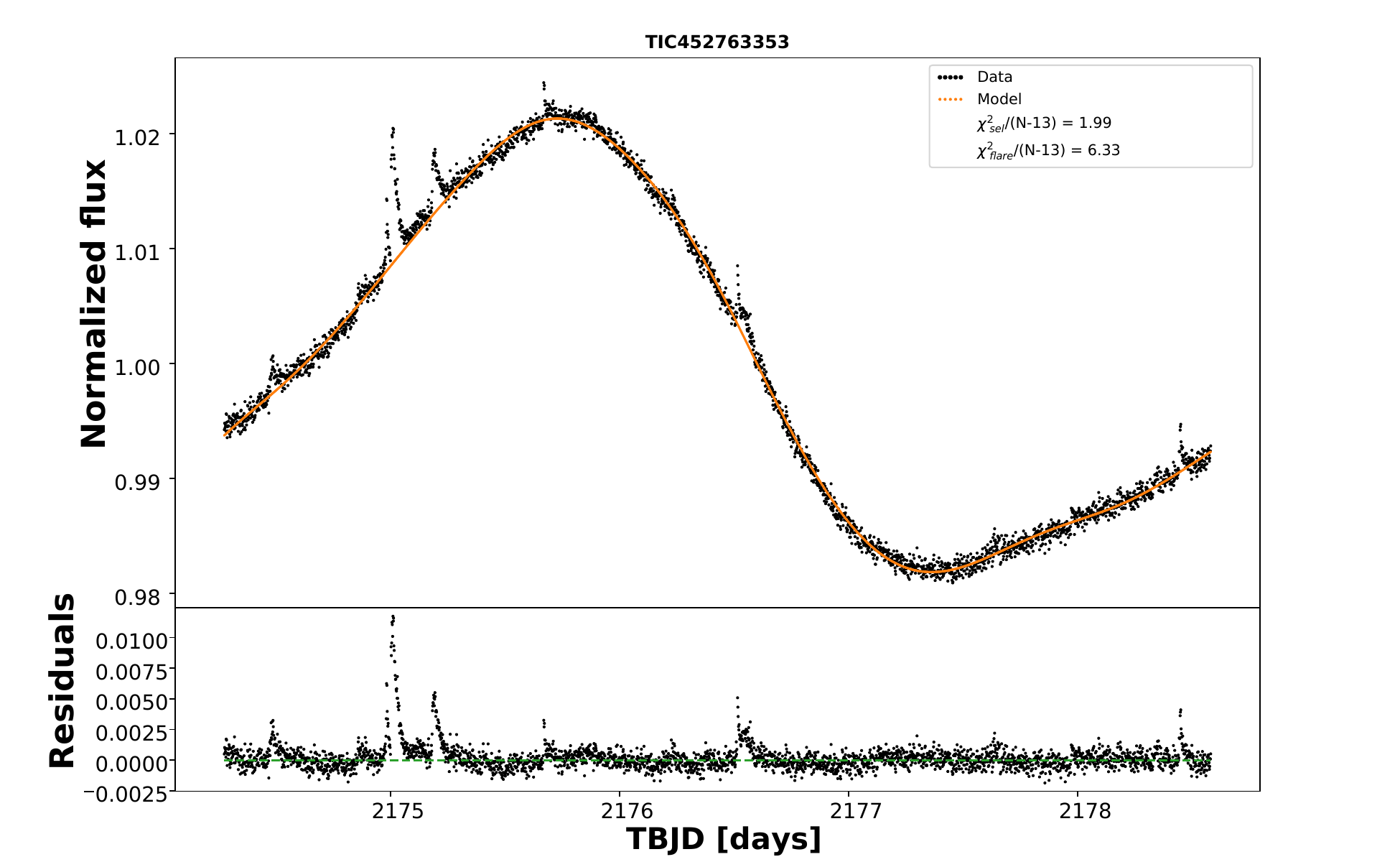}
    
    \includegraphics[width=0.45\linewidth]{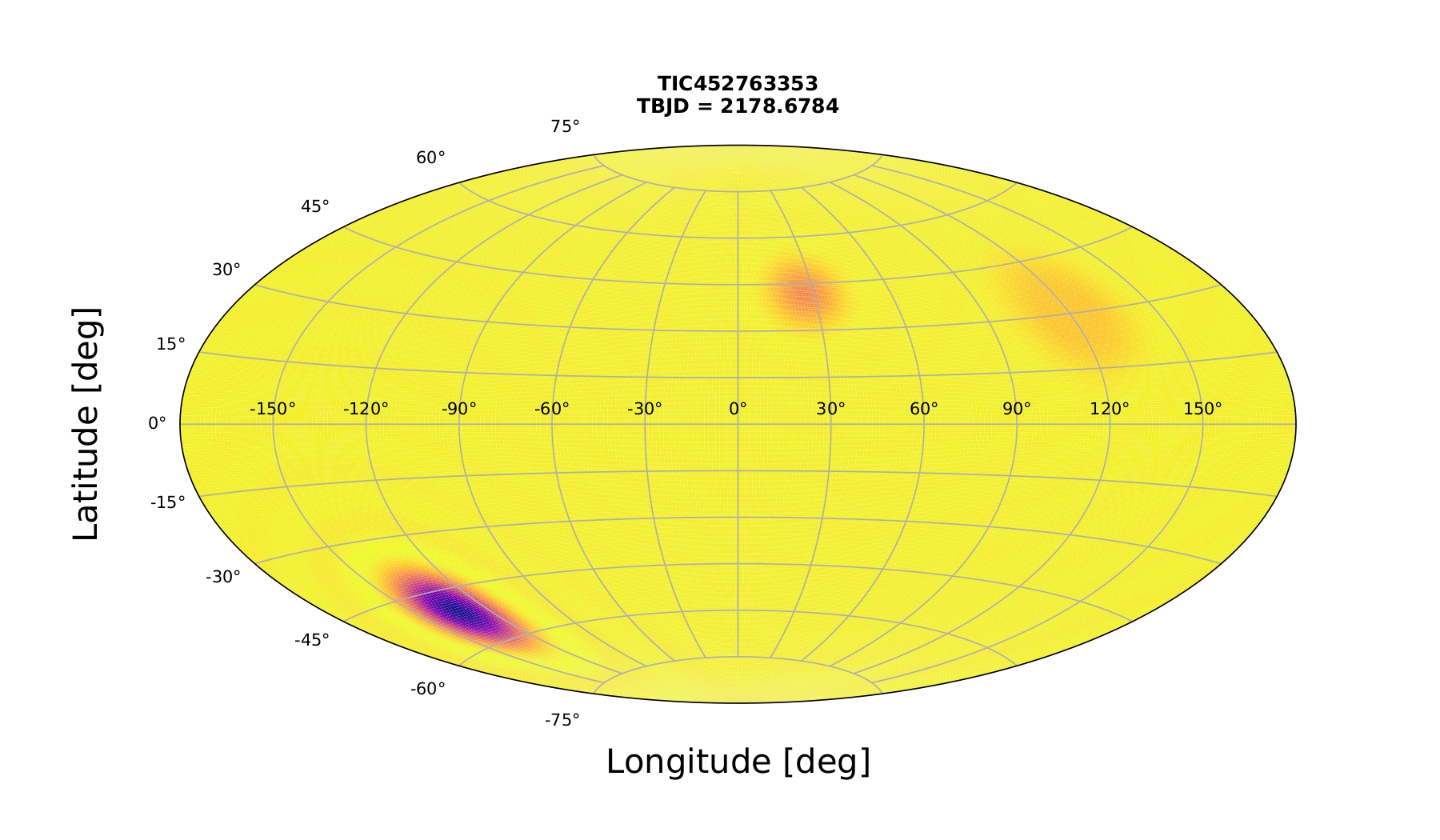}
    \quad\includegraphics[width=0.45\linewidth]{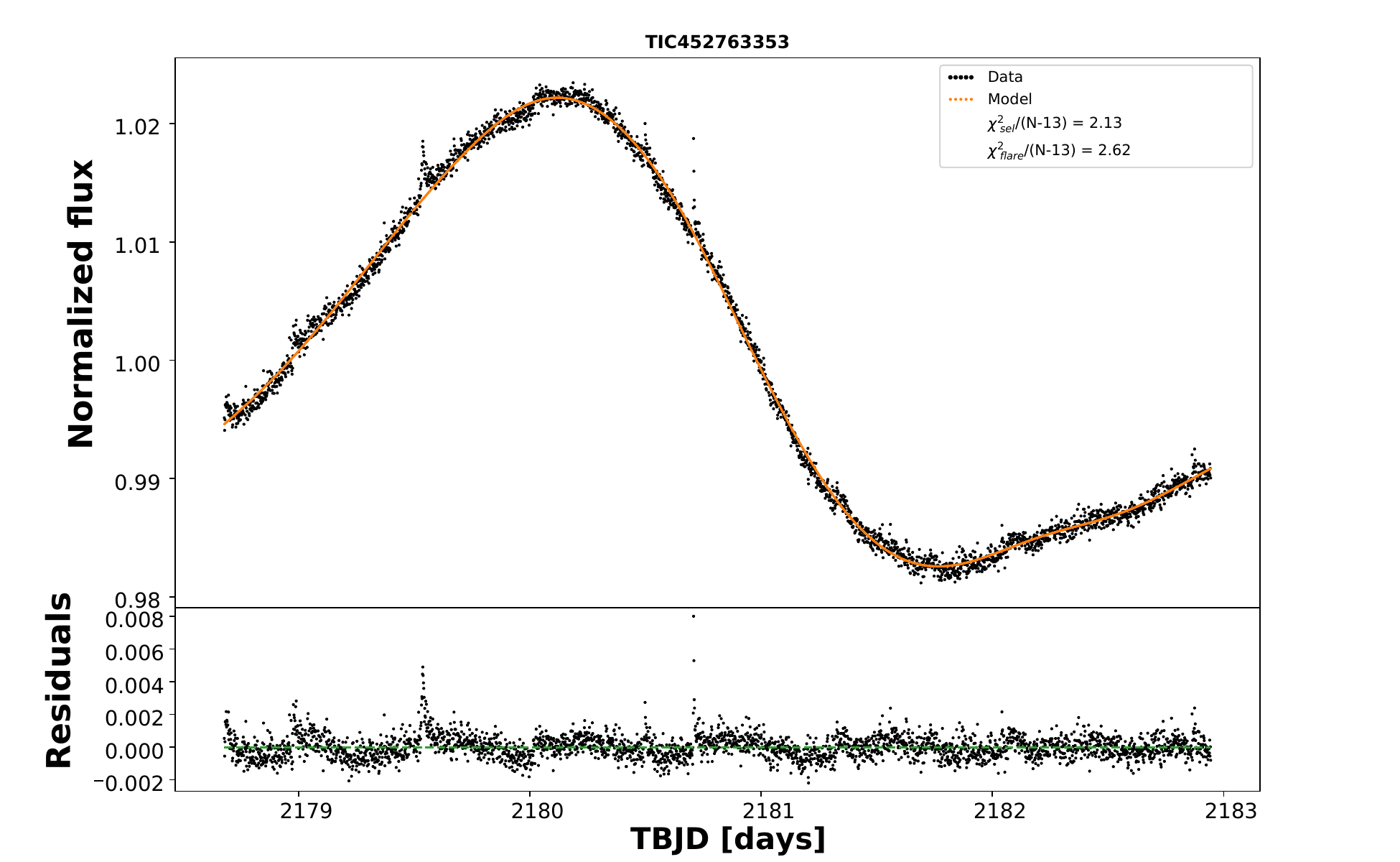}
    
    \includegraphics[width=0.45\linewidth]{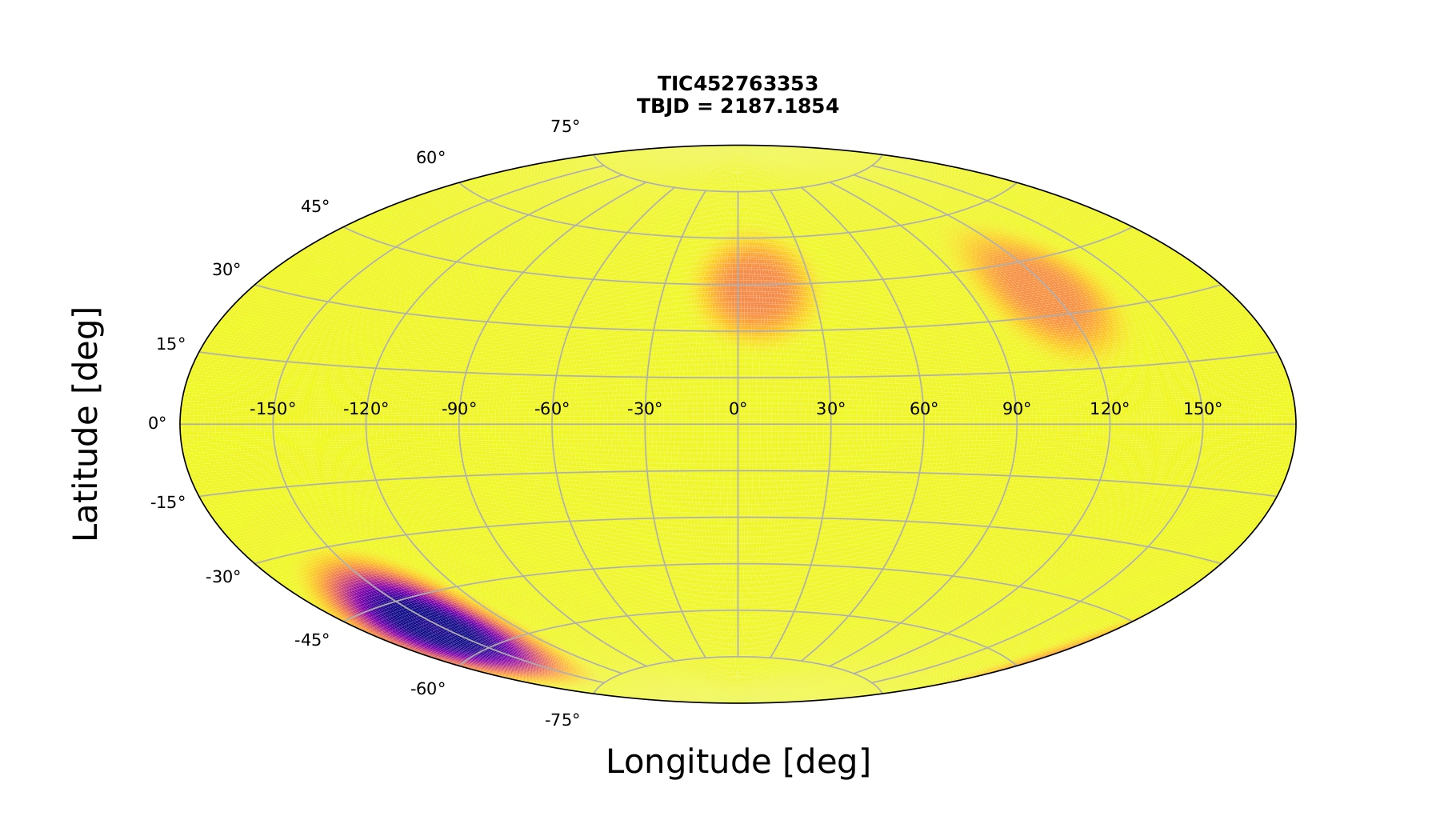}
    \quad\includegraphics[width=0.45\linewidth]{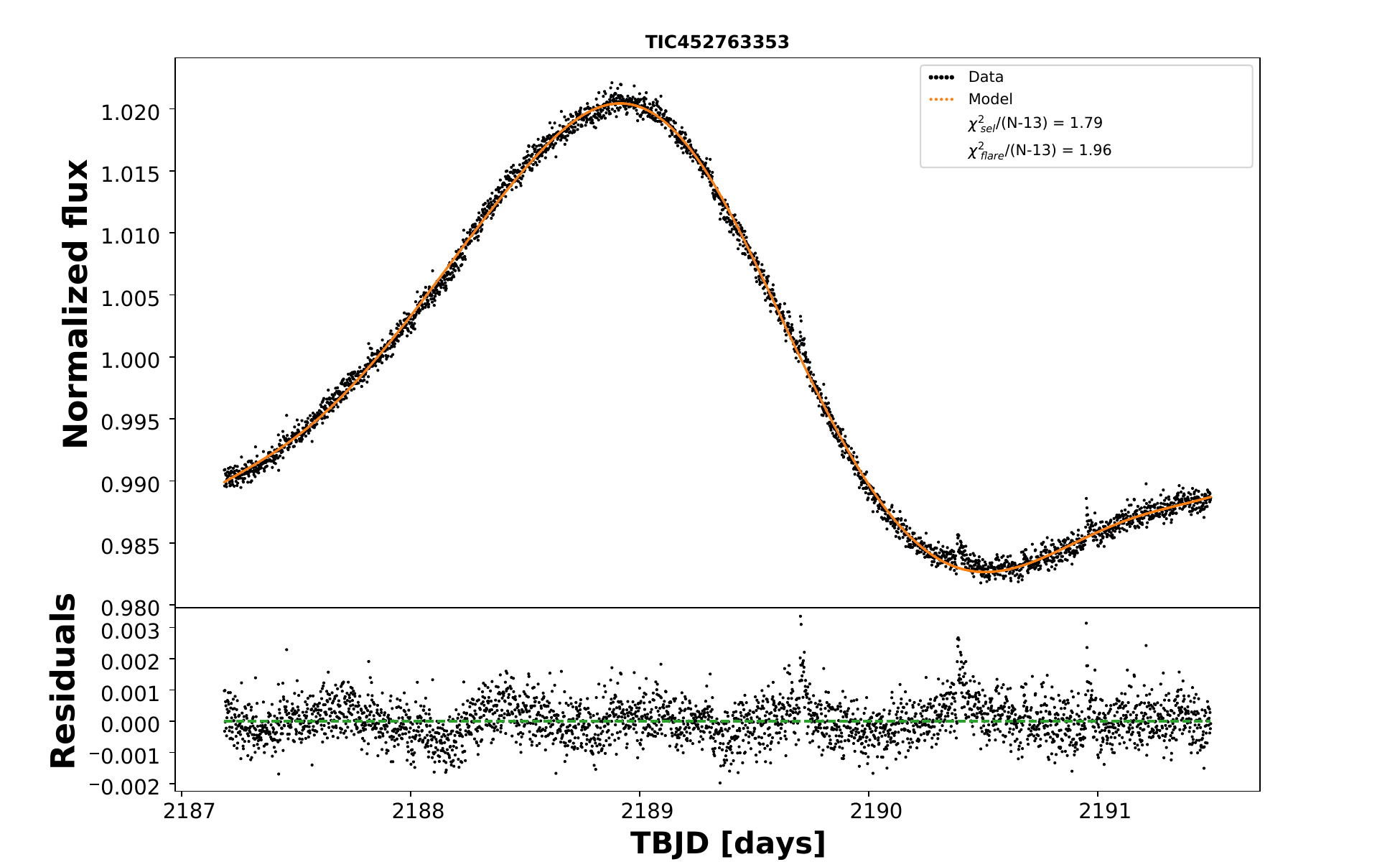}

    \includegraphics[width=0.45\linewidth]{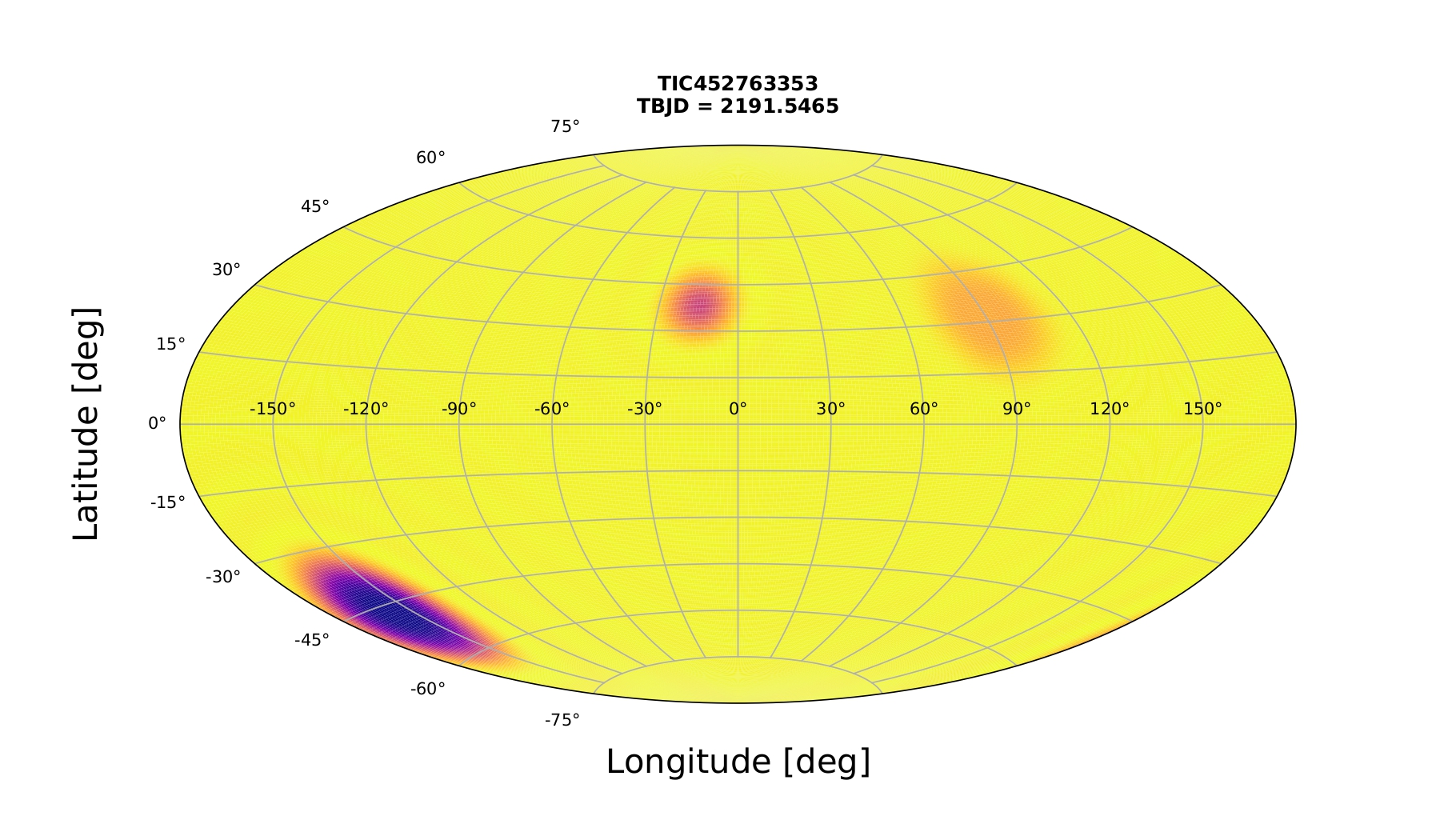}
    \quad\includegraphics[width=0.45\linewidth]{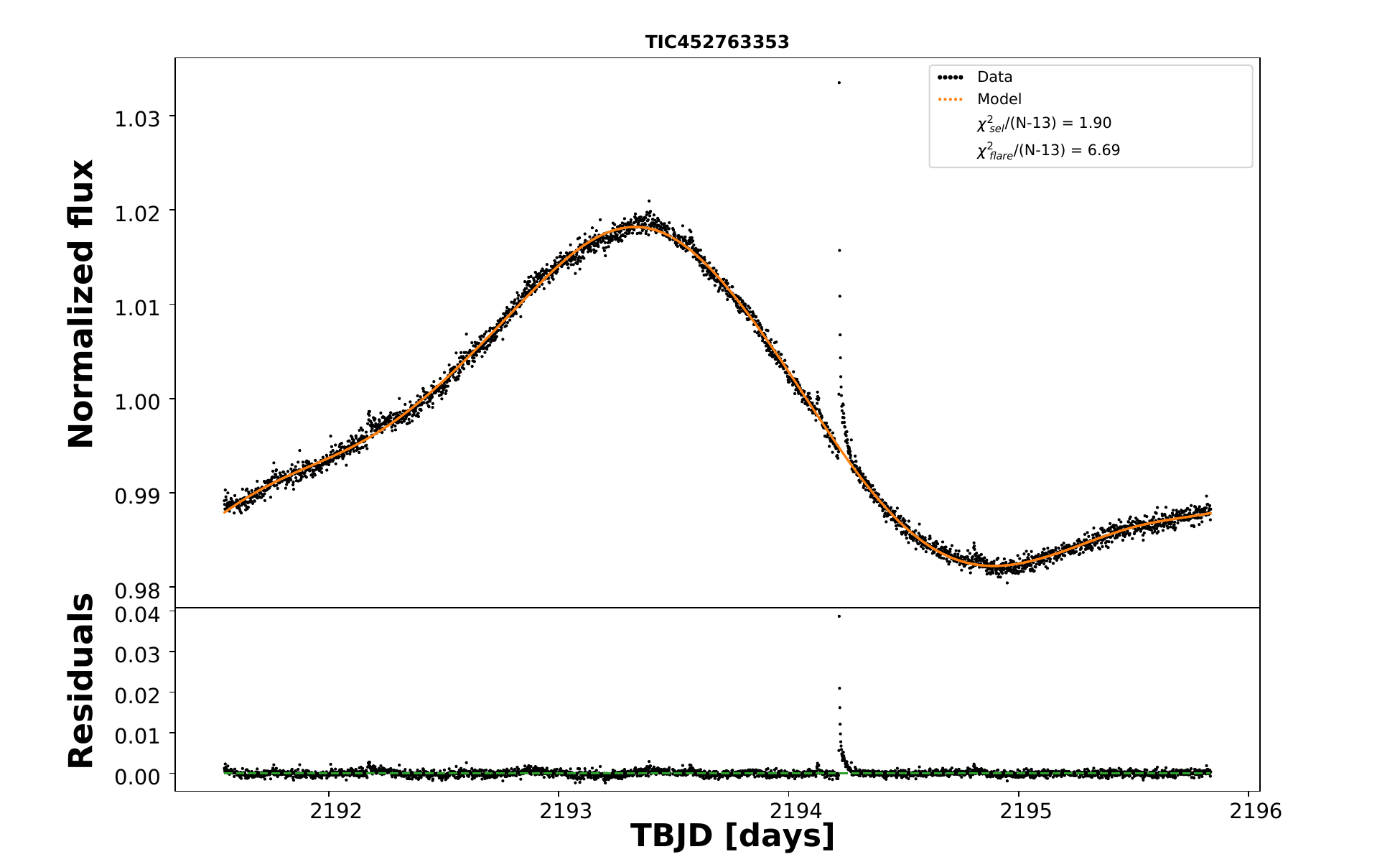}
    
    \caption{The left column represents the positions, sizes and contrast of spots in the Aitoff projection (at phase = 0) for GJ~182 in sector~32. The right column shows the observed light curve from TESS (black dots) and the model-fitted light curve (orange curve) along with their residuals and the reduced chi-square written in the right corner of the plot.  The first, second, fourth, and fifth rotations are represented in the first, second, third, and fourth rows, respectively.}
    \label{fig:gjspot_s32}
\end{figure*}

\begin{figure*}
    \includegraphics[width=0.45\linewidth]{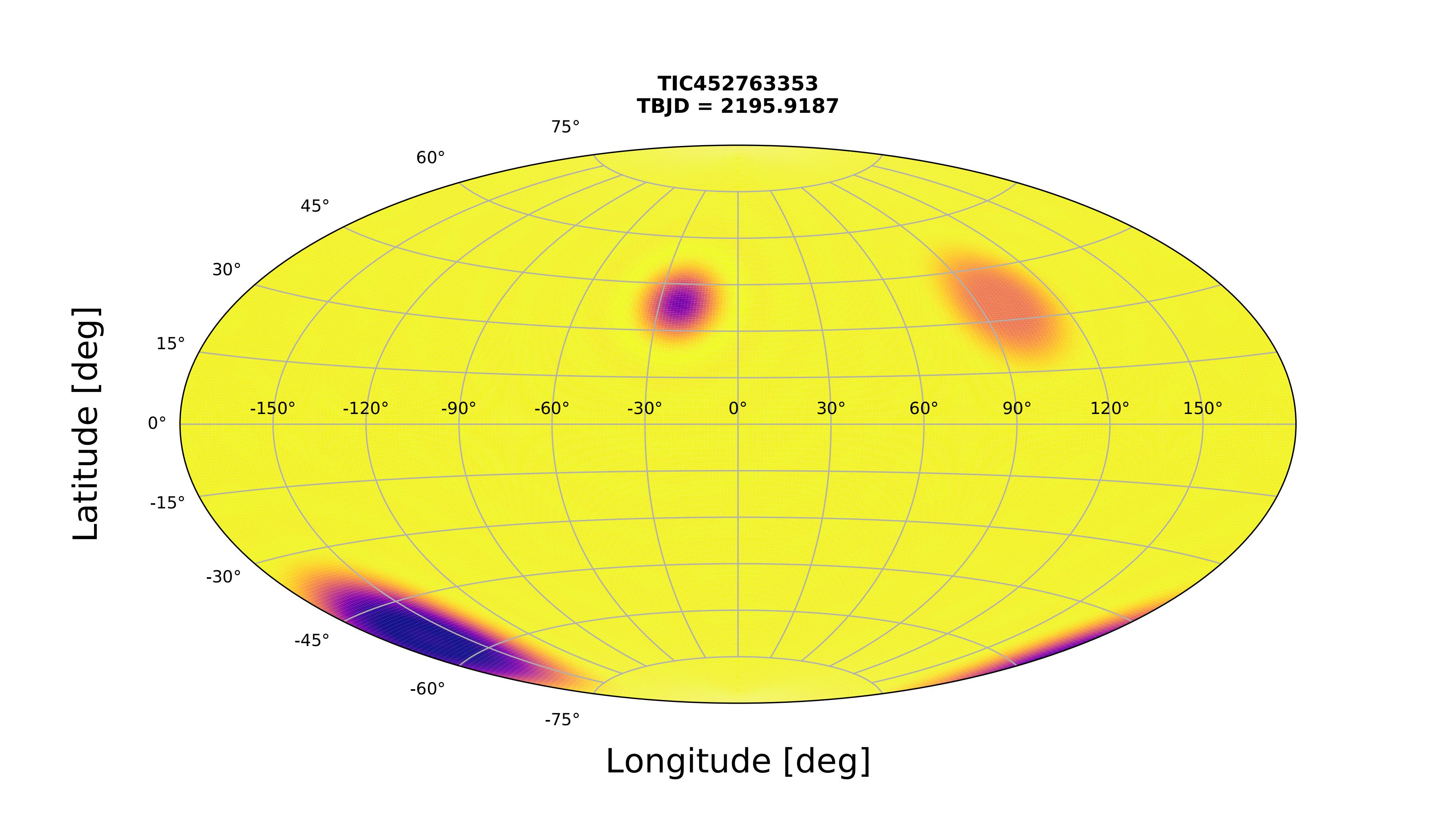}
    \quad\includegraphics[width=0.45\linewidth]{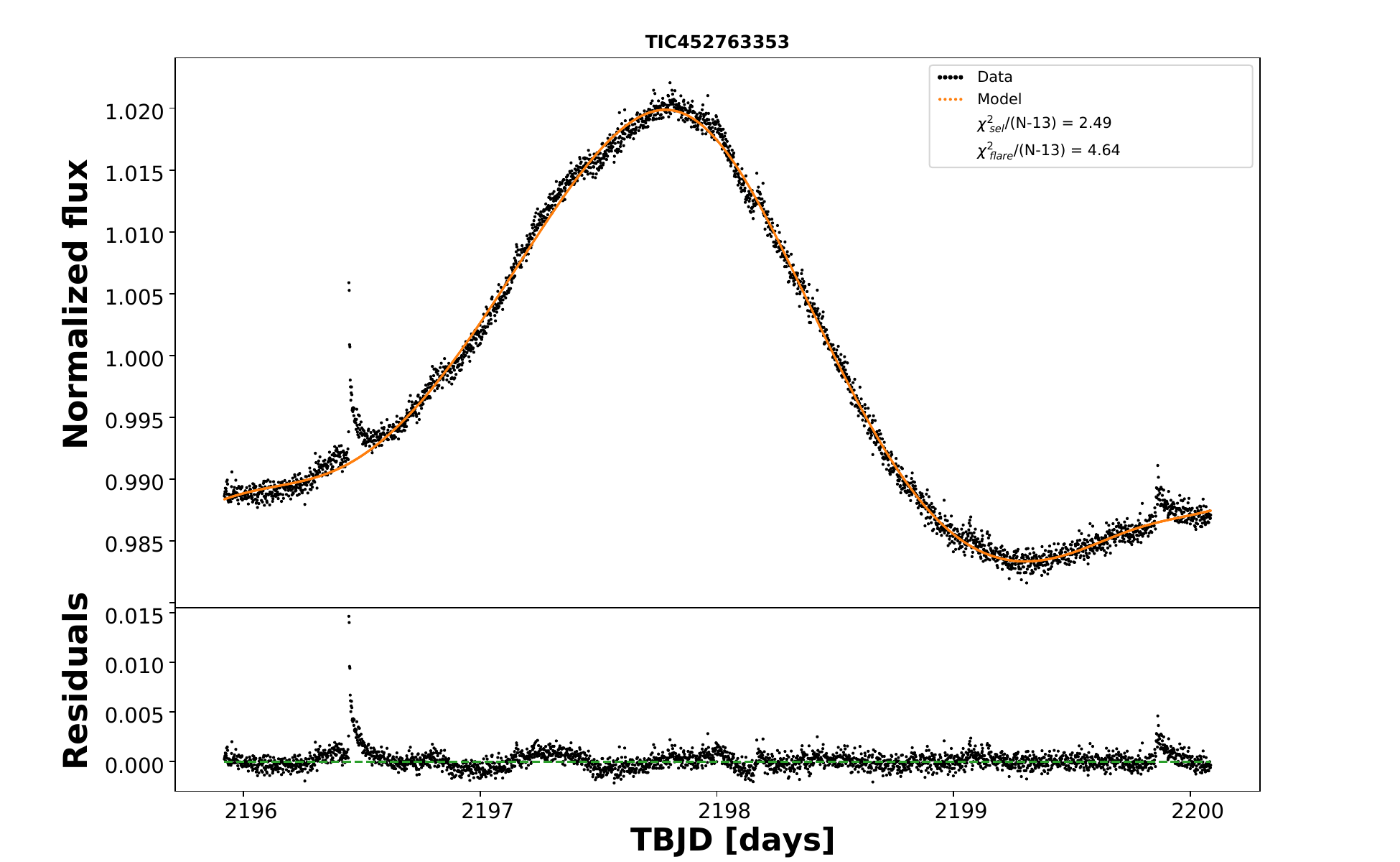}
    \caption{ (Continued) This figure represents the sixth rotation of GJ 182 in sector 32.}
    \label{fig:gjspot_s32_}
\end{figure*}

\begin{longrotatetable}
\setlength{\tabcolsep}{3.5pt} 
\renewcommand{\arraystretch}{1.2} 
\begin{deluxetable}{ccccccccccccccc}
\tablecaption{Estimated spots parameters of GJ~182 (sector~5) in this work.}
\label{tab:spots_pars_sec5}
\tablehead{
\colhead{Modulation} & \colhead{Start} & \colhead{Stop} & \colhead{Lat1} & \colhead{Lat2} & \colhead{Lat3} & \colhead{Initial} & \colhead{Initial} & \colhead{Initial} &  \colhead{T$_{spot1}$} &  \colhead{T$_{spot2}$} & \colhead{T$_{spot3}$} & \colhead{Size1} &  \colhead{Size2}  & \colhead{Size3}\\
\colhead{} & \colhead{Time} &  \colhead{Time} &  \colhead{($\phi_{1}$)} &  \colhead{($\phi_{2}$)} & \colhead{($\phi_{3}$)} & \colhead{Long1} &  \colhead{Long2} & \colhead{Long3 } & \colhead{} &  \colhead{} & \colhead{} &  \colhead{[$\%$ area} & \colhead{[$\%$ area} & \colhead{[$\%$ area} \\
\colhead{} & \colhead{[BTJD]} &  \colhead{[BTJD]} &  \colhead{[deg]} &  \colhead{[deg]}  &  \colhead{[deg]} & \colhead{$(\Lambda_{1})$ [deg]} &  \colhead{$(\Lambda_{2})$ [deg]}  & \colhead{$(\Lambda_{3})$ [deg]} & \colhead{[K]}  &\colhead{[K]} & \colhead{[K]}  & \colhead{of star]} &  \colhead{of star]} & \colhead{of star]} \\
}
 
\startdata
Rotation-1 & 1438.09 & 1442.12 & 0.06 & 40.90 & 30.82 & 115.23 & 150.76 & -72.22 &  2656 $\pm$ 1089 & 3030 $\pm$ 775 &  3746 $\pm$ 173 & 1.02 & 1.04 & 4.97 \\
Rotation-2 & 1442.43 & 1446.64 & -6.31 & 30.00 & 33.39 & 111.54 &  158.23 & -57.08 & 3374 $\pm$ 523 &  2775 $\pm$ 1002 &  3063 $\pm$ 750 & 3.06 &  1.06 & 1.04 \\
Rotation-4 & 1451.67 & 1455.61 & -8.84 & 13.10 & 67.67 & 146.00 & 160.19 & -10.33 & 3129 $\pm$ 645 &   3137 $\pm$ 642 &  3731 $\pm$ 203 & 1.52 & 1.28 & 4.05 \\
Rotation-5 & 1455.93 & 1460.18 & -25.13 & 18.27 & 61.48 & 179.47 & 173.68 & -39.22 & 2657 $\pm$ 1107 & 3746 $\pm$ 127 &  3742 $\pm$ 182 & 1.09 & 3.13 &       4.27 \\ 
\enddata
\end{deluxetable}

\begin{deluxetable}{cccccccc}
        \tablecaption{ Comparision between model estimated spot parameters and the analytically estimated spot parameters of GJ~182 in sector~5 in each modulation.}
        \label{tab:gj_spot_comp_s5}
        \tablehead{
        \colhead{Modulation} & \colhead{Analytic Mean Spot} & \colhead{Model Mean Spot} & \colhead{Analytical Spot Size} & \colhead{Model Spot Size} & \colhead{SNR} & \colhead{log-Probability} & \colhead{Reduced} \\
        \colhead{} & \colhead{Temperature [K]} & \colhead{Temperature [K]} & \colhead{[\% area of star]} & \colhead{[\% area of star]} & \colhead{} &\colhead{} & \colhead{chi-square}
        }
         
        \startdata
        Rotation-1 & 3176 $\pm$ 68 & 3550 & 8.98 $\pm$ 1.28 & 7.03 & 1298 & 6974 & 3.14  \\
        Rotation-2 & 3176 $\pm$ 68 & 3214 & 8.98 $\pm$ 1.28 & 5.15 & 1483 & 7502 & 2.40  \\
        Rotation-4 & 3176 $\pm$ 68 & 3522 & 8.33 $\pm$ 1.19 & 6.85 & 1519 & 7094 & 2.29  \\
        Rotation-5 & 3176 $\pm$ 68 & 3650 & 7.93 $\pm$ 1.13 & 8.49 & 1558 & 7819 & 2.18  \\
        \enddata
\end{deluxetable}
\end{longrotatetable}

%
\begin{longrotatetable}
\setlength{\tabcolsep}{3.5pt} 
\renewcommand{\arraystretch}{1.2} 
\begin{deluxetable}{ccccccccccccccc}
\tablecaption{Estimated spots parameters of GJ~182 (sector~32) in this work.}
\label{tab:spots_pars_sec32}
\tablehead{
\colhead{Modulation} & \colhead{Start} &  \colhead{Stop} &  \colhead{Lat1} &  \colhead{Lat2} & \colhead{Lat3} & \colhead{Initial} &  \colhead{Initial} & \colhead{Initial} &  \colhead{T$_{spot1}$} &  \colhead{T$_{spot2}$}  & \colhead{T$_{spot3}$} & \colhead{Size1} &  \colhead{Size2}  & \colhead{Size3}\\
\colhead{} & \colhead{Time} &  \colhead{Time} &  \colhead{($\phi_{1}$)} &  \colhead{($\phi_{2}$)} & \colhead{($\phi_{3}$)} & \colhead{Long1} &  \colhead{Long2} & \colhead{Long3 } & \colhead{} &  \colhead{} & \colhead{} &  \colhead{[$\%$ area} & \colhead{[$\%$ area} & \colhead{[$\%$ area} \\
\colhead{} & \colhead{[BTJD]} &  \colhead{[BTJD]} &  \colhead{[deg]} &  \colhead{[deg]}  &  \colhead{[deg]} & \colhead{$(\Lambda_{1})$ [deg]} &  \colhead{$(\Lambda_{2})$ [deg]}  & \colhead{$(\Lambda_{3})$ [deg]} & \colhead{[K]}  &\colhead{[K]} & \colhead{[K]}  & \colhead{of star]} &  \colhead{of star]} & \colhead{of star]}\\
}
 
      \startdata
      Rotation-1 & 2174.27 &       2178.58 &                   37.98 &                  -49.05 &                   25.14 &                          16.44 &                        -136.56 &                         104.69 &   $3334 \pm 559$ &  $2554 \pm 1110$ &  $3223 \pm 684$ &       2.57 &       1.09 &       2.52 \\
      Rotation-2 & 2178.68 &       2182.94 &                   41.26 &                  -50.90 &                   29.62 &                          25.92 &                        -132.76 &                         119.31 &  $2693\pm 1100$ &  $2546\pm 1108$ &  $3269 \pm 642$ &       1.09 &       1.13 &       2.72 \\
      Rotation-4 & 2187.19 &       2191.49 &                   43.06 &                  -52.27 &                   35.88 &                           6.70 &                        -152.74 &                         116.26 &   $3211 \pm 672$ &  $2926 \pm 1106$ &  $3260\pm 631$ &       2.36 &       2.55 &       2.65 \\
      Rotation-5 & 2191.55 &       2195.84 &                   37.70 &                  -47.31 &                   30.73 &                         -15.23 &                        -157.76 &                          89.93 &  $2674 \pm 1092$ &  $2771 \pm 1127$ &  $3285 \pm 619$ &       1.03 &       2.02 &       2.77 \\
      Rotation-6 & 2195.92 &       2200.09 &                   38.25 &                  -53.87 &                   33.75 &                         -22.66 &                        -163.62 &                          97.89 &  $2679\pm 1085$ &  $2946\pm 1102$ &  $3124\pm 768$ &       1.03 &       2.71 &       2.30 \\
      \enddata
\end{deluxetable}
\begin{deluxetable}{cccccccc}
        \tablecaption{Comparision between model estimated spot parameters and the analytically estimated spot parameters of GJ~182 in sector~32 in each modulation.}
        \label{tab:gj_spot_comp_s32}
        \tablehead{
        \colhead{Modulation} & \colhead{Analytic Mean Spot} & \colhead{Model Mean Spot} & \colhead{Analytical Spot Size} & \colhead{Model Spot Size} & \colhead{SNR} & \colhead{log-Probability} & \colhead{Reduced} \\
        \colhead{} & \colhead{Temperature [K]} & \colhead{Temperature [K]} & \colhead{[\% area of star]} & \colhead{[\% area of star]} & \colhead{} &\colhead{} & \colhead{chi-square}
        }
         
        \startdata
        Rotation-1 & 3176 $\pm$ 68 & 3185 & 7.56 $\pm$ 1.08 & 6.18 & 1618 & 7889 & 1.98  \\
        Rotation-2 & 3176 $\pm$ 68 & 3029 & 7.43 $\pm$ 1.06 & 4.95 & 1564 & 8294 & 2.13  \\
        Rotation-4 & 3176 $\pm$ 68 & 3142 & 7.42 $\pm$ 1.06 & 7.55 & 1705 & 8812 & 1.78  \\
        Rotation-5 & 3176 $\pm$ 68 & 3036 & 7.49 $\pm$ 1.07 & 5.82 & 1657 & 8817 & 1.90  \\
        Rotation-6 & 3176 $\pm$ 68 & 2980 & 7.30 $\pm$ 1.04 & 6.04 & 1444 & 7902 & 2.50\\
        \enddata
\end{deluxetable}
\end{longrotatetable}

\subsection{TIC~5800708 (2M0516+2214)}
\begin{figure*}
    \includegraphics[width=0.45\linewidth]{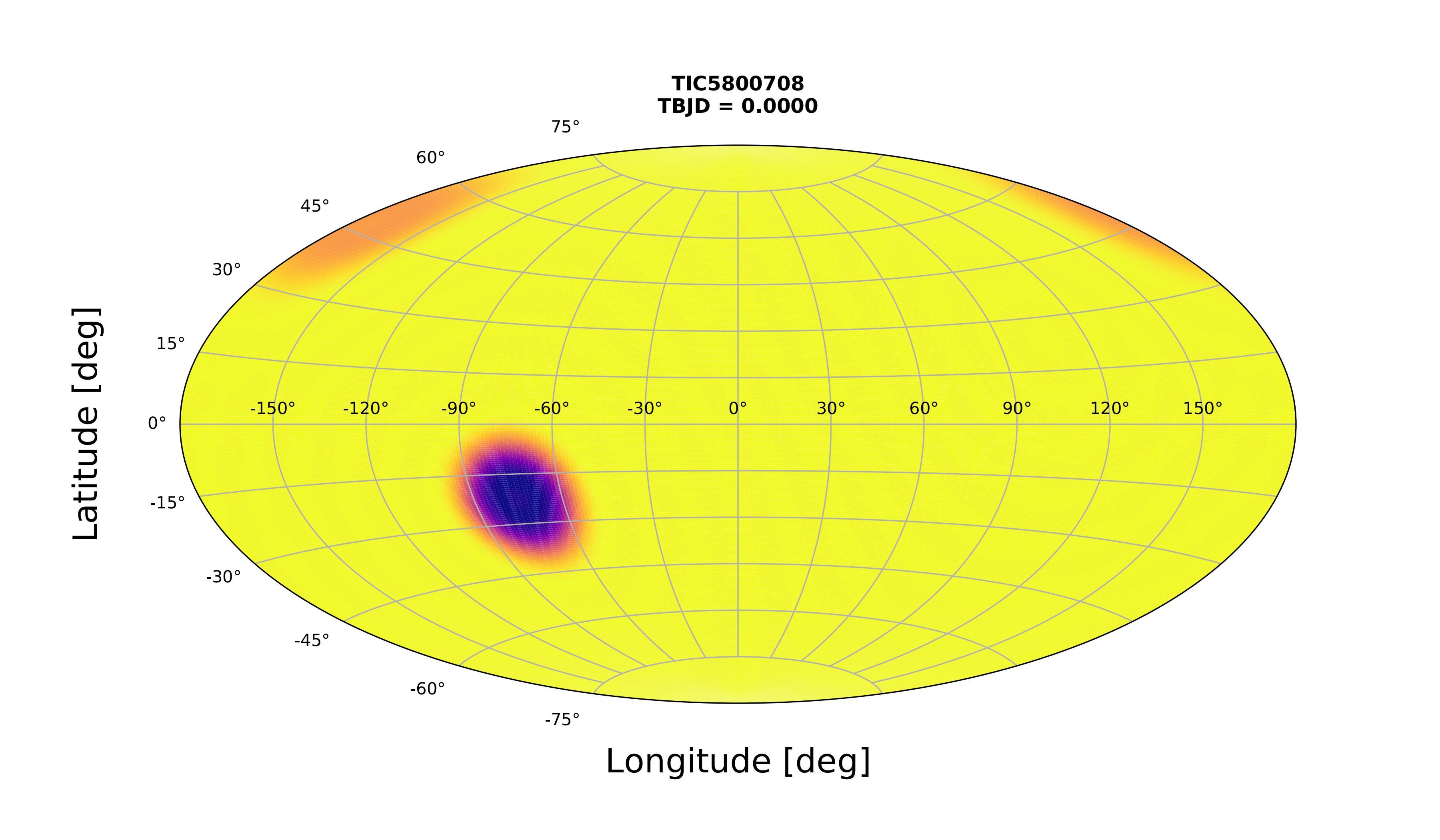}
    \quad\includegraphics[width=0.45\linewidth]{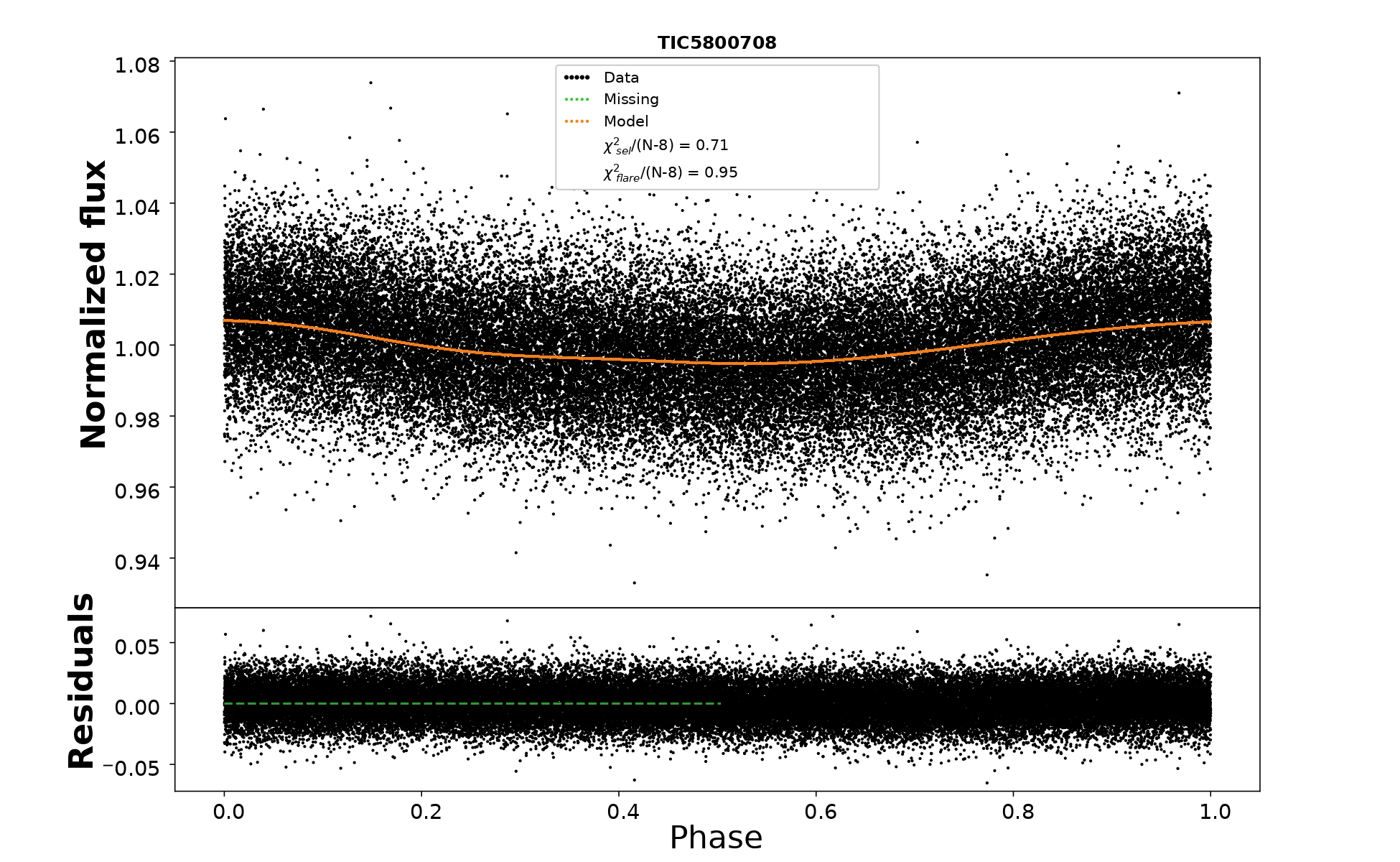}
   
    \quad\includegraphics[width=0.45\linewidth]{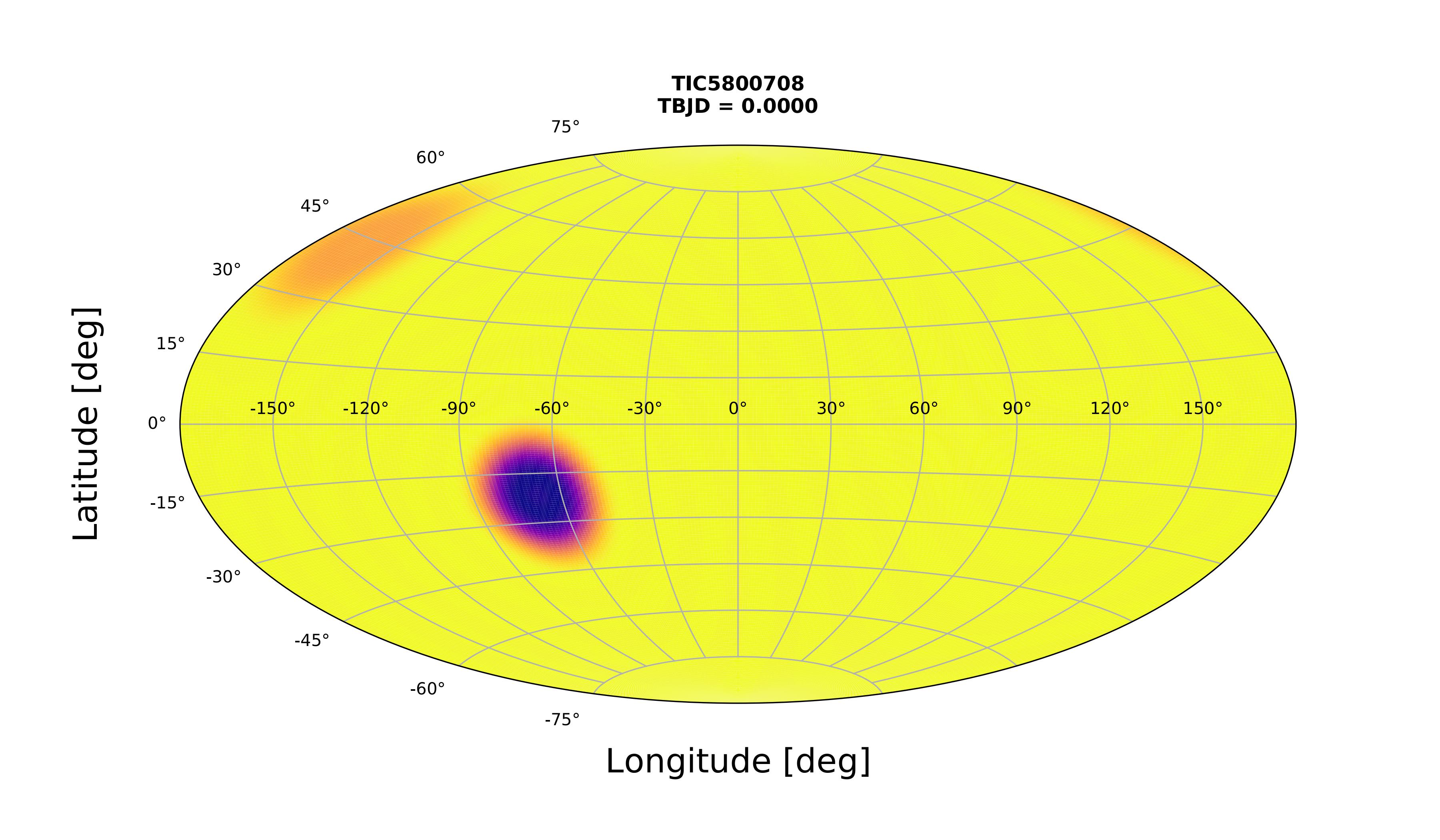}
    \quad\includegraphics[width=0.45\linewidth]{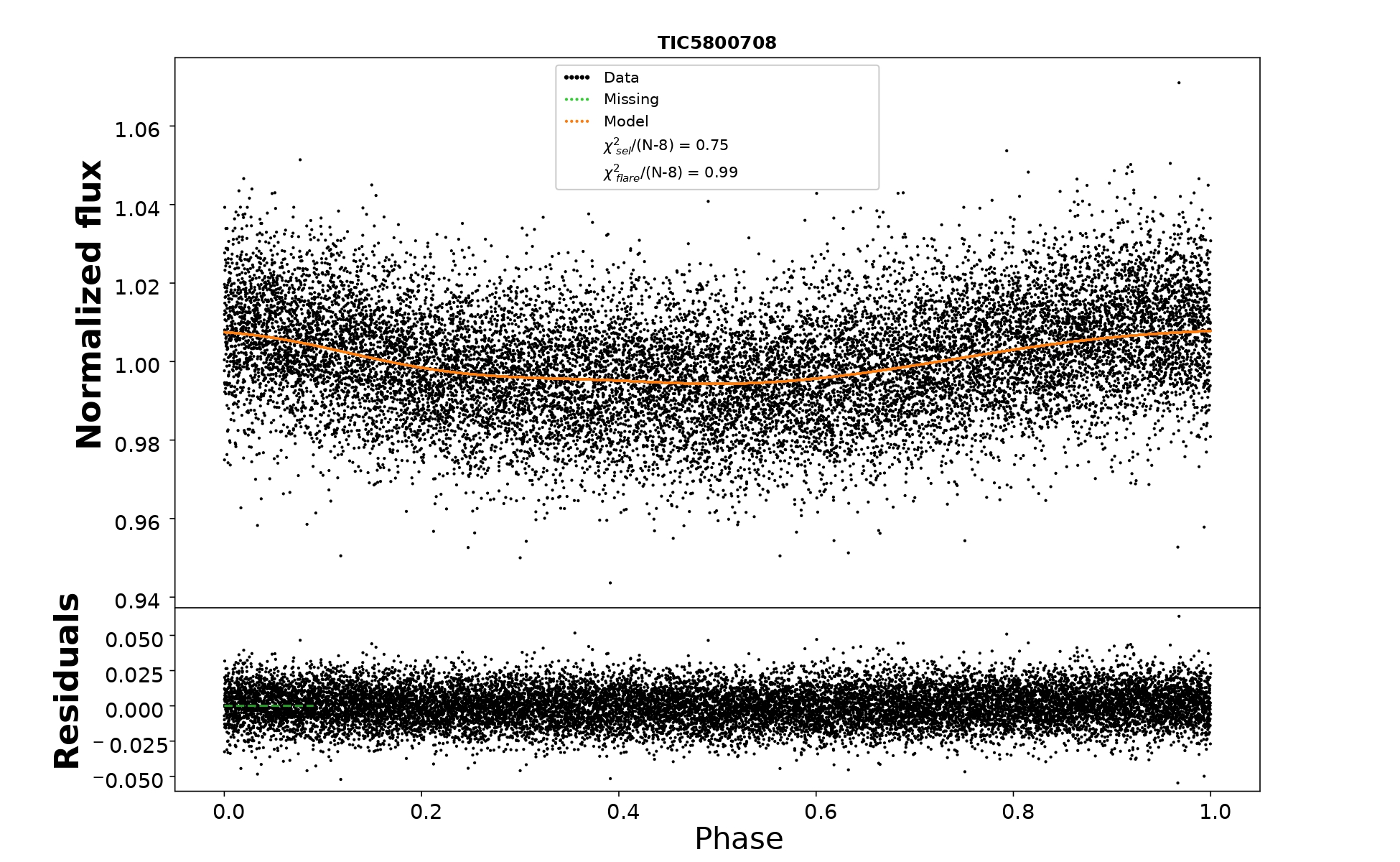}
    
    \quad\includegraphics[width=0.45\linewidth]{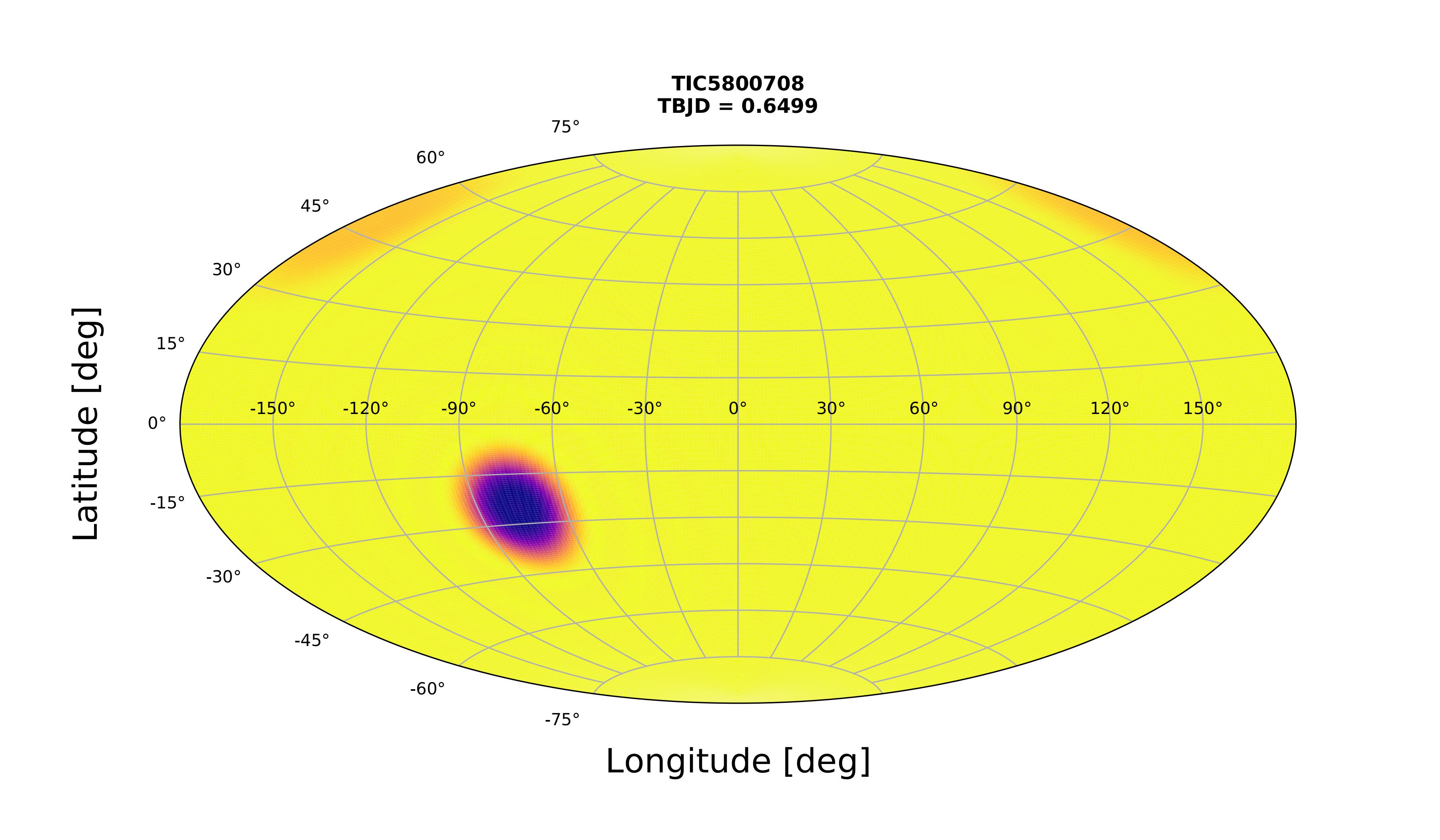}
    \quad\includegraphics[width=0.45\linewidth]{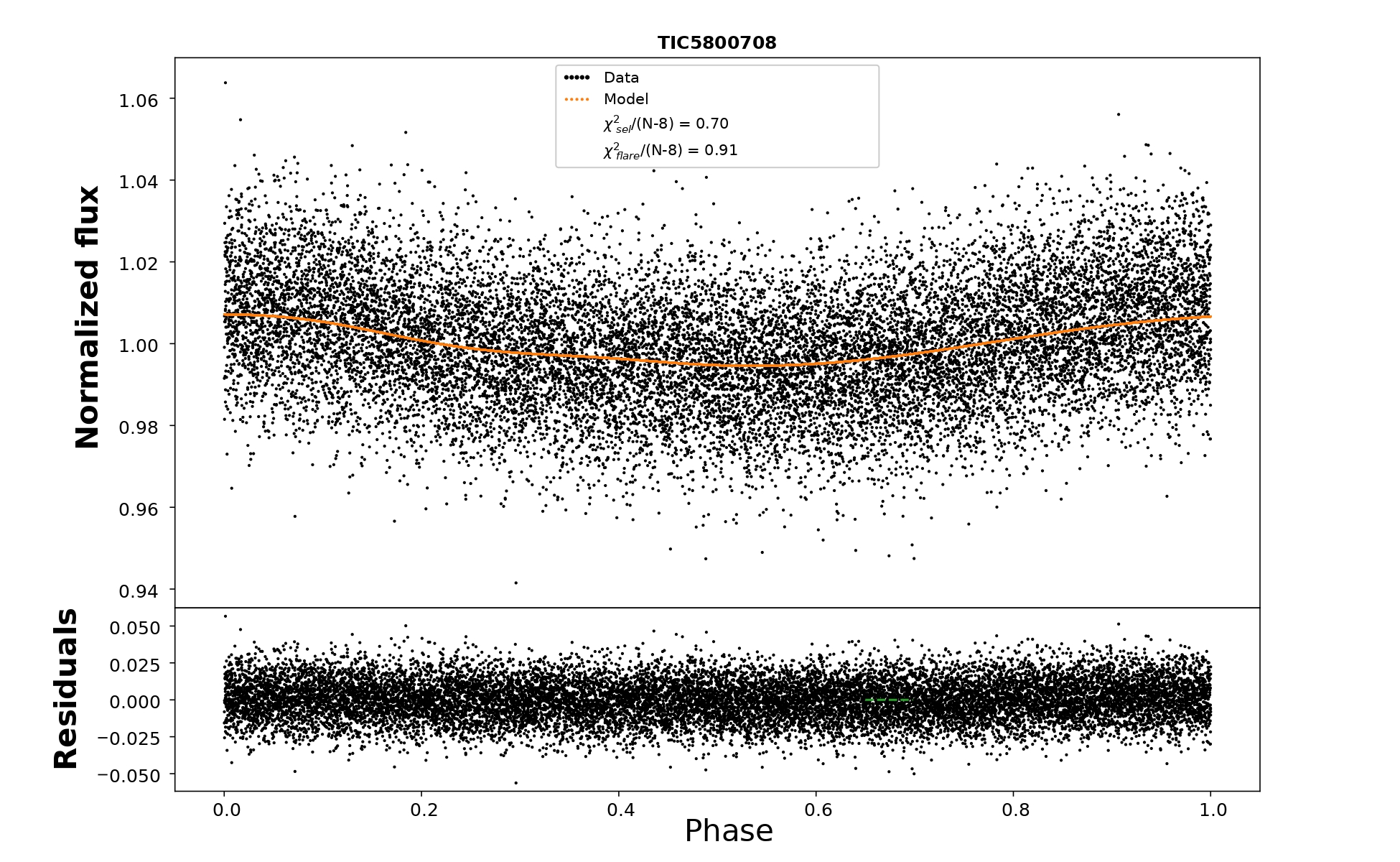}

    \quad\includegraphics[width=0.45\linewidth]{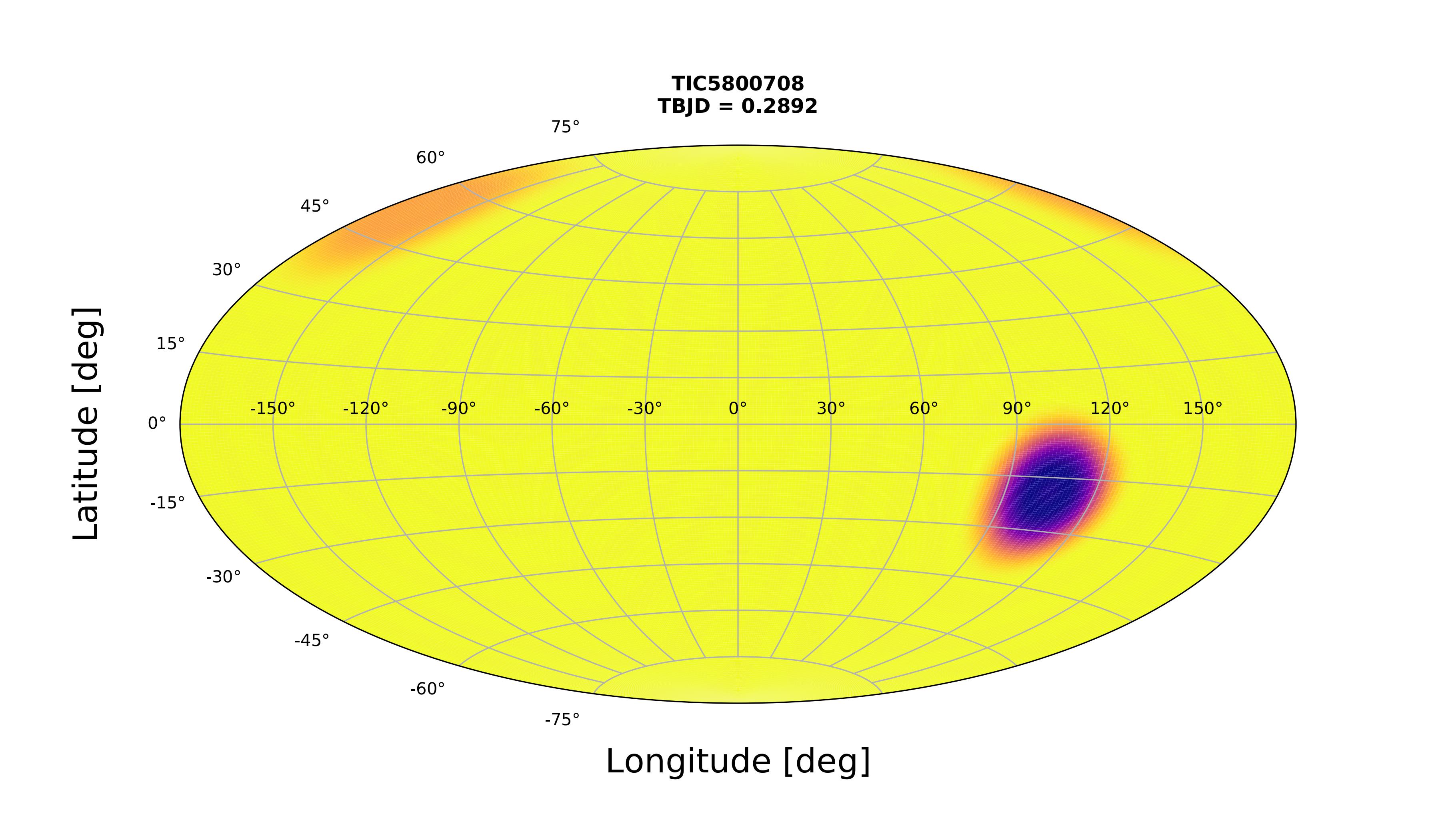}
    \quad\includegraphics[width=0.45\linewidth]{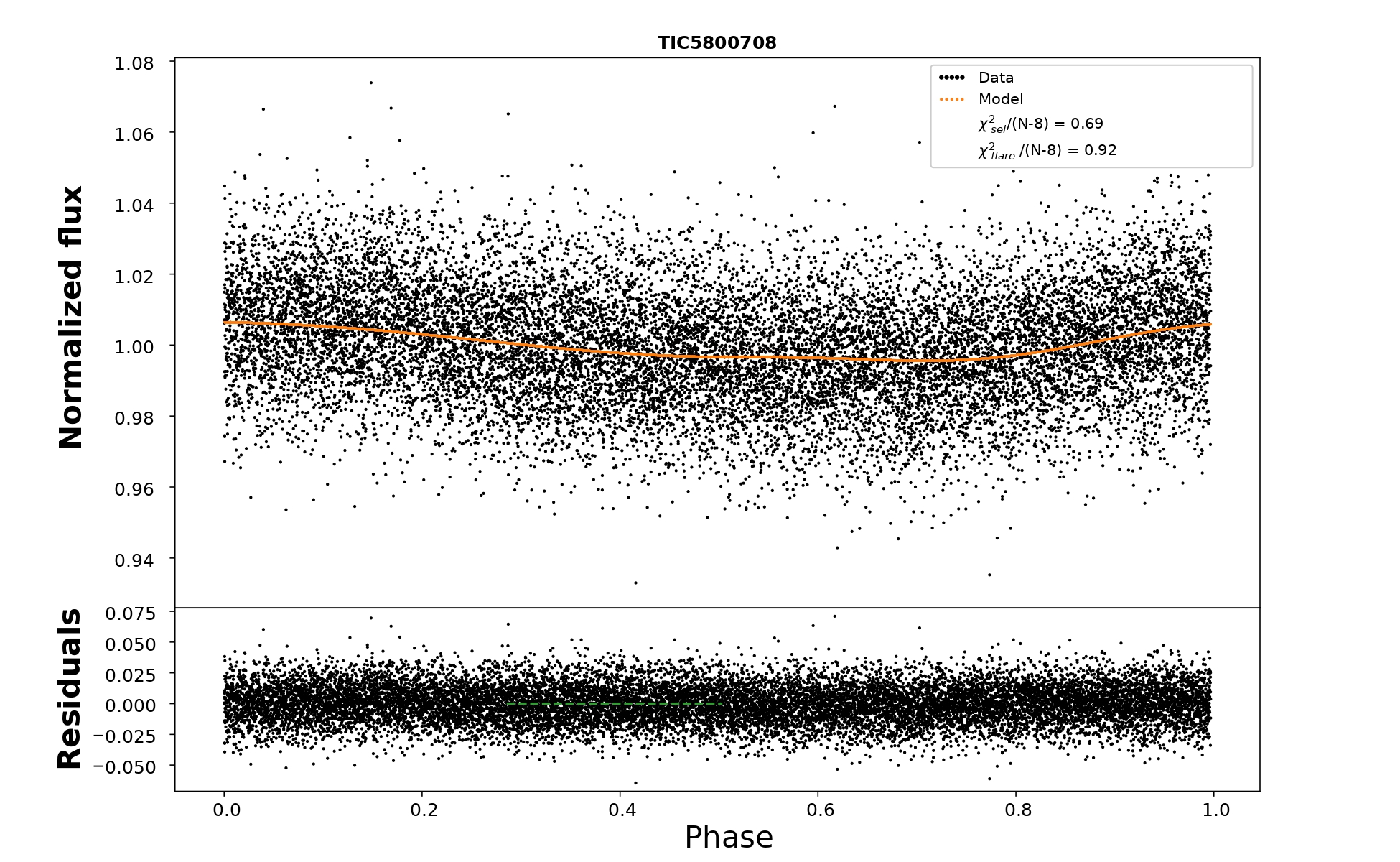}
    \caption{ Left column represents the positions, sizes, and contrast of spots in the Aitoff projection (at phase =0) of 2MASS J05160212+221452. The right column shows the phased light curve from TESS (black dots) and the model-fitted light curve (orange curve) along with their residuals and the reduced chi-square written in the corner of the plot. The upper panels are for combined phase light curve, 2nd panel shown for sector~43, 3rd panels are for sector~44 and the below panels displayed for sector~45 of 2MASS J05160212+221452.}
    \label{fig:2m0512spot_s43_44_45}
\end{figure*}

 2M0516+2214 object has been observed in TESS  sector 43 (September 2021; Camera 4; CCD 2), sector 44 (October 2021; Camera 2; CCD 1) and sector 45 (November 2021; Camera 1; CCD 4). The light curves obtained from each Sector varied slightly in amplitude, while displaying approximately similar nature in variability (Figure \ref{fig:2214_lc_ls_phs}). 
The estimated rotation periods from the LS periodogram are 1.10~days across all sectors. The Gaussian process methods provide rotation periods of 1.101 $\pm$ 0.002~d, 1.101 $\pm$ 0.001~d, and 1.101 $\pm$ 0.004~d of sector~43, 44, and 45 respectively that were precisely similar to the LS periodogram (refer Table \ref{tab:rot_per}). This is the first time we have estimated the rotation period for this object. The power spectrum for 2M0516+2214 demonstrated a significant peak centered at this rotation period. Moreover, the phase-folded light curves also appear sinusoidal while folding it on this significant peak, supporting the presence of starspot activity on the stellar surface, depicted in figure \ref{fig:2214_lc_ls_phs} which makes it a suitable object for this study.  But the LCs of 2M0516+2214 in each sector had large scatter due to statistical noise, making it challenging to construct a reliable spot model. To address this, we first constructed a combined phased LC from all three sectors (43,44, and 45), then modeled the combined phased LC with a two-spot configuration and it strongly suggested the presence of two primary spots (see 1st row of Fig \ref{fig:2m0512spot_s43_44_45}). As a result, one spot is located in the lower mid-latitude region (-22$^{\circ}$), and the other one is at a higher latitude around 47$^{\circ}$, separated by 99.2$^{\circ}$. Combining the phased LC from all sectors leads to long-term spot features with a slightly higher signal-to-noise ratio of around 83. Using this combined model as a starting point, we modeled the phased LCs of each sector (43,44, and 45) individually and we obtained a two-spot model for better fit to the LCs.  Therefore, the other sectors (43,44 and 45) also showed similar kind of spot distribution. 
During the model of combined LC, sector~43,44 and 45, the spot1 was situated at  47.5$^{\circ}$, 42.9$^{\circ}$, 46.5$^{\circ}$ and 52.3$^{\circ}$ in latitude respectively whereas spot2 was located at  -22.7$^{\circ}$, -22.0$^{\circ}$, -25.3$^{\circ}$ and -19.3$^{\circ}$ in latitude respectively. The spot distribution of the combined phased LC and LC in sector~43,44 and 45 were almost similar.
Furthermore, we also recreated the light curve with a three-spot model and obtained a different model and analytical solution of spots' parameters. Apart from that, we also checked for the log-probability (logP) value and found that in the two-spot model logP is relatively higher than in the three-spot model and the model estimated parameters are also close to the analytical solutions. So, we choose the two-spot model for the best fit for 2M~J0512+2214. The estimated parameters of starspots have been seen in Table \ref{tab:2m0516_spots_pars} and the mean spot model was listed in Table \ref{tab:2m0516spot_comp}.

\begin{longrotatetable}
\begin{deluxetable}{cccccccccccc}
\label{tab:2m0516_spots_pars}
\tablecaption{\edit1{Estimated Spots parameters of 2M0516+2214 of all sectors.}}
\tablehead{
\colhead{Sector} & \colhead{Start} & \colhead{Stop} & \colhead{Lat1} & \colhead{Lat2} & \colhead{Initial} & \colhead{Initial} & \colhead{T$_{spot1}$} &  \colhead{T$_{spot2}$} & \colhead{Size1} & \colhead{Size2}  \\
\colhead{} & \colhead{Time} & \colhead{Time} & \colhead{($\phi_{1}$)} &  \colhead{($\phi_{2}$)} & \colhead{Long1 $(\Lambda_{1})$} & \colhead{Long2 $(\Lambda_{2})$ } & \colhead{} & \colhead{} & \colhead{[$\%$ area} & \colhead{[$\%$ area}  \\
\colhead{} & \colhead{[BTJD]} & \colhead{[BTJD]} &  \colhead{[deg]} &  \colhead{[deg]} & \colhead{[deg]} & \colhead{[deg]} & \colhead{[K]}  &\colhead{[K]} & \colhead{of star]} & \colhead{of star]}\\
}
 
\startdata
          Combined & 2474.17 & 2550.63 &  47.46 & -22.63 & -172.72 & -75.18 &  2787$\pm$249 & 2045$\pm$1017 & 2.64 & 2.93 \\
         43 & 2474.17 &  2498.47 & 42.91 & -22.02 & -163.05 & -68.15 & 2785$\pm$ 258 & 2217$\pm$ 913 &  2.72 & 2.87 \\
         44 &2500.19 & 2524.44&  46.51 &  -25.34 &  -175.09 &  -76.77 &   2782$\pm$ 254 &    2276$\pm$ 870 &  2.63 &   2.28 \\
         45 & 2516.19 &     2550.63  &  52.31&  -19.29 &   -167.52 &  103.77&   2833$\pm$203 &    2150$\pm$935 &  2.79 &  3.03 \\
\enddata
\end{deluxetable}
\begin{deluxetable}{cccccccc}
        \tablecaption{ Comparision between model estimated spot parameters and the analytically estimated spot parameters of 2M0516+2214 in the combined sector as well as sector~43,44 and 45.}
        \label{tab:2m0516_spot_comp}
        \tablehead{
        \colhead{Modulation} & \colhead{Analytic Mean Spot} & \colhead{Model Mean Spot} & \colhead{Analytical Spot Size} & \colhead{Model Spot Size} & \colhead{SNR} & \colhead{log-Probability} & \colhead{Reduced} \\
        \colhead{} & \colhead{Temperature [K]} & \colhead{Temperature [K]} & \colhead{[\% area of star]} & \colhead{[\% area of star]} & \colhead{} &\colhead{} & \colhead{chi-square}
        }
         
        \startdata
        Combined & 2752 $\pm$ 65 & 2480 & 5.40  & 5.57 & 82.7 & 33882 & 0.71  \\
        43 & 2752 $\pm$ 65 & 2541 & 5.54 & 5.59 & 89.19 & 11665 & 0.75\\
        44 & 2752 $\pm$ 65 &  2584 &  5.47  &  4.92 &  83.8 &  11417 & 0.70\\
        45 & 2752 $\pm$ 65 &  2546 &  5.15  &  5.83 &  75.4 &  10742 & 0.69 \\
        \enddata
        \tablecomments{}
        \label{tab:2m0516spot_comp}
\end{deluxetable}
\end{longrotatetable}

\subsection{Flare Analysis}\label{sec:FFD}
\begin{figure*}
    \centering
    \includegraphics[width=0.45\linewidth]{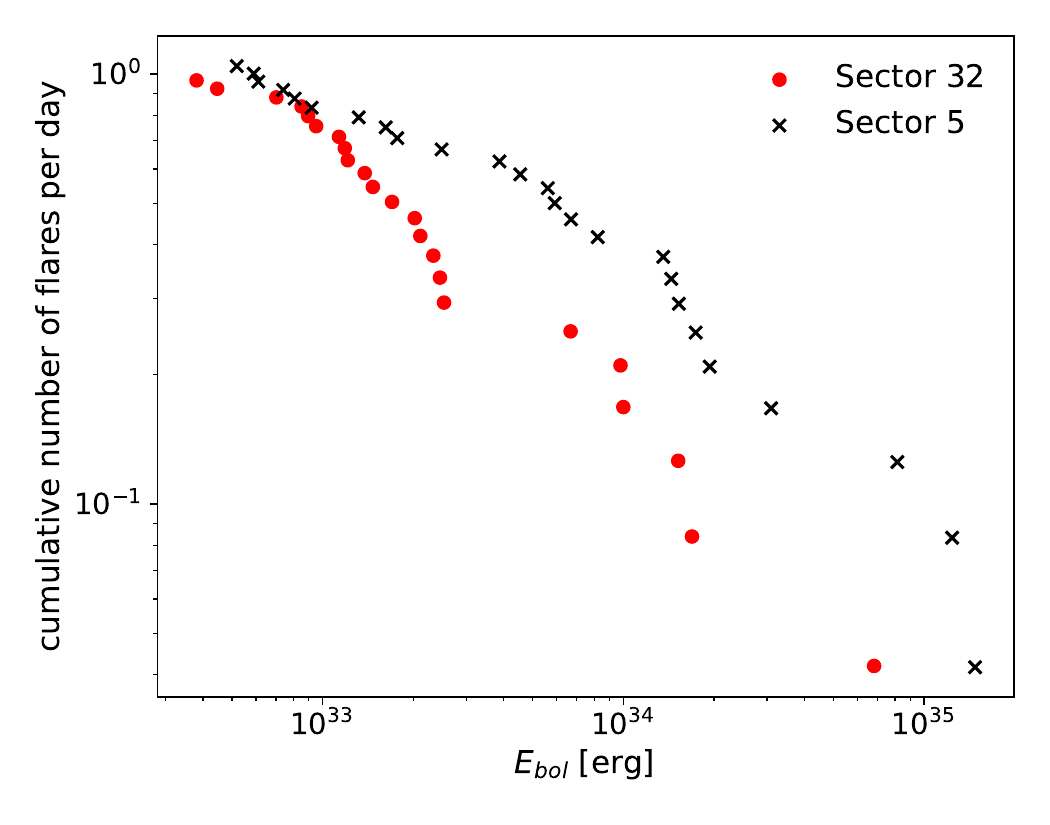}
    \quad\includegraphics[width=0.45\linewidth]{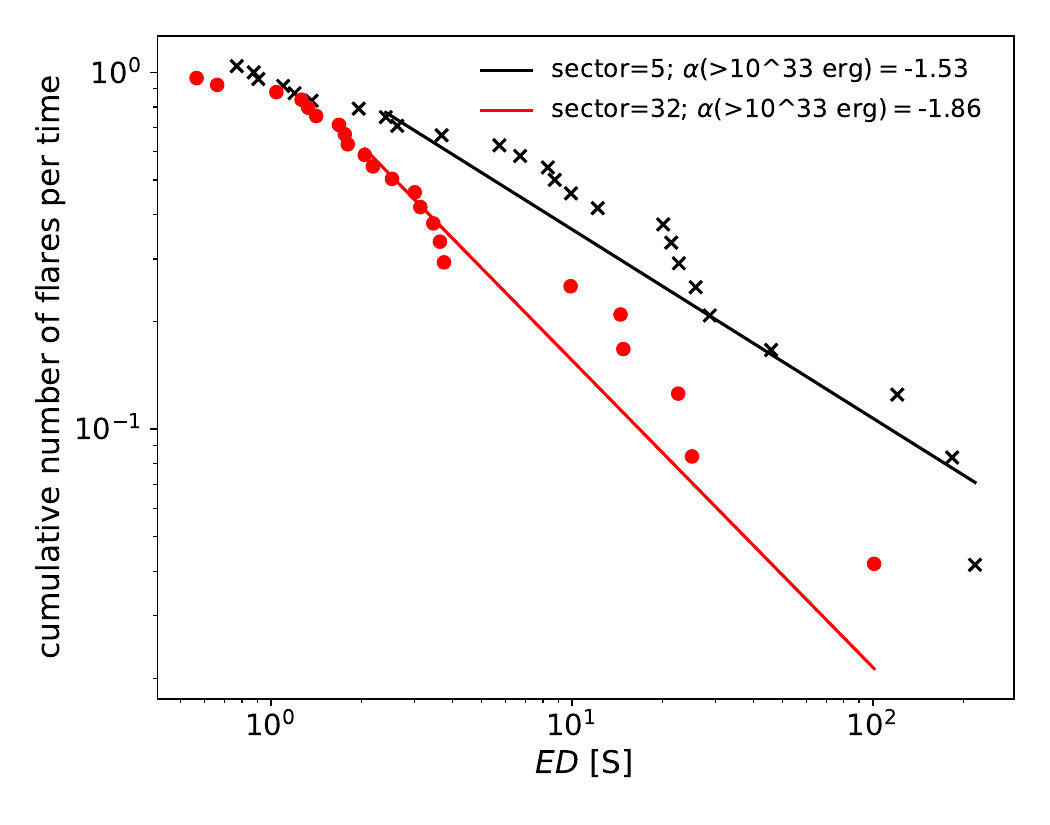}
    \quad\includegraphics[width=0.45\linewidth]{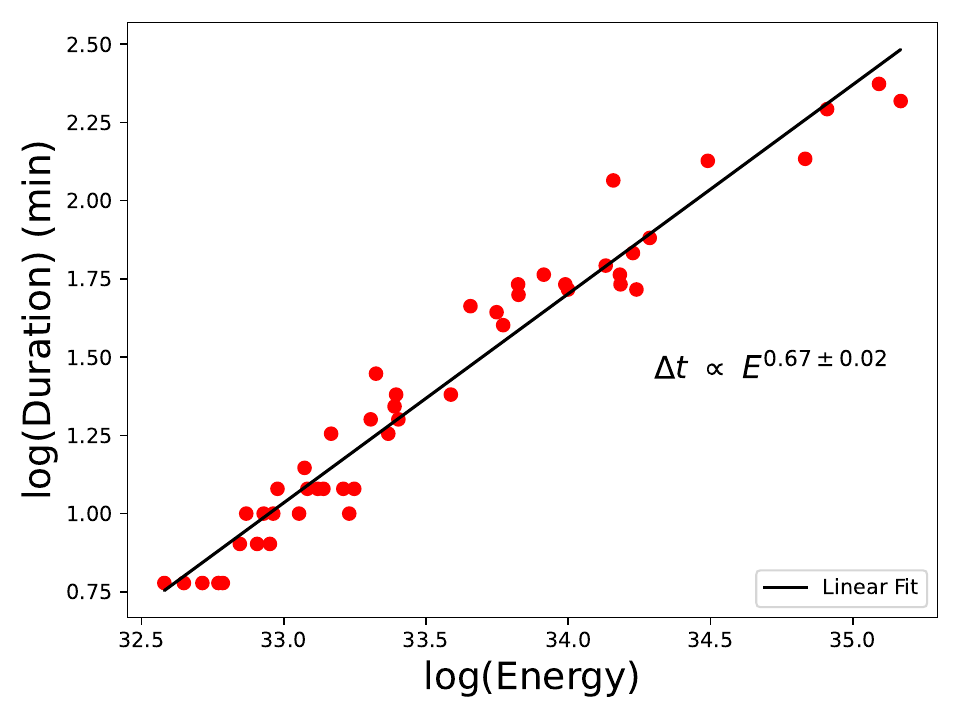}
    \caption{Cumulative flare frequency distributions vs. energy (left panel) and equivalent duration (right panel) for GJ 182 in sectors 5 and 32 are shown here. For a given energy (or equivalent duration) on the x-axis, the cumulative number of flares per day is given on the y-axis. Bottom: Flare duration versus total energy of the flare is shown. }
    \label{fig:ffd}
\end{figure*}
\begin{figure*}
    \centering
    \includegraphics[width=0.45\linewidth]{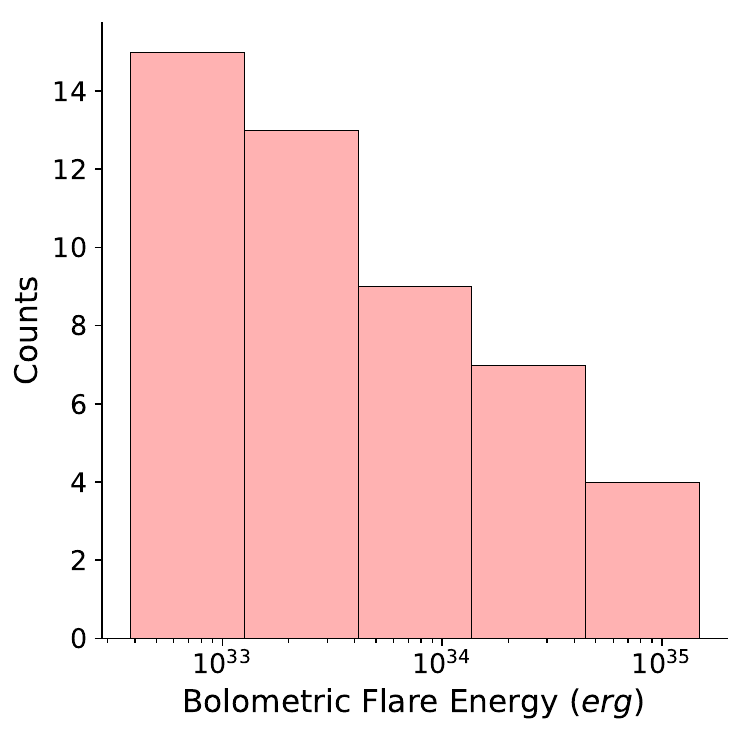}
    \quad\includegraphics[width=0.45\linewidth]{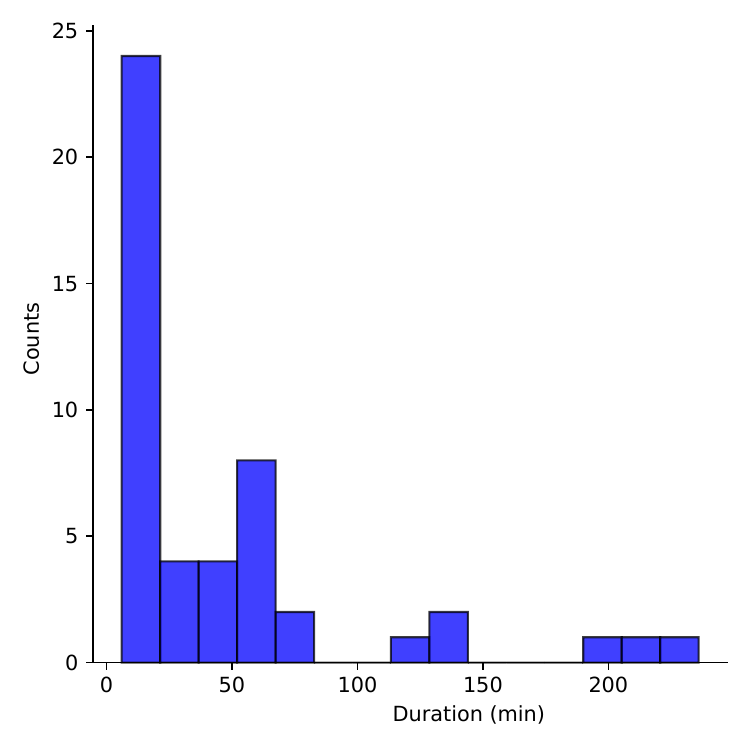}
    \caption{Histograms for the energy (left) and duration (right) of flares of GJ~182 are shown.}
    \label{fig:hist_plot}
\end{figure*}

 In this study, we examined the flare events of GJ~182 across two sectors. In sector~5, we identified 25 confirmed flare events and 5 probable events while in sector~32 there were 23 confirmed events and 2 probable events.
The parameters of confirmed flare events of GJ~182 are summarised in Table \ref{tab:flr_pars_sec5} and \ref{tab:flr_pars_sec32} and the flare energies and duration of the flares are plotted as a histogram in Figure \ref{fig:hist_plot}. The bottom panel of Figure \ref{fig:ffd} illustrates the relationship between flare duration and total energy in the log-log plot. The duration was derived by subtracting the start of the flare time from the end of the flare time and the total energy was estimated as described in section \ref{sec:flr_enr}. The plot showed a strong correlation (correlation coefficient$\sim$ 0.97) between these properties which implied that stronger superflares have much longer duration. As a result, a linear positive correlation was found between duration ($\Delta t$) and bolometric energy of flares ($E_{bol}$) i.e. $ \Delta t \propto E_{bol}^ {0.67 \pm 0.02}$.  Generally, the theoretically predicted value of the power-law index of E was  1/3 for the flares on solar-type stars i.e. for magnetic reconnection. \citet{Maehara2015EP&S...67...59M} got $0.39 \pm 0.03$. Later, \citet{Namekata2017ApJ...851...91N} found  $0.38 \pm 0.06$ for solar white-light flares. \citet{Maehara2021_10.1093/pasj/psaa098} again found a positive correlation for YZ CMi with an index of E was 0.21 $\pm$ 0.04  which satisfied the magnetic reconnection theory. But for superflares on  GJ 182 (M0.5), we obtained a slightly higher value i.e. $ 0.67 \pm 0.02$.  Previous studies e.g., \citet{Jackman2021MNRAS.504.3246J}, \citet{Tu2021ApJS..253...35T} have also found slightly diverse values of 0.6 for mid-M dwarfs using NGST and 0.42 for solar-type stars using TESS respectively. For a K2V object, Kepler-411, \citet{Araujo2021ApJ...922L..23A} have reported 0.86 $\pm$ 0.03 which is larger than our estimated value.  

In addition, the flares occurrences in M-dwarfs follow power-law distribution in energy \citep{Lacy1976ApJS...30...85L}. This distribution is described by the equation,
\begin{equation}
    dN (E) = \beta E^{-\alpha} dE dt 
\end{equation}
where N is the number of flares that occurred in the observational period $dt$, E is the total flare energy, $\beta$ is a proportionality constant, and $\alpha$ is the power-law index. The power law index $\alpha$ also denotes the slope of the cumulative flare frequency distribution (FFD). 
Here, we investigated this distribution by plotting the cumulative FFD of all flare events of GJ~182 and fitted it to a power-law model\footnote{https://altaipony.readthedocs.io/en/latest/tutorials/ffds.html} using Markov Chain Monte Carlo (MCMC) method to find out $\alpha$ and $\beta$.   For many stars, flare frequency distribution follows the simple power law. Initially, we attempted to fit the flare frequency distribution plot with the whole energy range of the flare but the fit appeared inaccurate at the low energy region.  This may be due to the redder wavelength of TESS observation,  where low energy flares are less reliably detected in the presence of photometric noise  \citep{Doyle2019MNRAS.489..437D, Tu2021ApJS..253...35T, Vida2024Univ...10..313V}. Then, the $\alpha$ index may be affected by this energy range used for fitting \citep{Yang2023A&A...669A..15Y}. To address this we avoided the initial phase of the horizontal trend at the low energy range of the FFD and used 10$^{33}$ to 10$^{35}$ erg energy range for better fit. As a result, in sector 5, we obtained the power-law index ($\alpha$) as  1.53 $\pm$ 0.12 while  1.86 $\pm$ 0.22 for sector 32 in the range of energy from  10$^{33}$ to  10$^{35}$ erg. For comparison, \citet{Lin2019ApJ...873...97L} reported the power-law index as $1.82 \pm 0.02$ and \citet{Yang2019ApJS..241...29Y} obtained $2.13 \pm 0.05$  of M-type stars. Similarly, \citet{Maehara2021PASJ...73...44M} found $\alpha$ to be approximately $1.75 \pm 0.04$ for the flares on M-dwarfs in the energy range between $10^{32}$ to $10^{34}$ erg while \citet{Yang2023A&A...669A..15Y}  also gave $\alpha$ $\sim$ $1.85 \pm 0.13$. So, our estimated value of $\alpha$ is almost similar within error bar to the obtained value from earlier studies and as $\alpha$$ >$~-2 which suggested that the total energy of the flares is mainly dominated by high-energy flare events rather than low-energy \citep{Paudel2018ApJ...861...76P, Jackman2021MNRAS.504.3246J,Gao2022AJ....164..213G}

\section{Discussion}\label{sec:diss}
We analyzed the power spectrum of two young M-dwarfs, GJ~182 and 2M0516+2214 visually and we found that they exhibited a prominently dominated peak (Fig \ref{fig:lc_ls_phs}, Fig \ref{fig:2214_lc_ls_phs}). But apart from this large peak, there are relatively weak secondary peaks, also present in a few sectors.
In the case of GJ~182, there is a small peak present in sector 5 and it is diminishing in sector 32 LC (middle panel in figure \ref{fig:lc_ls_phs}). We detected that this secondary peak appears at the harmonic of the primary peak. Similar case for 2M0516+2214, as we go from sector 43 to 45 the power of the second small peak increases. We also calculated the false-alarm probability (FAP) of the main peak for these objects, as determined by the LS algorithm, which was found to be very small $\sim$ 0. Subsequently, folding the light curves with this main peak, a clear periodic nature emerged (Figure \ref{fig:lc_ls_phs}, \ref{fig:2214_lc_ls_phs}). Folding the light curves at the secondary peak does not show any periodic nature; rather it is more scattered. This behaviour was often attributed to the presence of cool spots or spot groups on its surface, which co-rotating with the objects and periodically came in and out of our line of sight \citep{Rebull2016AJ....152..113R}. GJ~182 was observed with an interval of almost two years and it is observed that the shape and amplitude of the light curve also changed significantly (figure \ref{fig:lc_ls_phs}). In particular, the light curve of GJ~182 changes two to one local minima in one rotation period. In sector 5, GJ~182 showed a double-dip structure in the phase light curve, while in sector 32 the dip disappeared. As a result, the strength of the secondary peak in the LS periodogram (Fig \ref{fig:lc_ls_phs}) in sector 32 decreases compared to sector 5. Moreover, the shape of the LC changes over the sector but the rotation period remains the same. Such a scenario might arise due to spot/spot group evolution and/or latitudinal differential rotation \citep{Rebull_II2016AJ....152..114R}. Previously, \citet{Davenport2015ApJ...806..212D} also suggested that such double-dip light curves possibly arise due to the spots which are well-separated in longitudes.

We further analyzed and modeled the light curves of these two objects in each sector to study the starspot distribution on the stellar surface, as their brightness variation shows an almost periodic nature. We run the model for each sector individually to check any spot evolution on the surface.  In our analysis of the LC in sector~5, we observed a significant change in the amplitude of the LC during each rotation of the object. To investigate the variation we divided the LC into a few segments for modeling and each segment contains one full rotation of the object (discussed in section \ref{sec:result}). We found that our modeling on GJ~182 strongly indicates the presence of three spots on the surface. Notably, the spot2 in sector~5 exhibited a consistent shift in longitude from 150.8$^{\circ}$ to 173.7$^{\circ}$ across each rotation. While spot1 demonstrated a more significant longitudinal shift, moving from 115.3$^{\circ}$ to 179.5$^{\circ}$. The shiftiness of spot1 is much more than spot2 because of the differential rotation of this object as spot1 is placed near the equator while spot2 is at a higher latitude. Mainly, this spot1 produced the local minima of the LC and the other two spots (mainly spot3) were responsible for the secondary minima. As spot3 moves towards the higher latitude region (30$^{\circ}$ to 67$^{\circ}$) during each modulation, consequently the dip in the secondary minima vanishes. Furthermore, the total size of the spot is also varied from 5.15\% to 8.49\% of the stellar surface during each rotation. Similarly, in sector~32, the LC of each rotation was also reconstructed by a three-spot model. Meanwhile, spot2 migrated to the lower latitude region while the other two spots remained at the mid-latitude region (around 25$^{\circ}$ to 43$^{\circ}$). These mid-latitude spots mainly contribute to the local minima of LC in sector~32. As spot2 goes to the lower latitude region, it can not contribute much to produce minima in the LC. So, the shape of the LC also changes from sector~5 and the dip in the secondary minima completely vanished in sector~32. The total size of the spot also varying 5\% to 7.5\% during each rotation in sector~32. These findings regarding differential rotation, shifting in longitudes, migrating the spots, as well as dynamic change of spot coverage with time, provide valuable insight into the magnetic activity, revealing a complex interplay of factors that influence the surface features of this object. 
The spectropolarimetric measurement and surface mapping using Zemman-Doppler imaging of GJ~182 revealed that it has a strong axisymmetric toroidal magnetic field and non-axisymmetric poloidal component \citep{Donati2008MNRAS.390..545D, Lund2021MNRAS.502.4903L}. The non-axisymmetric poloidal field and 41$^{\circ}$ deviation of the magnetic axis from the rotation axis \citep{Lang2012MNRAS.424.1077L} can cause less stability of stellar spots which might lead to shorter lifetime or exhibit more erratic migration patterns across the stellar surface.  The separation between the spots in longitudes also varies in both sectors during each rotation, reflecting the differential rotation for this object.
Previously, GJ~182 was confirmed as a differential rotator with a surface angular rotation shear between the equator and poles is $d\Omega$=0.06 $\pm$ 0.03 rad d$^{-1}$ \citep{Donati2008MNRAS.390..545D}.  Furthermore, we identified 25 flares in sector~5 while 23 flares in sector~32 and we observed that in sector~5 most of the flare events had greater energy (19 flares had energy$\ge$ 10$^{33}$ erg ). In the case of sector~32, 17 flares had energy $\ge$ 10$^{33}$ erg.
This could be possible due to the change in the magnetic field strength which was already suggested by \citet{Donati2008MNRAS.390..545D}. Using Stokes V data \citet{Donati2008MNRAS.390..545D} observed the longitudinal field variation for this object. \citet{Reiners2009A&A...496..787R} also reported the average magnetic field around 2.5~kG from unpolarized spectroscopy. The shape and amplitude of the light curve have changed over two years for this object and can be explained by the evolution of spots and their sizes  along with their magnetic field strength. This kind of scenario was also observed in AU Mic and YZ CMi objects reported in \citep{Ikuta2023ApJ...948...64I}.  The total spot area relative to the stellar hemisphere varied significantly in two years and the flare frequency remained nearly constant in both sectors (25/25.987 = 0.96~day$^{-1}$ in sector~5 and 23/25.589=0.90~day$^{-1}$ in sector~32) although the shape and amplitude of the light curve changed. we obtained a mean spot temperature of approximately  3484~K and 3074~K in sector~5 and 32 respectively and an average spottedness  varied from 5\% to 8.5\% across sectors 5 and 32 for three-spot modelling (Table \ref{tab:gj_spot_comp_s5} and \ref{tab:gj_spot_comp_s32}). 

 For 2M0516+2214, the visual inspection of the smooth periodic variation in the LCs of all sectors strongly indicated the presence of two primary spots on the surface. However, the LCs exhibited significant scatter due to statistical noise. Two-spot model for this object reveals that it had a high-latitude spot along with a spot in the lower mid-latitude region. However we observed a significant shift in the longitude of  spot2 in sector~45 , although their size remained consistent at approximately 5.4\%. As the inclination angle of this object is 24.48$^{\circ}$, the local minima of the LC can primarily be attributed to the mid-latitude spot. The signal-to-noise ratio (SNR) of the analyzed LCs is 89,83 and 75 for sectors 43, 44, and 45 respectively. These values are slightly lower than the ideal SNR of 86 required for accurate reconstruction using the BASSMAN algorithm \citep{Bicz2022ApJ...935..102B}. This low value of the SNR complicates the analysis for finding the accurate position of the starspots. Although we got a slightly higher SNR value from the borderline SNR for the modeling in combined phased LC and the sector~43  and we obtained a similar kind of spot distribution in  across the sectors. Lower SNR value might be caused of the shift of the position of starspots \citep{Bicz2022ApJ...935..102B}. The fully understand the behavior of shift in longitudes/latitudes of the spots, we need high-resolution temporal data with less statistical noise. From the two-spot modeling of 2M0516+2214, we obtained a mean spot temperature of approximately  2631~K and average spottedness of around  5.42\% in sectors 43,44, and 45. The estimated spot parameters and comparison with the analytical solution are listed in Table \ref{tab:2m0516_spots_pars}, \ref{tab:2m0516_spot_comp}.

\subsection{Flare Analysis}\label{sec:diss_flr_analy}
In addition, we have detected  48 flare events of GJ~182 and about  most of the flares (35 out of 48 events) are in the superflare category ( greater than 10$^{32}$ erg) which indicates GJ~182 is a magnetically active object. From visual Inspection among  48 flares,  we identified one ``flat-top" flare event, 6 classical flares, and  17 flares that exhibited a complex nature by showing a ``peak-bump" profile. Additionally, the other 7 flares showed complex substructures during the rise phase. The remaining flare may have shown less clear features that might be an issue with insufficient temporal resolution. To better understand these structures, 20s cadence data would be beneficial, as it could reveal more details of these events. Apart from these several flares showed multiple peak emissions. However, we could not categorize these events due to their complexity. The ``flat-top" flare exhibited a fast rise and then relatively constant emission levels at peak (see Figure \ref{fig:flr_flat_top}) and then again exponential decay \citep{Howard2022ApJ...926..204H}. The duration of the constant peaks lasts from a few seconds to minutes. Previously, \citet{Jackman2021MNRAS.504.3246J} reported a few flat-top flares for low-mass stars using 13-sec cadence data from the Next Generation Transit Survey (NGTS). Using a 20-s cadence of TESS data of low-mass flare stars, \citet{Howard2022ApJ...926..204H} classified 24 flares exhibiting flat-top morphology. Such flares might result from the superposition of multiple unresolved peaks within a single event \citep{Howard2022ApJ...926..204H} or when quasiperiodic pulsations are unresolved, they can contribute to the flat-top morphology in some flares (e.g., \citet{Jackman2019MNRAS.482.5553J}. They can also arise from a prolonged-emission event \citep{Howard2022ApJ...926..204H}. In the peak-bump morphology of flares, there is a steeper rise and shallower decay but hump/humps were present in the decay phase \citep{Davenport2014ApJ...797..122D} (see Figure \ref{fig:flr_peak_bump}). Peak-bump flare emission is possible due to a cascade of smaller reconnection events or random superposition of two sympathetic flares from the same or nearby active regions \citep{Hawley2014ApJ...797..121H, Davenport2016IAUS..320..128D, Howard2022ApJ...926..204H}. Such peak-bump flares have been previously observed e.g., \citep{Gunther2020AJ....159...60G, Jackman2021MNRAS.504.3246J, Howard2022ApJ...926..204H}. Moreover, these complex types of flares have a longer duration (52 min-165 min) as well as higher energy (10$^{33}$ - 10$^{34}$).  A few flares also showed complex emission during the rise phase. Such rise-phase complexity in flare is also observed on the Sun, where accelerated electrons heat the lower atmospheric layers in nearby emission regions. But these processes do not occur simultaneously or with identical intensity \citep{Veronig2010ApJ...719..655V, Howard2022ApJ...926..204H}. The histogram plot of the flare energy distribution (Figure \ref{fig:hist_plot})  of GJ~182 showed that most of the flare emitted in the higher energy range, ie $ 10^{33} - 10^{34}$ erg.  As the \textsc{TESS} filter is more sensitive to objects with low temperatures, we obtained the overall energy distribution in the superflare region. In Section \ref{sec:FFD}, we obtained the value of the power-law index $\alpha$ of GJ~182 in both sectors through cumulative FFD. Consistent with previous work for M-dwarfs (mentioned in section \ref{sec:FFD}), our obtained value of $\alpha$ ( 1.53 $\pm$ 0.12 in sector 5 and  1.86 $\pm$ 0.22 in sector 32)  matches quite well. Figure \ref{fig:ffd}; (bottom) showed the strong correlation between the durations and the energy of the superflares in our data set in the energy range $ 10^{32}~to~10^{35}$ erg. Here, the observational value of the power-law index in the duration vs energy plot from TESS was $ 0.67 \pm 0.02$ which is slightly larger than the theoretical prediction ($\beta$ $\sim$ 1/3). Since this value is derived from the theory of magnetic reconnection for the solar-type flares, it may not precisely illustrate the superflares in the M-dwarfs \citep{Tu2020ApJ...890...46T}. Other literature (mentioned in section~\ref{sec:FFD}) also reported a slightly higher value and pointed out that one reason may be different spectral types or their different coronal magnetic field strength \citep{Maehara2015EP&S...67...59M}. Another reason may be due to the length of the flare loop of the largest flare is comparable to the solar radius as flare loop length is correlated with the electron temperature and emission measure of the stellar flares \citep{Shibata1999ApJ...526L..49S}. This strong correlation between duration and flare energy also indicates that strong superflares last longer than their lower energy counterparts \citep{Araujo2021ApJ...922L..23A}. 

 \subsection{Starspot Area and Stellar Flares}\label{sec:starspot_flr_rel}
\begin{figure*}
    \centering
    \includegraphics[width=.48\linewidth]{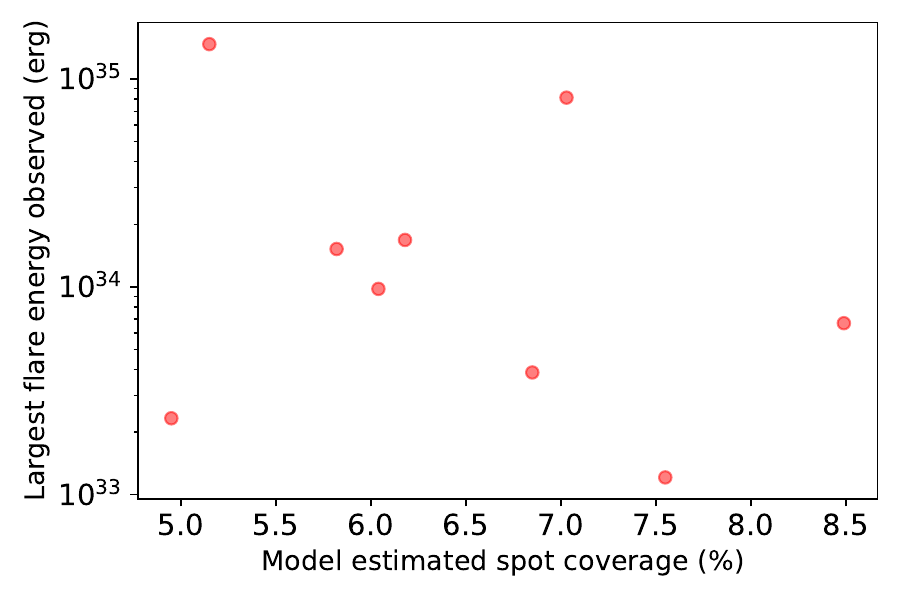}
    \quad
    \includegraphics[width=.48\linewidth]{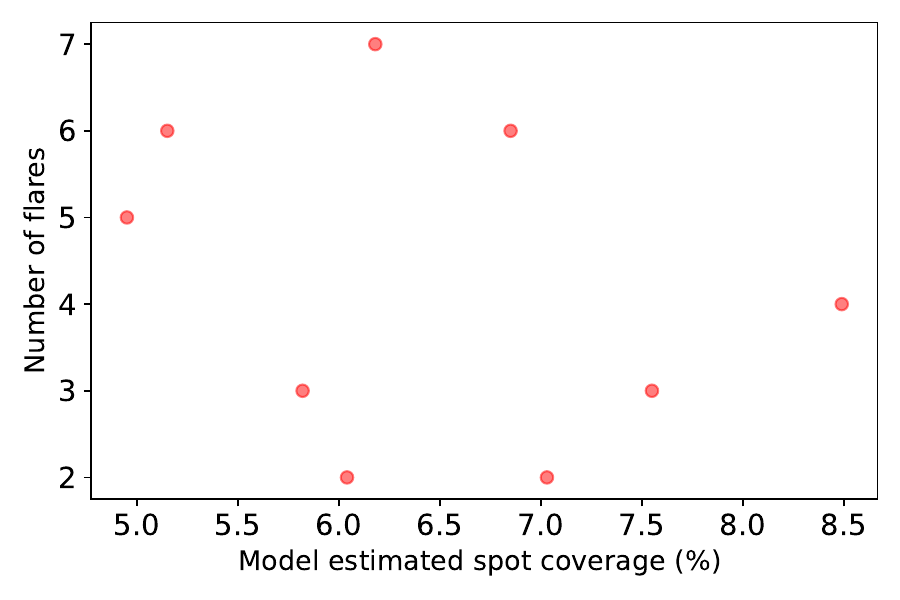}   
    \caption{The relationship between rotational phase and flare energy for sector~5 (left) and sector~32 (right) for GJ~182. It is to be noted that no correlation was found between them.}
    \label{fig:spot_area_flare_rel}
\end{figure*}

\begin{figure*}
    \centering
    \includegraphics[width=.48\linewidth]{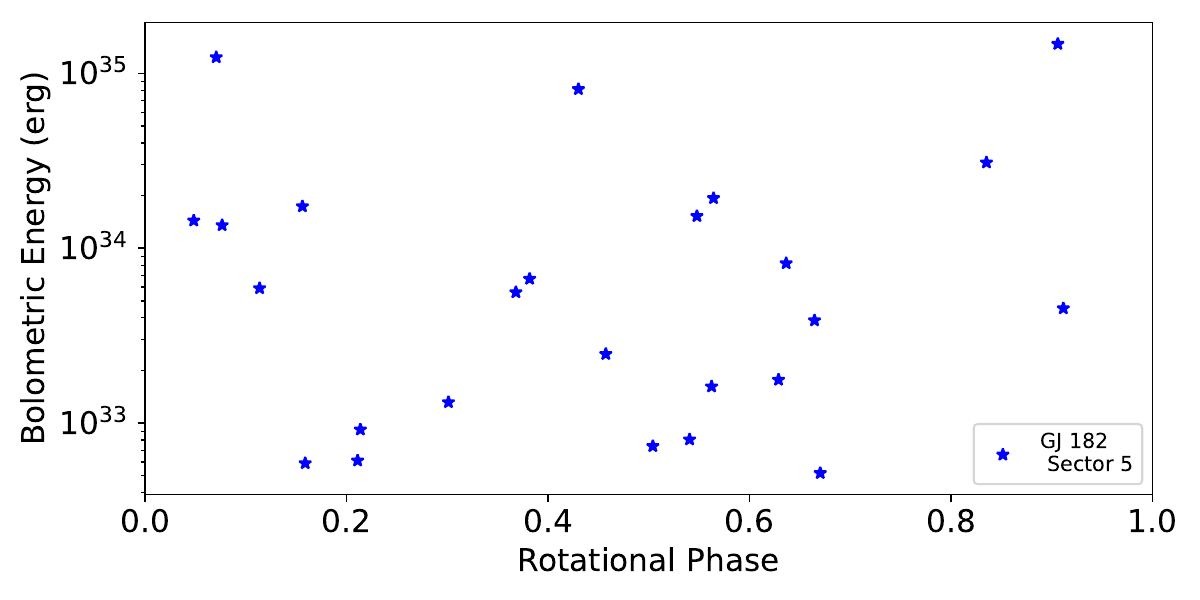}
    \quad
    \includegraphics[width=.48\linewidth]{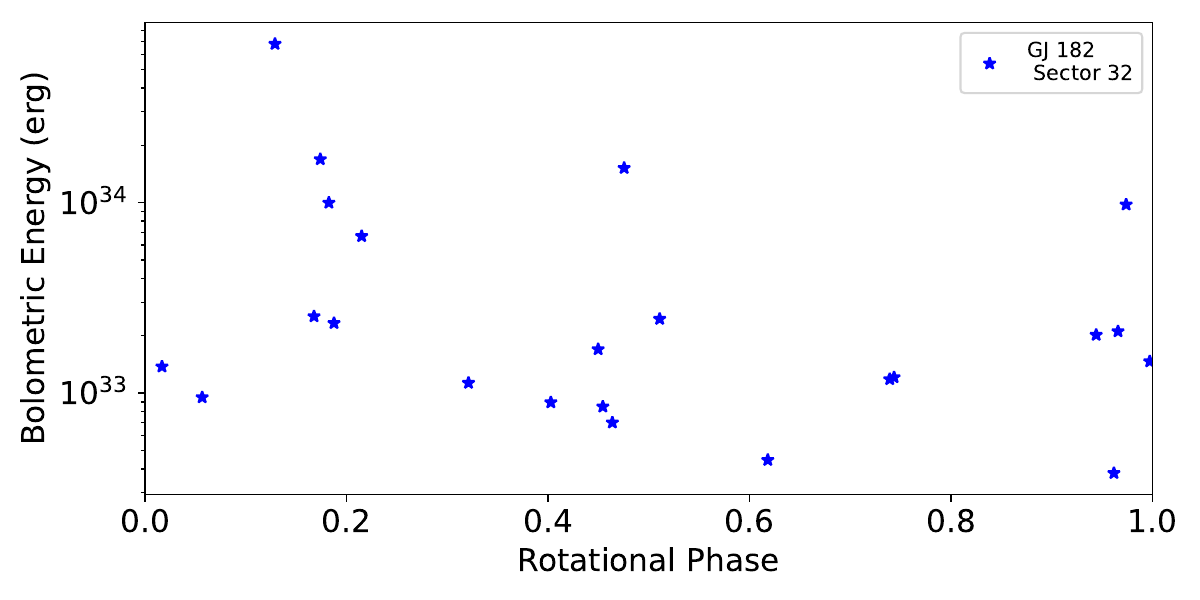}   
    \caption{The relationship between rotational phase and flare energy for sector~5 (left) and sector~32 (right) for GJ~182. It is to be noted that no correlation was found between them.}
    \label{fig:phs_rot_energy}
\end{figure*}

 It is generally believed that the nature of solar flares and superflares on solar-type stars are caused by the same physical process i.e. magnetic reconnection \citep{Maehara2015EP&S...67...59M}. This suggests that the largest flare may be associated with the largest starspots. Therefore, we examine the correlation between stellar spot coverage and the flares' number and energy to investigate whether the fact is observed. From the spot modeling of GJ~182, we observed that the area of the spots are changing during each rotation. To check the correlation we identified the highest-energy flare and the total number of flares occurring during each rotation. Initially, we plotted our model's estimated spot coverage in each rotation with the largest flare energy observed during each modulation. However, there is no relation found and there is an overall large scatter among them (see figure \ref{fig:spot_area_flare_rel}). We further test statistically using the Pearson correlation coefficient (P) which indicates a lack of any correlation with a value, of P=-0.33. Further, we look at the number of flares as a function of spot coverage. Again tested with Pearson coefficient correlation and did not find any linear correlation among them (P=-0.29). As there is less number data coverage on the spot area, this may contribute to these discrepancies. A more dataset on the time-series photometry of GJ~182 could reveal a deeper insight into this relationship. Furthermore, we again examined the correlation between the bolometric energy of flares and the rotational phase of GJ 182. It could be expected that more and larger flares will occur where there is a concentration of spots, particularly at the phase corresponding to the minimum flux. However, as illustrated in Figure \ref{fig:phs_rot_energy}, no significant relationship was apparent.
 The lack of correlation between spots and flares is suspected to occur in fully convective stars \citep{Roettenbacher2018ApJ...868....3R}. Additionally, for such stars, an anticorrelation between flares and spots has also been shown \citet{Bicz2022ApJ...935..102B}, where the magnetic field may be potential. A potential magnetic field has no free magnetic energy to be released during the flare \citep{Aschwanden2005psci.book.....A}. Furthermore, \citet{Hawley2014ApJ...797..121H} and \citet{Morin2008MNRAS.390..567M} demonstrated that flares on active M dwarfs appear randomly across many independent active regions.

\section{Summary}\label{sec:summ}
We conducted starspot modeling of \textit{TESS} light curves of two young M-dwarfs, GJ~182 (M0.5) and 2M0516+2214 (M4.5), using \textit{BASSMAN} software to investigate the starspots distribution on the stellar surface and find out the spot properties. Our main results can be summarized as follows:

\begin{enumerate}

\item We measured the rotation period of the two objects using the Lomb-Scargle periodogram and Gaussian regression process. GJ 182 has a rotation period of  4.348 $\pm$ 0.016  days in sector~5 and 4.384 $\pm$ 0.042 days in sector~32, which agrees with the previous studies within the error bar. For the first time, we have estimated the rotation period 2M0516+2214 of 1.102 $\pm$ 0.004 days and the rotation period estimated from both methods matches well.

\item The TESS light curves of our selected two young M-dwarfs i.e. GJ 182 and 2MASS
J05160212+2214528 were reconstructed by BASSMAN software.  For GJ 182, a three-spot model effectively described the light curves, while a two-spot model was suitable for 2MASS J05160212+2214528. To understand better visualization of starspot evolution we divided the LCs of GJ~182 into several segments,  each representing full rotation of the object and all segments were well reconstructed by the three-spot model.

\item For GJ 182, we estimated the mean spot temperature to be approximately  3484~K in sector~5 and 3074~K in sector~32  and
spottedness varying from  5 \% - 8.5 \% of the stellar surface and for 2MASS~J05160212+2214528 we obtained mean spot temperature around  2541~K - 2759~K and average spottedness about  5.4 \%.

\item A total of  48 flare events with bolometric energies between $ 3.81 \times10^{32}~and~1.47\times 10^{35}$ erg from GJ~182 were detected. Further, we have estimated the flare energy within the TESS band, ranging from 8.27$\times$ 10$^{31}$ to 6.81$\times$10$^{33}$ erg. Moreover, to produce such flare events we have also calculated the lower limit of the magnetic field from  12~G to 232~G. Among  48 flare events, we identified six flares that had a classical shape,  only one flare showed flat-top structure, and most of them (17) were categorized as peak-bump flares. Moreover, we also identified 7 flares that showed rise-phase complexity. Moreover, we also conduct an analysis of the starspot area and flare, but could not find any kind of significant correlation among them.

\item For the flares, a strong correlation was found between bolometric flare energy and the duration of the flare with a slope=$ 0.67 \pm 0.02$, showing that larger flares last longer and support the magnetic reconnection theory like solar flares. 

\item The slope of FFD for GJ~182 is measured to be $\alpha$= -1.53$\pm$0.12 in sector~5 and $\alpha$= -1.86$\pm$0.22 in sector~32   in the energy range 10$^{33}$ to 10$^{35}$ erg and agress well with previous findings for other M-dwarfs. This value also indicated that the total energy of the flare was dominated by high-energy flare events. Further, the rotational phase of GJ~182 did not show any correlation with the flare energy.

\end{enumerate}

\section{acknowledgments}
\begin{acknowledgments}
 The authors would like to thank the anonymous referee for the
helpful comments and suggestions that significantly improved the
paper.This research work is supported by the S N Bose National Centre For Basic Sciences under the Department of Science and Technology, Govt. of India. This paper includes data collected by the TESS mission. Funding for the TESS mission is provided by the NASA's Science Mission Directorate. Funding for the TESS mission is provided by the NASA Explorer Program. STScI is operated by the Association of Universities for Research in Astronomy, Inc., under NASA contract NAS 5-26555. R.K. is grateful to the Department of Science and Technology (DST), Govt. of India, for their INSPIRE Fellowship scheme. RK is thankful to Kamil Bicz for valuable discussions regarding \textsc{BASSMAN} software.
\end{acknowledgments}

\vspace{5mm}
\facilities{TESS}

\software{Python 3 \citep{Van10.5555/1593511},
          astropy \citep{2013A&A...558A..33A,2018AJ....156..123A},  
          BASSMAN \citep{Bicz2022ApJ...935..102B}, 
          Altaipony \citep{Ilin2021JOSS....6.2845I},
          Starspot \citep{Angus2018MNRAS.474.2094A},
          matplotlib \citep{Hunter2007CSE},
          numpy \citep{Harris2020Natur.585..357H},
          lighkurve \citep{Lightkurve_Collaboration2018ascl.soft12013L}.
          }
.
 \section{Data Availability}
The TESS data presented in this article were obtained from the Mikulski Archive for Space Telescopes (MAST) at the Space Telescope Science Institute. The specific observations analyzed can be accessed via \dataset[doi: 10.17909/jp7k-qy50]{https://doi.org/10.17909/jp7k-qy50}.

\appendix

\section{Appendix information}

\begin{figure*}
    \centering
    \includegraphics[width=.31\linewidth]{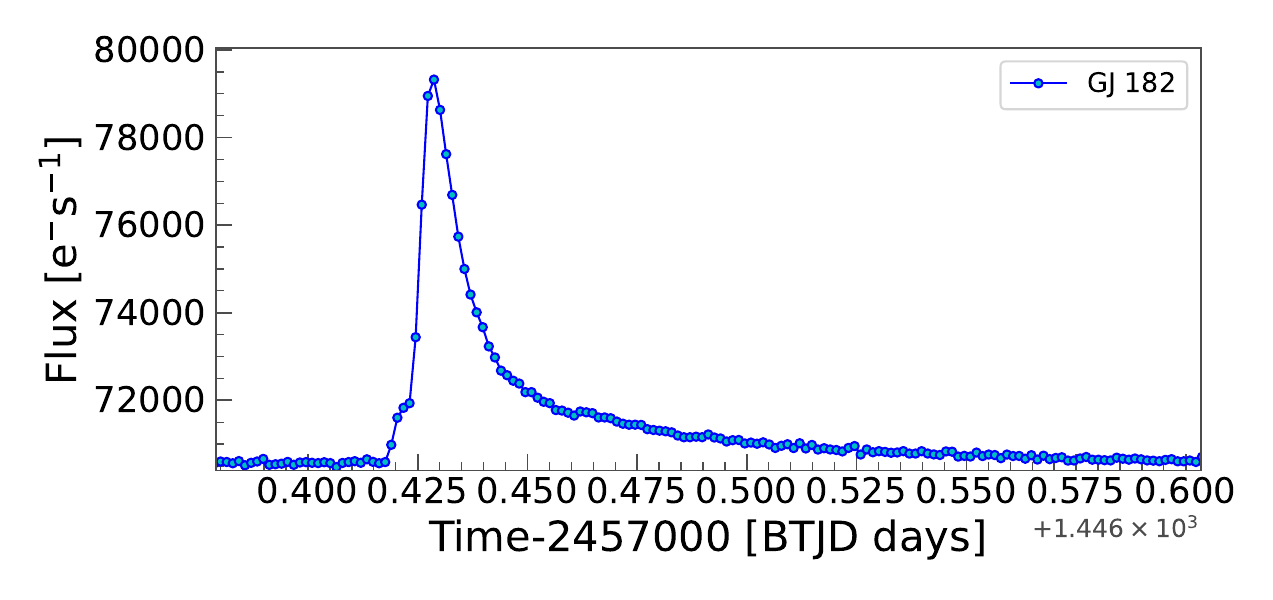}
    \quad
    \includegraphics[width=.31\linewidth]{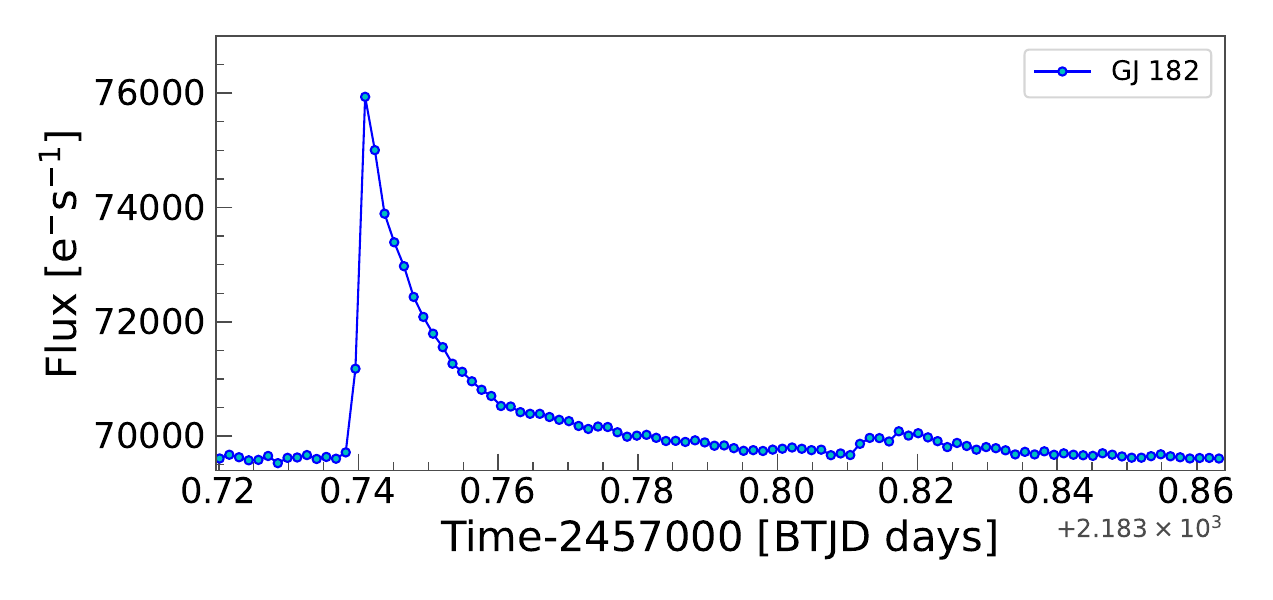}  
    \quad\includegraphics[width=.31\linewidth]{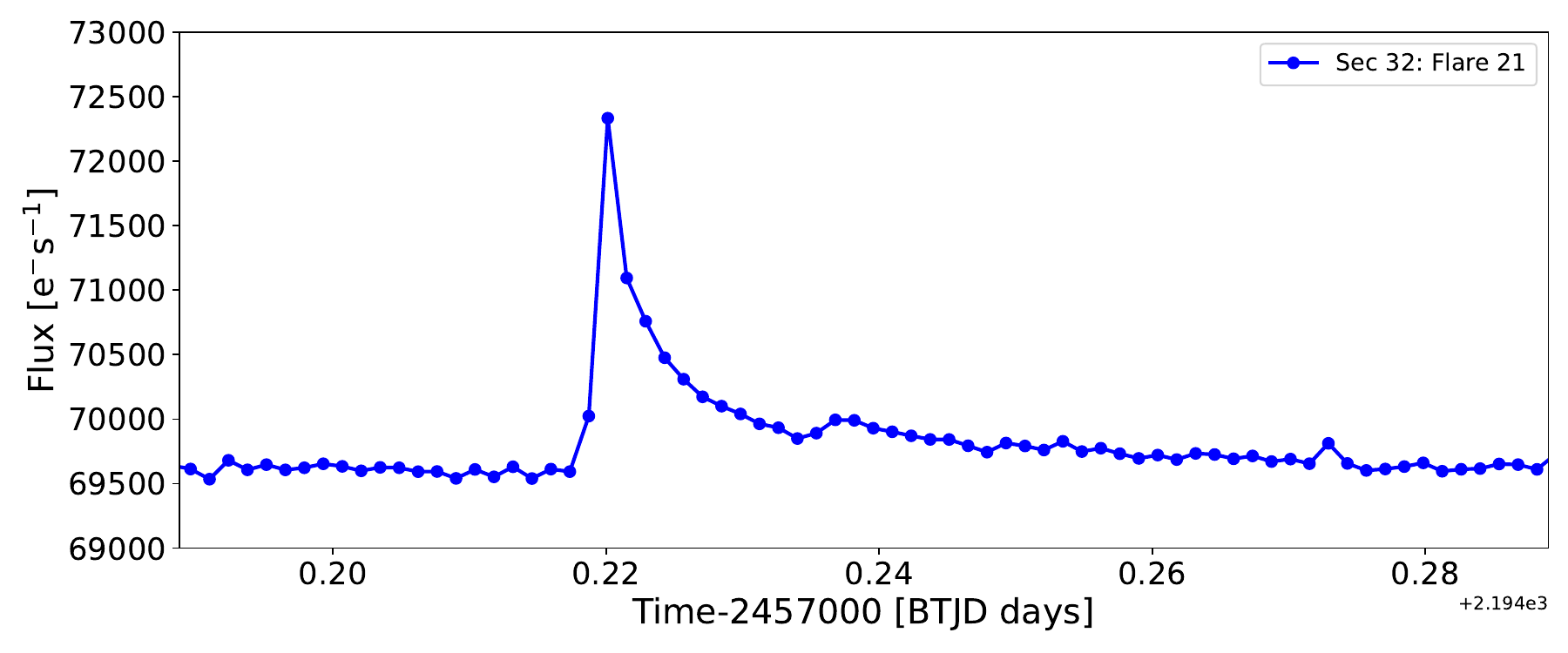}
    \caption{ Examples of classical flare events are shown here which have fast rise and slow exponential decay. Out of 48 flares, 6 are in classical shape.}
    \label{fig:flr_classical}
\end{figure*}

\begin{figure}
    \centering
    \includegraphics[width=.31\linewidth]{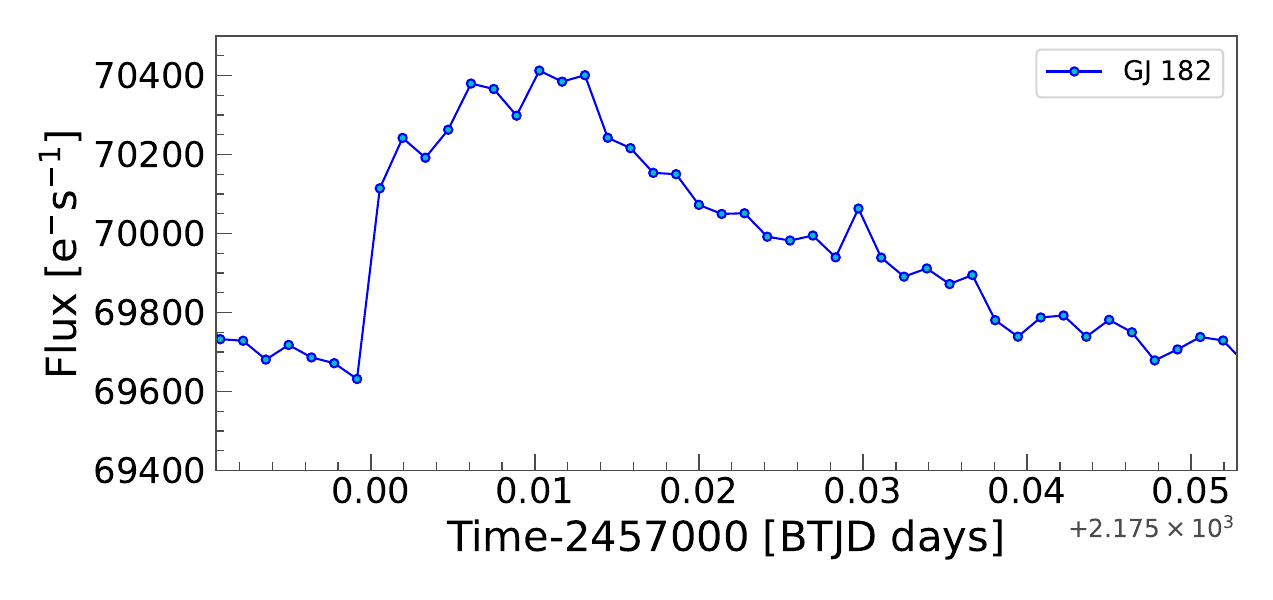}
    \quad
    \includegraphics[width=.31\linewidth]{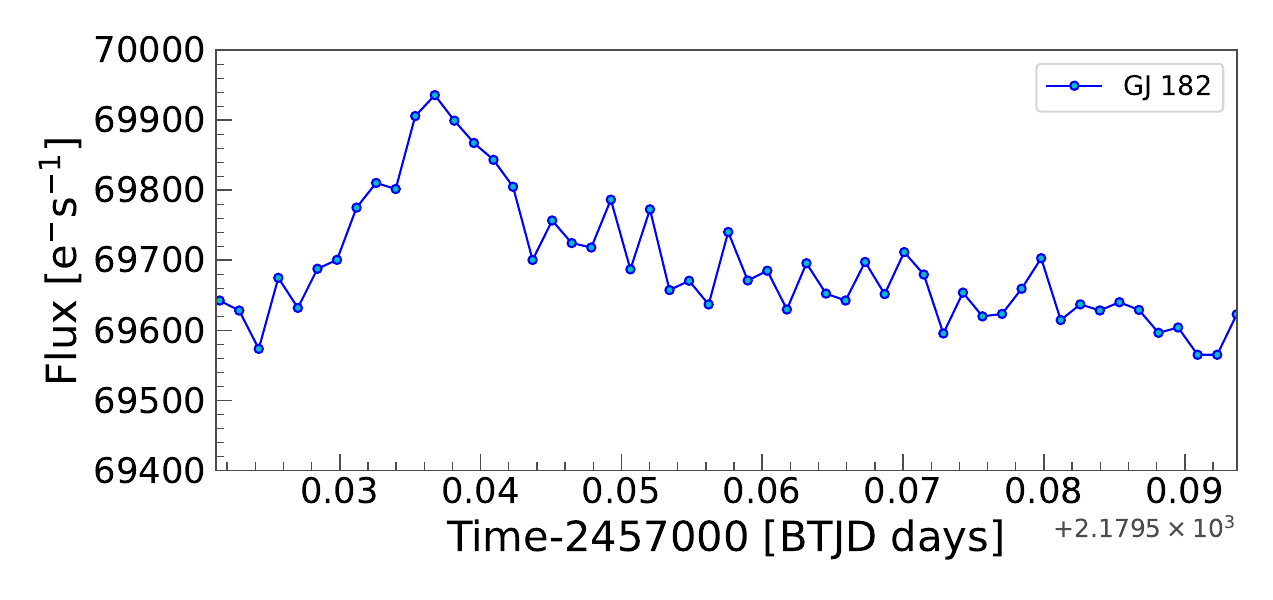}  
    \quad\includegraphics[width=.31\linewidth]{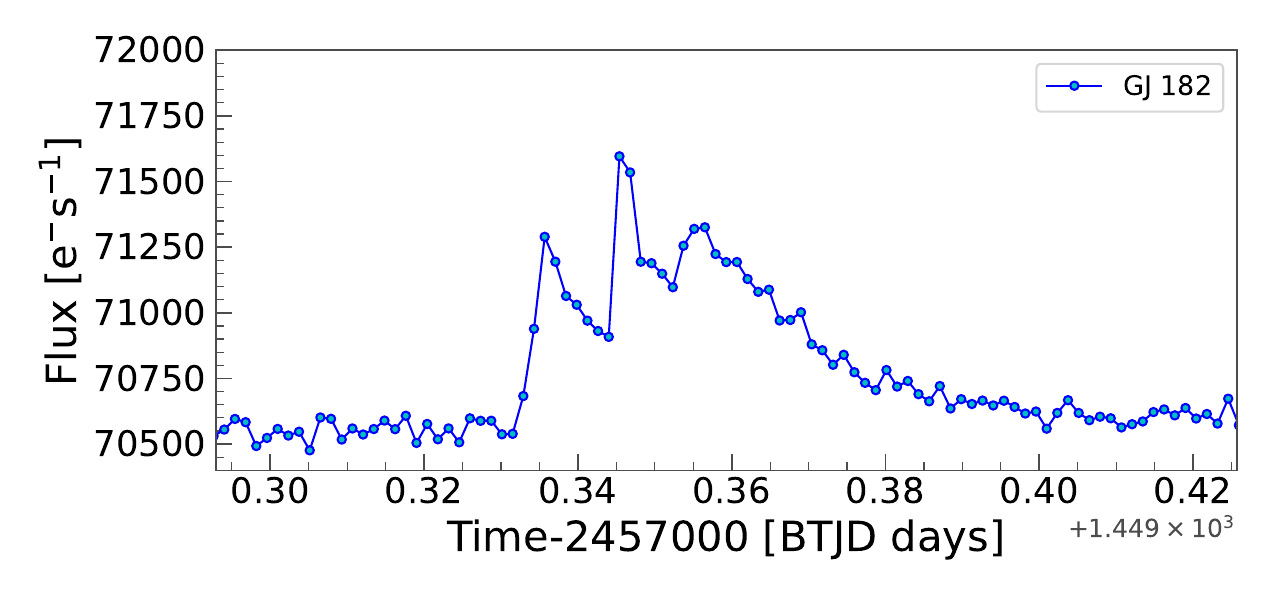}
    \caption{ Examples of rise phase complexity flare events with a complex nature in the rise phase are shown here.}
    \label{fig:rise_phs_comp_top}
\end{figure}

\begin{figure}
    \centering
    \includegraphics[width=.31\linewidth]{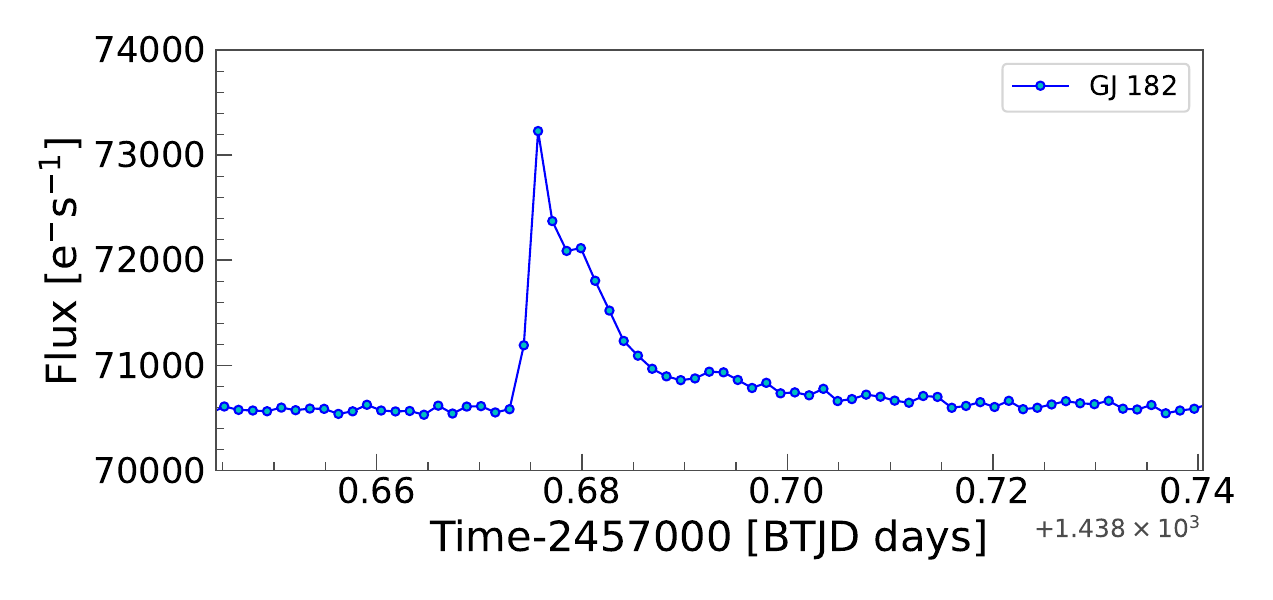}
    \quad\includegraphics[width=.31\linewidth]{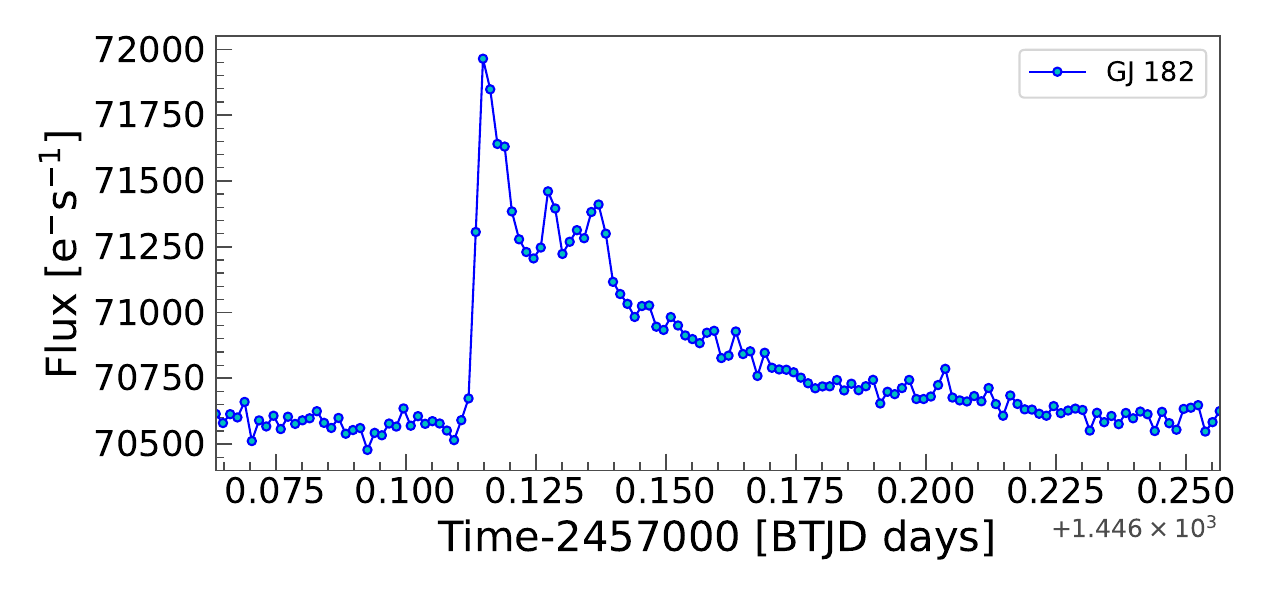}
    \quad\includegraphics[width=.31\linewidth]{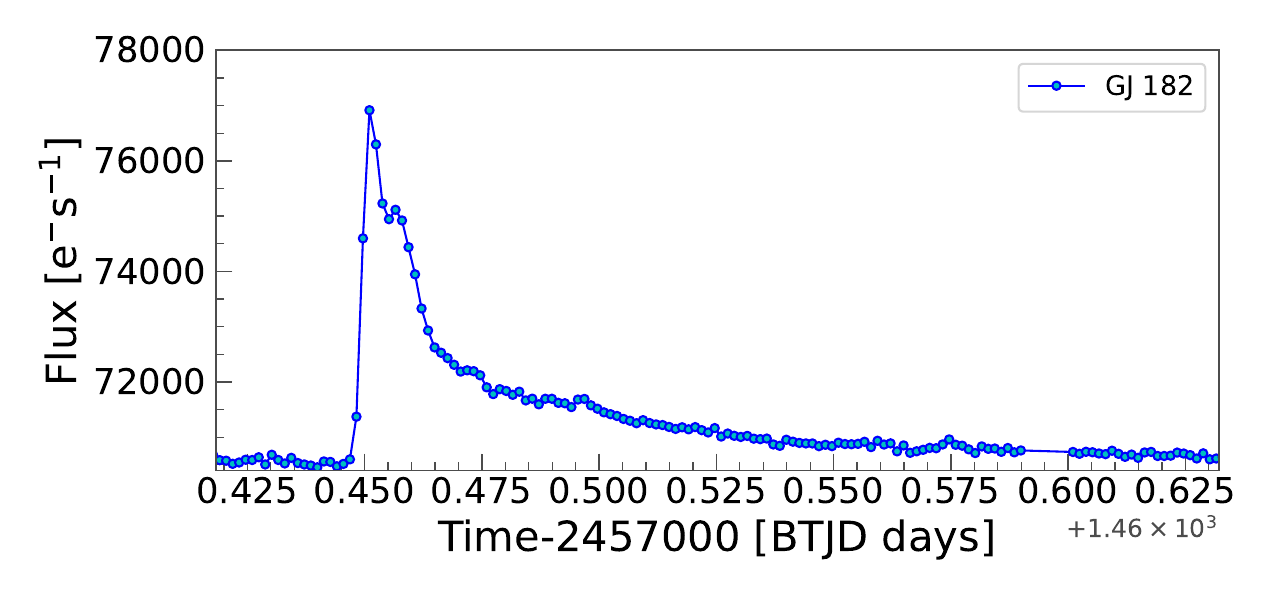}
    \caption{ Nearly 17 out of 48 flare events show complex substructure during decay. They are categorized as peak-bump flares.  For example we have shown three of them.}
    \label{fig:flr_peak_bump}
\end{figure}

\begin{figure}
    \centering
    \includegraphics[width=.31\linewidth]{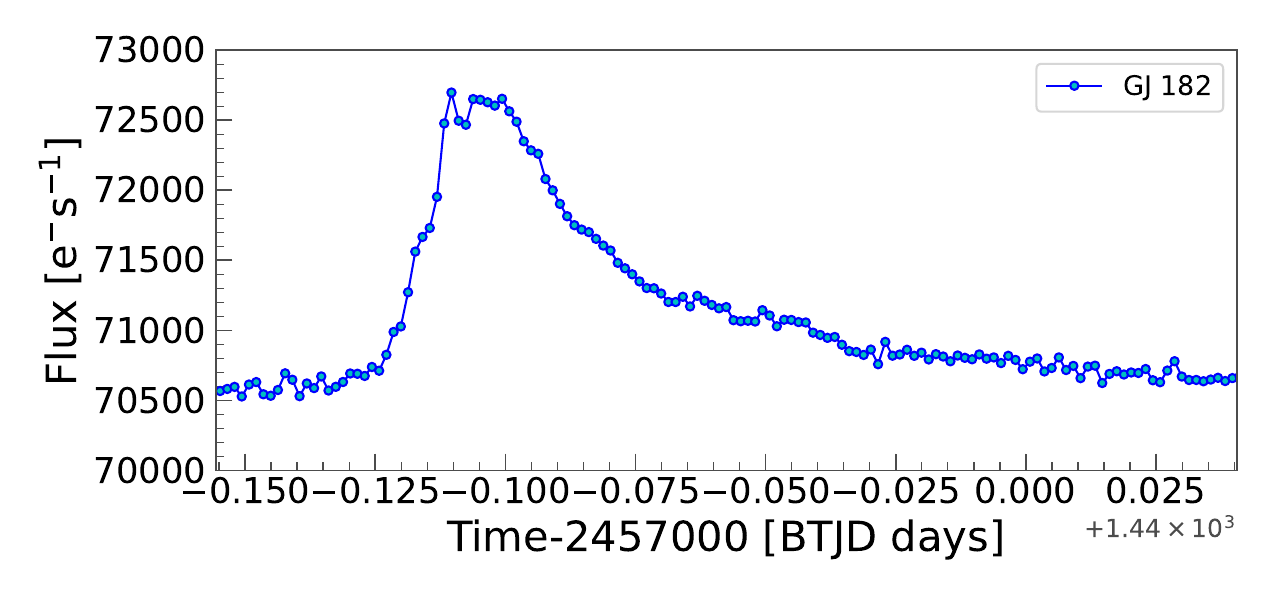}
    \caption{ Only one flare has a flat-top structure with a high level of impulsive rise phase and before decaying there is a constant emission level at the peak brightness.}
    \label{fig:flr_flat_top}
\end{figure}

\begin{figure}
    \centering
    \includegraphics[width=.31\linewidth]{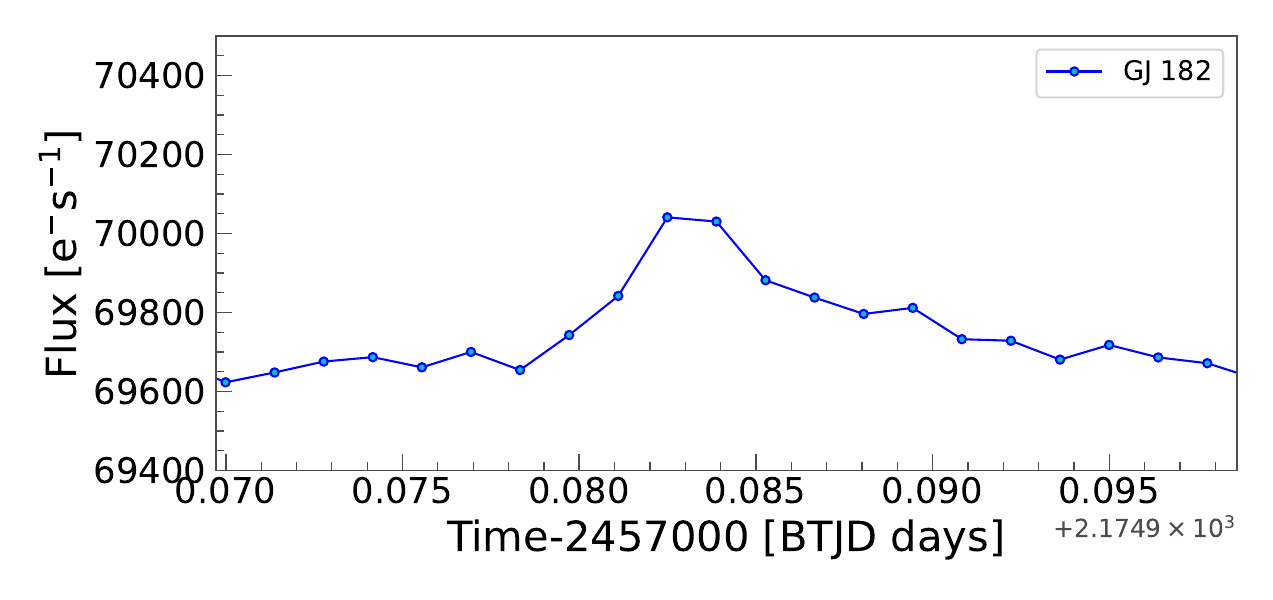}
    \quad\includegraphics[width=.31\linewidth]{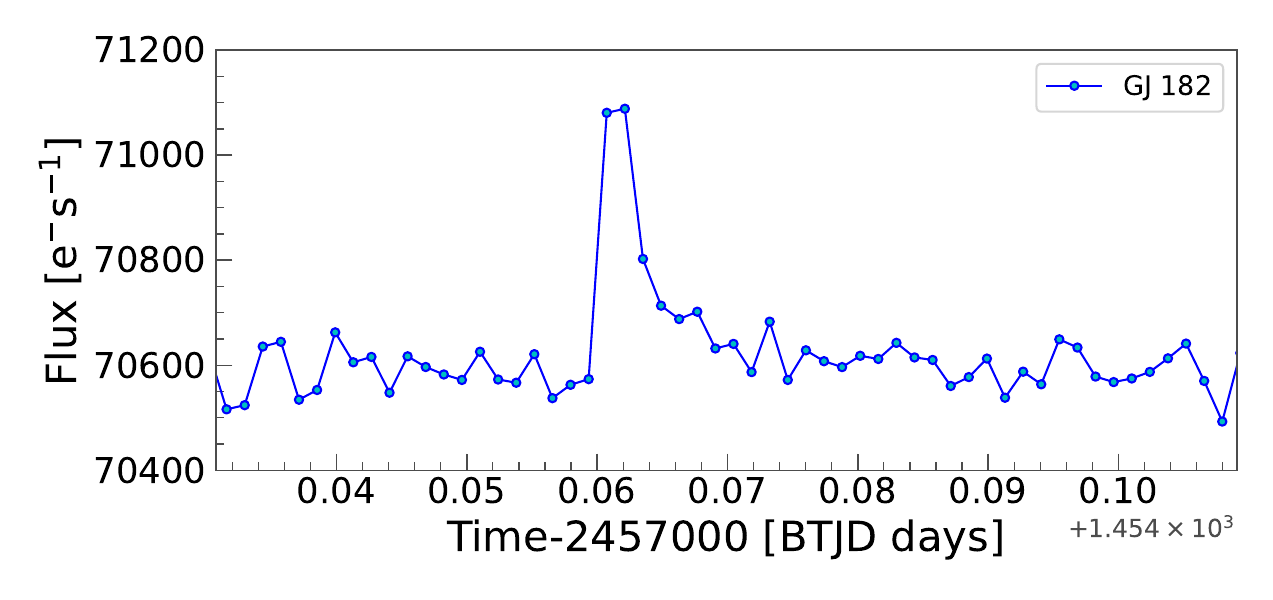}
    \quad\includegraphics[width=.31\linewidth]{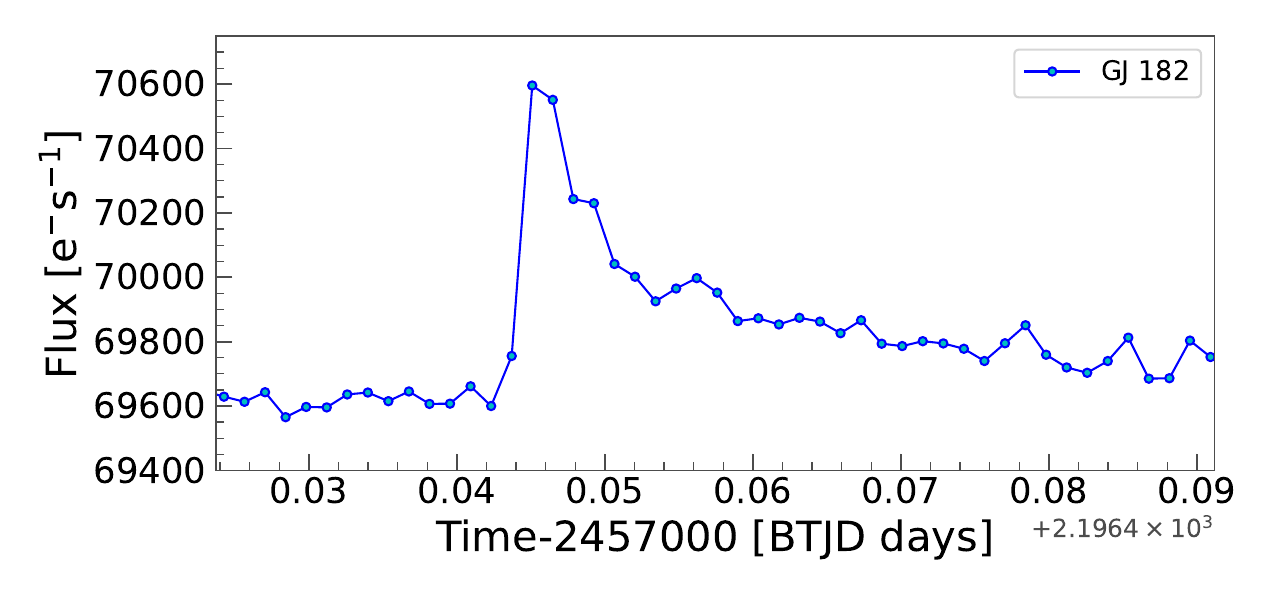}
    \quad\includegraphics[width=.31\linewidth]{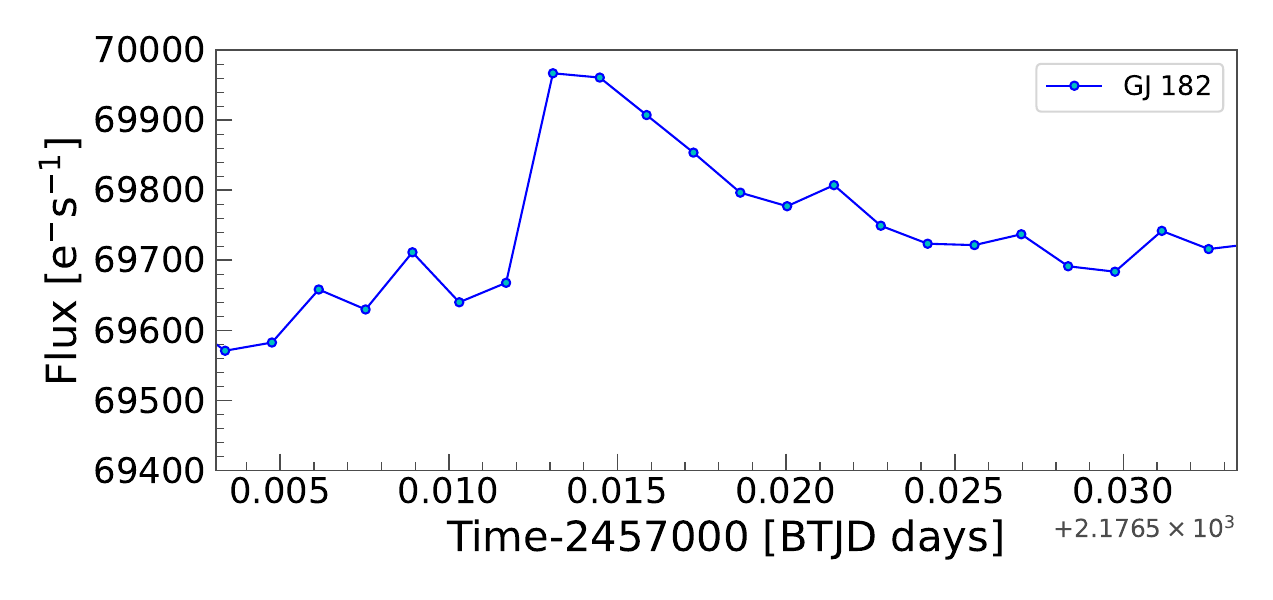}
    \quad\includegraphics[width=.31\linewidth]{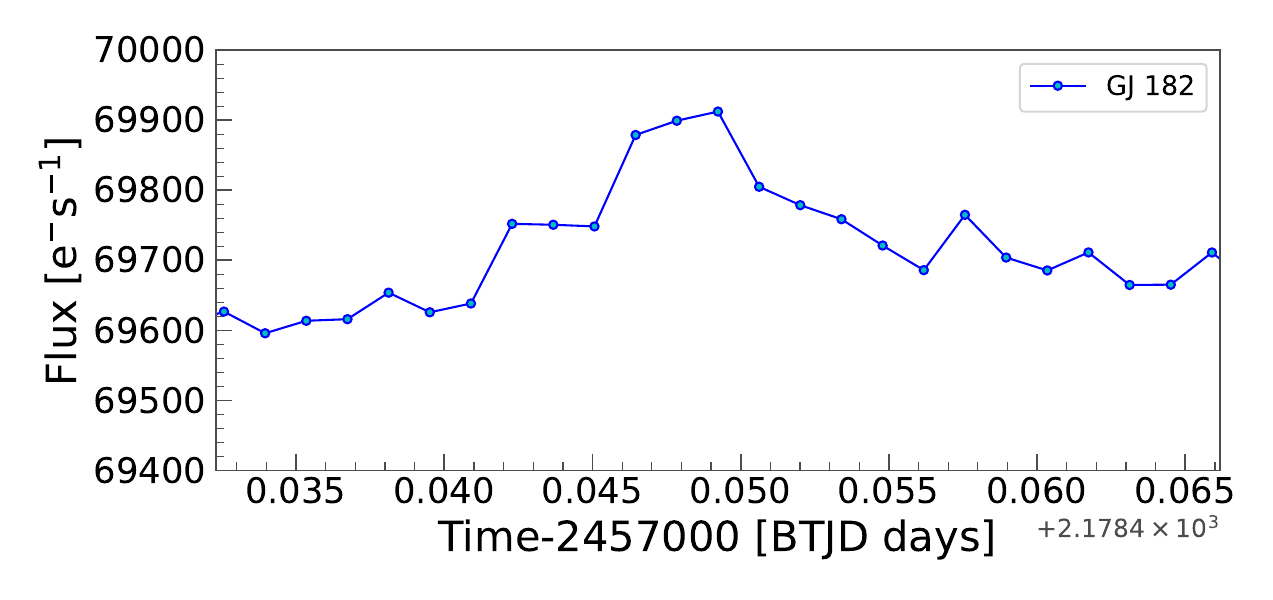}
    \caption{ A few flares have insufficient temporal resolution.}
    \label{fig:temp_res_flr}
\end{figure}

\bibliography{sample631}{}
\bibliographystyle{aasjournal}

\end{document}